\documentclass[12pt]{article}

\usepackage[utf8]{inputenc}
\usepackage[T1]{fontenc}

\usepackage{setspace} 
\usepackage[vmargin = 1.25in, hmargin =1.25in]{geometry}

\usepackage[usenames,dvipsnames,svgnames]{xcolor}
\usepackage{amsmath, amssymb, amsfonts, graphicx, tikz,pdflscape, mathtools, amsthm, upgreek, bm,pgfplots,csvsimple,multirow, multicol, booktabs}
\usepackage{verbatim}
\usepackage[longnamesfirst]{natbib}
\setcitestyle{citesep={,}}
\usepackage{accents}
\newcommand{\ubar}[1]{\underaccent{\bar}{#1}}
\usepackage{needspace}

\usepackage{comment}
\usepackage[inline]{enumitem}
\usepackage{tocvsec2}
\usepackage{threeparttable}
\usetikzlibrary{calc, cd, arrows}
\usepackage[skip = 10pt]{caption}
\allowdisplaybreaks[1]
\allowdisplaybreaks[1]

\usepackage{array,booktabs,tabularx}
\newcolumntype{Y}{>{\raggedright\arraybackslash}X}

\definecolor{Blue}{RGB}{86,180,233}
\definecolor{Orange}{RGB}{230,159,0}
\definecolor{Green}{RGB}{0,158,115}
\definecolor{GmailBlue}{RGB}{42, 93, 176} 
\usepackage[
	pagebackref,
	colorlinks=true,
	citecolor= GmailBlue,
	linkcolor=GmailBlue,
	urlcolor = GmailBlue
]{hyperref}

\newcommand{\bibtexorder}[1]{}

\usepackage{pgfplots}
\usepgfplotslibrary{groupplots,colorbrewer}
\pgfplotsset{compat=newest}
\pgfplotsset{cycle list/Set1}
\usepackage{tikz}
\usetikzlibrary{matrix,calc,shapes,arrows.meta,positioning}
\tikzset{
    vertex/.style = {shape=circle,draw, minimum size = 1.8em, inner sep = 0pt},
    edge/.style = {->,> = latex}
}

\usepackage[capitalize,noabbrev]{cleveref}

\newtheoremstyle{break}
{}
{}
{\itshape}
{}
{\bfseries}
{}
{\newline}
{}

\theoremstyle{break}
\newtheorem{thm}{Theorem}
\newtheorem*{theorem*}{Theorem}
\newtheorem*{cor*}{Corollary}

\newtheorem{prop}{Proposition}
\newtheorem{lem}{Lemma}

\crefname{prop}{Proposition}{Propositions}
\crefname{thm}{Theorem}{Theorems}
\crefname{lem}{Lemma}{Lemmas}
\crefname{blem}{Lemma}{Lemmas}
\crefname{as}{Assumption}{Assumptions}

\theoremstyle{definition}

\newtheorem{rem}{Remark}
\newtheorem*{rem*}{Remark}
\newtheorem{claim}{Claim}

\newtheorem{as}{Assumption}

\def\a{\alpha}
\def\b{\beta}
\def\g{\gamma}
\def\d{\delta}

\def\h{\eta}
\def\th{\theta}

\def\k{\kappa}
\def\l{\lambda}

\def\f{\varphi}

\def\D{\Delta}

\def\R{\mathbf{R}}

\def\AA{\mathcal{A}}

\def\P{\mathbf{P}}

\DeclareMathOperator{\E}{\mathbb{E}}
\DeclareMathOperator*{\argmax}{argmax}

\DeclareMathOperator*{\miz}{minimize}
\DeclareMathOperator*{\argmin}{argmin}

\DeclareMathOperator{\cav}{cav}  

\DeclareMathOperator{\WS}{WS}
\DeclareMathOperator{\SW}{SW}
\DeclareMathOperator{\SWS}{SWS}

\newcommand{\Paren}[1]{\left( #1 \right)}

\newcommand{\Brac}[1]{\left[ #1 \right]}

\newcommand{\Set}[1]{\left\{ #1 \right\}}

\newcommand{\de}{\mathop{}\!\mathrm{d}}

\usepackage{datetime}
\newdateformat{specialdate}{\THEDAY~\monthname[\THEMONTH] \THEYEAR}

\title{Should the Timing of Inspections be Predictable?\thanks{We thank Yifan Dai for excellent research assistance. For comments, we thank Nageeb Ali, Dirk Bergemann, Olivier Compte, Brett Green, Deniz Kattwinkel, Nicolas Lambert, Lucas Maestri, Stephen Morris, Teemu Pekkarinen, Jacopo Perego, Jo\~{a}o Ramos, Ron Siegel, Ludvig Sinander, Curtis Taylor, and Alex Wolitzky. For helpful feedback, we thank audiences at VSET, Stanford, SMYE Orleans, ESEM,  Aalto, ESSET Gerzensee, UT Austin, Georgetown, BFI Economic Theory Conference, Penn State, Columbia Economic Theory Conference, Emory, Boston University, City University of Hong Kong, Carnegie Mellon/Pittsburgh, UCL, USC, Essex, Paris, LSE, Oxford, Queen Mary, EC, SAET, Warwick, Boston College, Tel Aviv, Ben Gurion, Hebrew University, and Bonn. Knoepfle acknowledges financial support from the Academy of Finland (Project  325218).}}

\date{\specialdate\today}

\author{Ian Ball\thanks{Department of Economics, MIT, \texttt{ianball@mit.edu}.} \and Jan Knoepfle\thanks{School of Economics and Finance, Queen Mary University of London, \texttt{j.knoepfle@qmul.ac.uk}.}}

\begin{document}


\maketitle 

\vspace{-0.5cm}

\begin{abstract}
A principal hires an agent to work on a long-term project that culminates in a breakthrough or a breakdown. At each time, the agent privately chooses to work or shirk. Working increases the arrival rate of breakthroughs and decreases the arrival rate of breakdowns. To motivate the agent to work, the principal conducts costly inspections. 
She fires the agent if shirking is detected.
We characterize the principal's optimal inspection policy. Predictable inspections are optimal if work primarily generates breakthroughs. Random inspections are typically optimal if work primarily prevents breakdowns. Crucially, the agent's actions affect the curvature of the project's survival probability, which determines his induced risk attitude over the timing of planned inspections. 
\end{abstract}


\newpage

\section{Introduction} \label{sec:introduction}

Inspections are widely used to provide incentives within long-term relationships. Venture capital investors conduct financial audits to ensure that entrepreneurs do not divert funds for their private benefit.\footnote{In 2017, the four accounting firms Deloitte, EY, KPMG, and PwC earned over 47 billion dollars from auditing alone \citep{economist2018audit}.} 
Research grants are extended only after researchers pass intermediate reviews. Firms and workers are inspected to ensure their compliance with health, safety, and environmental regulations. 

Some inspections are conducted at pre-announced times; others are surprises. In this paper, we study the optimal timing of costly inspections in a dynamic moral hazard setting. We show how the nature of the agent's task and the inspection technology determine whether predictable or random inspections are optimal. Under our main inspection technology, we establish the following. 
If the agent's main task is \emph{innovation}---think of an entrepreneur in a start-up working toward a technological breakthrough---then predictable inspections are always optimal. 
If the agent's main task is \emph{maintenance}---think of a worker following safety guidelines to prevent an accident (breakdown)---then random inspections are typically optimal.

We analyze the following continuous-time model. A principal hires an agent to work on a long-term project that culminates in a breakthrough or a breakdown. At each instant, the agent privately works or shirks.\footnote{In some applications, shirking represents an unproductive (or even fraudulent) activity, such as diverting funds.} Work increases the arrival rate of breakthroughs and decreases the arrival rate of breakdowns. During the project, the agent's flow payoff depends on whether he works or shirks. At the end of the project, the agent's continuation payoff depends on whether the project ended in a breakthrough or a breakdown. To make the inspection problem nontrivial, we assume that in the absence of inspections, the agent strictly prefers shirking to working. 

The principal commits to the timing of costly inspections. Each inspection yields a binary result---pass or fail---that (partially) reveals the agent's past actions, as described in detail below. 
If the agent fails an inspection, the principal terminates the project. 
We solve for the cheapest inspection policy that induces the agent to work continuously until the project ends (in a breakthrough or breakdown). 

If the principal inspects more often, then the agent has stronger incentives to work: shirking will be detected sooner, and the project will be terminated before the agent can enjoy shirking for very long. But inspecting more often is more costly for the principal, so it is optimal for the principal to inspect just often enough that the agent is willing to work. The question remains whether the timing of these inspections should be predictable or random. 

The inspection technology is as follows. Shirking by the agent leaves behind evidence stochastically, and this evidence is hidden to both players until it is publicly revealed at the next inspection. Once the agent begins shirking, he knows that there is some chance that his shirking has already left behind evidence. If evidence has been left behind, then the agent will fail the next inspection, whether or not he continues shirking. Thus, past shirking makes it more attractive to continue shirking. For this reason, the agent is tempted to \emph{globally} deviate by shirking for a positive duration. In practice, the appeal of these deviations is illustrated by the trend of ``quiet quitting.''\footnote{\label{ft:qq}A 2022 Gallup poll found that half of US workers are ``quiet quitters'' \citep{Gallop2022}. A partner at McKinsey describes the mindset of a quiet quitter as follows: ``My boss doesn't check in on me. HR takes six months to do a write-up. I bet I can stay in this gig for two years and not do much.'' \citep{mckinsey2022}.} 

The optimal inspection policy is designed to most cheaply deter certain global deviations. Crucially, global deviations affect the probability that the project ends (in a breakthrough or breakdown) before each planned inspection is carried out. The principal's expected cost from a planned inspection depends on the probability that the project survives until the inspection \emph{when the agent works}; the incentive power of the planned inspection depends on the probability that the project survives until the inspection \emph{when the agent deviates}. The project's survival probability is a decreasing function of time, and this function is more convex (in the Arrow-Pratt sense) if the project's hazard rate is higher. The effect of the agent's actions on the curvature of the survival probability determines whether random or predictable inspections maximize incentive power at a given expected inspection cost.

To cleanly illustrate the central force in the model, we first derive the principal's optimal policy when the inspection technology is \emph{perfect}. That is, each inspection perfectly reveals whether the agent has previously shirked.

If the agent's main task is innovation---work speeds up breakthroughs by more than it delays breakdowns---then it is optimal for the principal to inspect \emph{periodically}. That is, the time between consecutive inspections is constant (\cref{res:speeding_up_perfect}). Consistent with this result, venture capitalists commonly disburse funds in stages after checking periodically that the entrepreneur has invested previous funds appropriately \citep[p.~139]{gompers1999venture}. 
In the innovation regime, shirking prolongs the project by delaying breakthroughs. Thus, shirking makes the project's survival probability \emph{less convex} as a function of time. In this case, conducting each inspection at a nonrandom time is the cheapest way for the principal to deter the agent from shirking. 

Conversely, if the agent's main task is maintenance---work delays breakdowns by more than it speeds up breakthroughs---then it is optimal for the principal to conduct inspections \emph{randomly}. Under the optimal policy, inspections are conducted with a constant hazard rate (\cref{res:delaying_perfect}). Consistent with this result, workplace safety inspections, which aim to prevent accidents, are generally conducted without advance notice.
In the maintenance regime, shirking shortens the project by generating breakdowns. Thus, shirking makes the project's survival probability \emph{more convex} as a function of time. In this case, conducting inspections at random times is the cheapest way for the principal to deter the agent from shirking.  


Next, we derive the principal's optimal policy when the inspection technology is \emph{imperfect} in the sense that shirking by the agent does not always leave behind evidence. The longer the agent has shirked since the last inspection, the more likely he is to fail the next inspection. If the agent passes one inspection, then he is likely to pass another inspection conducted soon after, even if he shirks in between. To deter global deviations, it is wasteful for the principal to conduct imperfect inspections in short succession. Thus, the imperfect inspection technology creates a new motive for the principal to space apart inspections.

If the agent's main task is innovation, then it is optimal for the principal to inspect periodically (\cref{res:imperfect_inspections}.\ref{res:speeding_up_imperfect}), as in the case of the perfect inspection technology.  Indeed, periodic inspections are already spaced apart, so the imperfect inspection technology creates an additional force toward periodicity. If the agent's main task is maintenance but the rate at which work reduces breakdowns is below a threshold, then this spacing-apart motive dominates and periodic inspections remain optimal.

Conversely, if the agent's main task is maintenance and the rate at which work reduces breakdowns is above a threshold, then randomization is optimal (\cref{res:imperfect_inspections}.\ref{res:delaying_imperfect}). The optimal policy leverages the benefits of randomization while also spacing apart inspections. After each inspection, there is a fixed period without inspections.  At the end of this period, there is a positive probability that the next inspection is conducted immediately. Otherwise, the next inspection is conducted with a constant hazard rate thereafter. Once the inspection is conducted, the cycle repeats, beginning with an inspection-free period. An inspection policy with a similar form is used in an anti-corruption program in Brazil. Municipalities are randomly chosen for audit, but only after a fixed time has passed since their last audit \citep{AvisEtal2018}.


\begin{figure}[t]
\centering

\resizebox{\textwidth}{!}{%
\begin{tikzpicture}[
    scale=1,
    every node/.style={font=\small},
    policy/.style={align=center, font=\small},
    labelnode/.style={align=center, font=\small\bfseries}
]

\def\xleft{0}
\def\xarrow{2.2}
\def\xperf{5.2}
\def\ximperf{10.4}
\def\xsepone{2.6}
\def\xseptwo{7.8}
\def\ytop{1.15}
\def\ybot{-1.15}
\def\yhead{1.75}
\def\yheadrule{1.42}
\def\ymidperf{0}
\def\ymidimperf{-0.4}

\draw[gray!45] (\xsepone,\yhead+0.20) -- (\xsepone,\ybot-0.30);
\draw[gray!45] (\xseptwo,\yhead+0.20) -- (\xseptwo,\ybot-0.30);

\draw[gray!45] (\xleft-1.35,\yheadrule) -- (\ximperf+2.5,\yheadrule);

\node[labelnode] at (\xperf,\yhead) {Perfect technology};
\node[labelnode] at (\ximperf,\yhead) {Imperfect technology};

\def\xtext{\xleft-1.35}

\node[align=left, font=\small\bfseries, anchor=west] 
    at (\xtext,\ytop-0.2) {Innovation};
\node[align=left, font=\scriptsize, anchor=west] 
    at (\xtext,\ytop-0.75) {grant/funding reviews};

\node[align=left, font=\small\bfseries, anchor=west] 
    at (\xtext,\ybot+0.55) {Maintenance};
\node[align=left, font=\scriptsize, anchor=west] 
    at (\xtext,\ybot) {health/security inspections};


\draw[thick] (\xleft-1.35,\ymidperf) -- (\xseptwo,\ymidperf);
\node[policy] at (\xperf,0.75) {Periodic inspections};
\node[policy] at (\xperf,-0.75) {Random inspections};

\draw[thick] (\xseptwo,\ymidimperf) -- (\ximperf+2.5,\ymidimperf);
\node[policy] at (\ximperf,0.75) {Periodic inspections};
\node[policy] at (\ximperf,-1) {Delayed random inspections};

\end{tikzpicture}%
}
\caption{Optimal inspection policies according to the task and the inspection technology}
\label{fig:inspection-cases}
\end{figure}

\cref{fig:inspection-cases} summarizes the structure of the solution in each of the main cases. 

Finally, we generalize the inspection technology to allow the agent to recover from past shirking by working before an inspection. This technology nests the inspection technology in our main model and the reputation technology in \cite{board2013reputation}. If the recovery rate is small, then for a range of parameters, the optimal inspection policy is the same as under the main specification. 
If the recovery rate is sufficiently large, then it is optimal to inspect the agent with a constant hazard rate (\cref{res:high_recovery_exponential}), regardless of whether the agent's task is innovation or maintenance. Intuitively, randomization is necessary to deter the agent from shirking for a while and then returning to work shortly before the next inspection. This extension illustrates how the optimal inspection policy depends on the interaction between the inspection technology and the nature of the agent's task.

The rest of the paper is organized as follows. \cref{sec:lit} discusses related literature.  \cref{sec:model} presents the model and gives a few leading interpretations. \cref{sec:no_inspections} studies the agent's behavior without inspections. \cref{sec:recursive} formulates the principal's inspection problem recursively. Next, we solve for the optimal policy with perfect inspections (\cref{sec:perfect_inspections}) and imperfect inspections (\cref{sec:imperfect_inspections}). In \cref{sec:recovery}, we consider an inspection technology that allows for recovery. The conclusion is in \cref{sec:conclusion}. The main proofs are in \cref{sec:proofs}. Additional results and proofs are in the online appendices (\cref{sec:online_appendix_results,sec:online_appendix_proofs}). 

\subsection{Related literature} \label{sec:lit}


Our paper considers inspections that reveal information about past actions. We study the optimal timing of these inspections for motivating an agent whose actions affect the project's survival probability. 




In much of the literature on the optimal allocation of monitoring resources, monitoring reveals the agent's \emph{current} action. In \cite{Lazear06}, the agent is punished if he is shirking at the moment he is monitored, so the agent's incentive to work depends only on the current monitoring intensity.\footnote{Most of the subsequent work on dynamic contracts analyzes monitoring of current actions; see \cite{ANTINOLFI2015105}, \cite{piskorski2016optimal}, \cite{chen2020optimal}, \cite{li2020optimal}, \cite{dai2022dynamic}, \cite{rodivilov2022monitoring},  \cite{wong2022dynamic}, and \cite{solan2021dynamic, solan2023not}.  In \cite{halac2016managerial} and \cite{dilme2015}, the principal's investment has a persistent effect on her monitoring capabilities, but monitoring still reveals information about current actions only. In dynamic adverse selection problems, the monitored state is distributed independently across periods in \citet{chang1990dynamic}, \cite{monnet2005optimal}, \cite{wang2005dynamic}, \cite{popov2016stochastic}, \cite{Malenko2013}, and \cite{li2023dynamics}; the state is serially correlated in \cite{ravikumar2012optimal} and \cite{Kim2015}.} As a result, the cheapest way for the principal to motivate the agent to work over a given period is to monitor with constant intensity over that period, thus creating constant incentives.\footnote{\cite{EeckhoutEtal2010} extend this insight to a setting with unobservable heterogeneity in agent payoffs. They test the theory using data on police crackdowns.}

Closer to our paper is the literature on inspections that reveal information about an agent's \emph{past} actions \citep{Varas2020,wagner2021relational,achim2024tension}. A key building block is the tractable model of firm reputation introduced in \cite{board2013reputation}. In their model, a firm's effort determines the Poisson transition rates of a binary \emph{quality state}. Consumers receive exogenous Poisson signals of quality and update their common belief about firm quality; this belief, interpreted as the firm's reputation, determines the firm's flow payoffs. \cite{Varas2020} endogenize the market's information by allowing the principal to design the timing of inspections that publicly reveal the quality state.\footnote{One difference is that in \cite{Varas2020}, even under maximal effort, the state is stochastic. This ensures that inspections reveal information even if the agent exerts maximal effort throughout.} This quality state is directly payoff-relevant: inspections create informational value by allocating consumers more efficiently. These inspections also indirectly motivate the inspected firm to exert effort to improve quality. 

\cite{Varas2020} solve for the inspection policy that maximizes consumer welfare net of the inspection costs, subject to the constraint that the policy induces continuous effort by the firm.  Under their inspection technology, the binding incentive constraints are local, so the cheapest way to induce continuous effort is to inspect with a constant hazard rate (see our \cref{res:high_recovery_exponential}). 
This creates a tension because spacing inspections apart generates greater informational value.

Our model isolates the incentive role of inspections. Each inspection reveals the \emph{evidence state}, which is not directly payoff-relevant but serves as the basis for punishing the agent. 
Unlike \cite{Varas2020}, our model incorporates breakthroughs and breakdowns. Moreover, under our inspection technology, the binding deviations are global. This is natural in applications where the central concern is extended noncompliance, e.g., diverting a substantial share of provided funding.
When the agent globally deviates, the nature of the task determines how the curvature of the project's survival probability changes. This in turn determines whether predictable or random inspections most cheaply motivate the agent to work.\footnote{\cite{Eilat2023} consider predictable and random inspections in a different context.  They study a game between two agents who choose when to pay a cost to check whether an opportunity has arrived. Depending on the payoff parameters, equilibrium may feature predictable or random inspections.} 

 A number of papers study monetary incentives for agents working toward breakthroughs, without inspections: \cite{BEHE98, BEHE05, HOSA13,Green2016}, and \cite{HAKALI16, halac2017contests}.\footnote{In \cite{manso2011motivating} and \cite{klein2016importance}, the agent chooses between an unknown technology and a safe technology, which generates successes at a lower but certain rate. To encourage use of the unknown technology, it is optimal to reward later success and, potentially, early failures. For an analysis of incentives in the presence of breakdowns, without inspections, see \cite{keller2015breakdowns, bonatti2017learning, horner2022freerider}, and \cite{klein2022strategic}.} 
A recurring finding is that a deterministic deadline is optimal.\footnote{In \cite{Green2016}, the agent has to complete two breakthroughs, and the deadline for the second breakthrough is deterministic. The deadline for the first breakthrough, however, is random. This randomness encourages the agent to truthfully report intermediate progress, which is privately observed by the agent. In our model, breakthroughs and breakdowns are public.} The deadline punishes the agent for not achieving a breakthrough soon enough, thus motivating the agent to work. We show that \emph{inspections} should be deterministic if the agent's task is innovation. Despite the similarity of these conclusions, the mechanism is different. At a deadline, the project ends regardless of the agent's action history; after an inspection, the project ends only if the agent fails, which depends on the action history. To illustrate this difference, we solve an auxiliary optimal deadline problem in \cref{sec:deadlines}, holding fixed the agent's compensation; deterministic and exponential deadlines are both optimal.

Finally, a central force in our model is that the agent's action path affects the curvature of the project's survival probability. This curvature determines the agent's induced risk attitude over the timing of planned inspections.\footnote{To be clear, our results are not driven by primitive time preferences. See \cite{dejarnette2020time} and \cite{DillenbergerEtal2025} for an axiomatic analysis of the connection between impatience and risk preferences over time-lotteries.} A similar force arises in several other contexts.  When designing information for an agent in a stopping problem, the relative patience of the principal and the agent determines whether disclosures are spread out over time \citep{ElySzydlowski2020,Liu2026,saeedi2024getting,chen2026accelerator}. In an adverse selection problem in which the agent's discount rate is his private information, \cite{OrtolevaSafonovYariv2022} show how the agent's induced risk preferences can be used to screen the agent.

\section{Model} \label{sec:model}

\subsection{Setting} \label{sec:setting}

\paragraph{Environment}
Time is continuous and the horizon is infinite. There are two players: a principal (she) and an agent (he).
The principal hires the agent to work on a project. During the project, the agent privately chooses at each time $t$ in $[0, \infty)$ whether to work ($a_t = 1$) or shirk ($a_t  = 0$). The principal commits to the timing of costly inspections. Each inspection reveals information about the agent's past actions, as described below. 

The project ends if there is a public breakthrough or a public breakdown, which arrive independently at Poisson rates 
\[
   a_t \l_G  \quad \text{and} \quad (1 - a_t)\l_B,
\]
where $\l_G$ and $\l_B$ are nonnegative parameters. The subscripts abbreviate \emph{good} (for a breakthrough) and \emph{bad} (for a breakdown). A breakthrough can arrive only when the agent is working, and a breakdown can arrive only when the agent is shirking.\footnote{This assumption can be relaxed. If the breakthrough and breakdown rates were instead $\ubar{\l}_G + \l_G a_t$ and $\ubar{\l}_B + \l_B (1 - a_t)$, then we could incorporate the baseline arrival rates by redefining the discount rate to be $r + \ubar{\l}_G + \ubar{\l}_B$. (We would also redefine the flow utility to $\bar{u}(a_t) + \ubar{\l}_G W_G + \ubar{\l}_B W_B$; the payoff  primitives are defined below.)} The principal can also terminate the project at any time prior to a breakthrough or a breakdown. The game ends when the project ends---in a breakthrough, in a breakdown, or by termination. 

\paragraph{Inspection technology} \label{sec:inspection_tech} There is an evolving \textit{evidence state} $\th_t \in \{0,1\}$ that is hidden to both players. The current state is publicly revealed whenever the principal conducts an inspection. Initially, $\th_0 = 0$. While in state $0$, transitions to state $1$ occur at Poisson rate $(1 - a_t) \d$, independently of breakthroughs and breakdowns. State $1$ is absorbing. Our interpretation is that the state $\th_t$ indicates whether the agent's past shirking has left behind evidence. This evidence is uncovered only at an inspection. Since state $1$ is absorbing, evidence does not disappear.\footnote{In \cref{sec:recovery} we allow for transitions from state $1$ to state $0$.} The detectability parameter $\d > 0$ measures the rate at which evidence is left behind when the agent shirks. 

Since the state is binary, an inspection has two possible results. Say that the agent \emph{passes} (respectively, \emph{fails}) an inspection if the state is revealed to be $0$ (respectively, $1$). If the agent follows an action path $a = (a_s)_{s \geq 0}$, then the probability $p_t(a)$ that the agent passes an inspection conducted at time $t$ is
\begin{align}\label{eq:passage_prob}
     p_t (a) = \exp \Set{ - \d \int_{0}^{t} (1 - a_s) \de s}.
\end{align}
The passage probability $p_t$ is a decreasing, convex function of the duration of shirking prior to time $t$. 
If the agent fails one inspection, then he will fail all subsequent inspections. Therefore, the agent's conditional probability of passing an inspection at time $t$, given that he passed an inspection at an earlier time $t'$, is 
\[
    \frac{p_t (a)}{p_{t'} (a)} = \exp \Set{ -\d \int_{t'}^{t} (1 - a_s) \de s }.
\]
This conditional probability depends only on the duration of shirking between times $t'$ and $t$. 

\paragraph{Payoffs} The principal and the agent discount future payoffs using the exponential discount factor $e^{-r t}$, where $r > 0$. At each time $t$ while the project continues, the agent receives flow utility $\bar{u} (a_t)$, which depends on his current action $a_t$. A breakthrough yields continuation payoff $W_G$, and a breakdown yields continuation payoff $W_B$.\footnote{We assume that the project ends when a breakthrough or breakdown arrives. The continuation values $W_G$ and $W_B$ could capture the agent's payoffs in whatever continuation game results after a breakthrough or a breakdown. The key assumption is that the payoffs in the continuation game depend only on whether the project ended in a breakthrough or breakdown, not on the current evidence state. This holds in particular if, after a breakthrough or breakdown, oversight becomes unnecessary or is governed by a new contractual agreement.} If the agent is terminated, he gets his outside option continuation payoff, which is normalized to $0$.\footnote{That is, the flow payoff $\bar{u}(a_t)$ and the continuation payoffs $W_G$ and $W_B$ are defined relative to the outside option.} Every time the principal inspects the agent, she pays a lump-sum cost, normalized to $1$. 

We study the principal's cost-minimization problem: What is the cheapest policy that induces the agent to work continuously on the project? Throughout, we will refer to an inspection policy that solves this problem as \textit{optimal} or \textit{cost-minimizing}. This cost-minimization problem is the problem faced by a principal who must ensure compliance by the agent and whose only incentive tool is inspection.

\subsection{Applications} \label{sec:Applications}

Our stylized setting captures a range of applications, with different interpretations of the payoff parameters $\bar{u}(0)$, $\bar{u}(1)$, $W_G$, and $W_B$. In particular, these parameters can capture incentive schemes with time-invariant wages and time-invariant bonuses. 

\paragraph{R\&D funding}
The principal is a venture capitalist or a government agency that provides funds to the agent, an entrepreneur or researcher, to carry out an R\&D project. At each time $t$, the agent chooses whether to invest  ($a_t = 1$) or divert ($a_t = 0$) the funds. Diversion could mean using the funds for private benefits, such as luxurious travel, or for other projects outside the scope of the agreement with the funding body. Let $\varphi$ denote the agent's flow benefit from diversion. The inspected state $\th_t$ represents whether the financial records contain evidence of fund diversion. 
 There are no breakdowns ($\l_B = 0$). Breakthroughs arrive at rate $\l_G > 0$ if and only if the agent works.
Since $\l_G > \l_B$, the task is innovation. When the agent achieves a breakthrough, he receives a reward of $R > 0$, stemming from financial or reputational gains.\footnote{Instead of a lump-sum reward, a breakthrough may lead to a promotion or new job opportunity with expected flow wage $w_G$ in perpetuity. This can be captured by setting $W_G =  w_G /r $.} 
This application fits in our setting with the following parameters: 
\begin{align*}
    \bar{u}(0) =  \f, \quad \bar{u}(1) = 0 \qquad \text{ and } \qquad  W_G = R, \quad  W_B = 0. 
\end{align*}

In \cref{sec:transfers_inspections}, we consider an extension in which the principal chooses the (financial) reward $W_G$ and the timing of inspections to minimize the total expected incentive cost.  We find that for a range of inspection costs, it is strictly optimal for the principal to motivate the agent using inspections as well as financial rewards. 



\paragraph{Software development}
Suppose the agent is a software contractor or development team whose task is to deliver a particular IT solution to a client, who is the principal. Working means designing the architecture and testing the code. Shirking means neglecting these tasks or reallocating developers to other projects.
The inspected state $\theta_t$ represents whether past shirking has left discoverable evidence such as neglected tickets or missed tests. A breakthrough corresponds to successful delivery of the agreed product. Breakthroughs arrive at rate $\l_G > 0$ if and only if the agent works. There are no breakdowns ($\l_B = 0$). The agent receives a flow payment $w$ while the project continues, incurs cost $c$ for the development effort, and receives a continuation reward $R$ after delivery. Thus,
\begin{align*}
    \bar u(0)=w,
    \quad
    \bar u(1)=w-c,
    \qquad \text{ and } \qquad  W_G = R, \quad  W_B = 0.
\end{align*}


\paragraph{Incentivizing cybersecurity}
The principal is a board that wants an organization to maintain adequate cybersecurity practices. The agent is an IT department or external service provider. 
Working means monitoring logs, patching vulnerabilities, and rotating credentials. Shirking means neglecting these tasks.
The inspected state $\theta_t$ represents whether past negligence has left behind discoverable evidence such as missing log reviews or ignored alerts.
A breakdown corresponds to a breach, ransomware event, or service outage. While the agent shirks, breakdowns arrive at rate $\l_B > 0$. There are no breakthroughs ($\l_G = 0$). Since $\l_B > \l_G$, this task is maintenance. The agent receives a flow wage or service fee $w$ while the relationship continues. Shirking provides a flow benefit of $b$ by saving effort or resources. Thus,
 \begin{align*}
    \bar u(0)=w+b,
    \quad
    \bar u(1)=w,
    \qquad \text{ and } \qquad  W_G = 0, \quad  W_B = 0.
\end{align*}

\paragraph{Workplace safety} The principal is a safety regulator, and the agent is a firm operating a factory that must comply with workplace safety protocols. Compliance prevents workplace accidents but reduces running profits from  $\pi$ to $\pi - c$.
There are no breakthroughs ($\l_G = 0$). If safety protocols are ignored, a workplace accident (breakdown) occurs at rate $\l_B>0$. In case of a breakdown, the firm pays a lump-sum penalty $P$, is closed down for a period $\tau$, and resumes compliant operations afterwards.
Thus,
 \begin{align*}
    \bar u(0)=\pi ,
    \quad
    \bar u(1)=\pi-c,
    \qquad \text{ and } \qquad  W_G = 0, \quad  W_B = - P + e^{-r \tau} (\pi - c)/r.
\end{align*}

\subsection{Principal's problem} \label{sec:principal_problem}

The principal commits to a dynamic, stochastic inspection policy. Formally, an inspection policy is a sequence $\mathbf{T} = (T_n)_{n=1}^{\infty}$ of random variables satisfying $0 < T_1 < T_2 < \cdots$. The $n$-th inspection is conducted at (random) time $T_n$ if and only if the project has not ended by time $T_n$. Whenever the agent fails an inspection, the project is terminated immediately. The interpretation of this termination policy is as follows. Because failing an inspection occurs only off path, it is optimal for the principal to impose the maximal punishment on the agent. Immediate termination is indeed the maximal punishment for the agent under \cref{as:nontrivial}, which we impose below. 

Given an inspection policy $\mathbf{T}$, the agent chooses an action process $A = (A_t)_{t \geq 0}$  adapted to $\mathbf{T}$ with right-continuous paths.\footnote{Formally, $A$ is adapted to the natural filtration generated by the counting process $N_t = |\{ n: T_n \leq t\}|$ associated with $\mathbf{T}$.}
At each time $t$ during the project, the agent takes action $A_t \in \{0,1\}$, which can depend on the realized timing of all previous inspections.\footnote{The agent also observes the results of these inspections, but the game continues only if all previous inspections were passed.} The principal chooses an inspection policy $\mathbf{T}$ to minimize the expected inspection cost, subject to the constraint that it is a best response for the agent to work continuously until the end of the project, i.e., to choose $A_t = 1$ for all $t$. Denote this action process by $A = \mathbf{1}$. 

To state the problem formally, first define the project's survival probability 
\[
S_t(a) = \exp \Set{- \l_G \int_{0}^{t} a_s \de s - \l_B \int_{0}^{t} (1-a_s) \de s}.
\]
In words, $S_t(a)$ is the probability that the project has not yet ended in a breakthrough or a breakdown by time $t$, given action path $(a_s)_{0 \leq s \leq t}$. Next, define the
effective discount factor
\[
    D_t (a) = e^{-rt} S_t(a). 
\]
To simplify notation, set $T_0 = 0$ and $p_0(a) = 1$ for all action paths $a$. Given an inspection policy $\mathbf{T}$, the agent's expected payoff from an action process $A$ adapted to $\mathbf{T}$ is given by
\begin{equation} \label{eq:IC}
   U(A,\mathbf{T}) = \E \Brac{ \sum_{n=1}^{\infty} p_{T_{n-1}}(A)  \int_{T_{n-1}}^{T_n} D_t(A) \Paren{ \bar{u}(A_t) + A_t \l_G W_G + (1 - A_t) \l_B W_B} \de t}.
\end{equation}
The expectation is over the inspection policy $\mathbf{T}$ and the adapted action process $A$. For each realization of $(A,\mathbf{T})$, the expression inside brackets equals the agent's conditional expected utility, where the expectation is over the random inspection results and the random arrival of breakthroughs and breakdowns.\footnote{We use the conditional independence of inspection results and breakthroughs and breakdowns to factor the expectation into the product of the passage probability and the effective discount factor.} In the summation, each term is the agent's expected utility over the inter-inspection interval $[T_{n-1}, T_n)$. The term $p_{T_{n-1}}(A)$, as defined in \eqref{eq:passage_prob}, is the probability that the agent passes the $(n-1)$-th inspection, which implies that the agent passed all earlier inspections as well. Inside $D_t(A)$, the expression $S_t(A)$ serves two distinct roles---as the survival probability,  which scales the flow payoff $\bar{u}(A_t)$, and as part of the probability density over the time of the end of the project, which scales the continuation payoffs. 

The principal chooses an inspection policy $\mathbf{T}$ to minimize the expected discounted inspection cost 
\[
    \E \Brac{ \sum_{n=1}^{\infty} p_{T_{n-1}} (\mathbf{1}) D_{T_n} (\mathbf{1}) }= \E \Brac{ \sum_{n=1}^{\infty}  e^{- (\l_G +r) T_n}},
\]
subject to the constraint that $U(\mathbf{1}, \mathbf{T}) \geq U(A, \mathbf{T})$ for all right-continuous action processes $A$ adapted to $\mathbf{T}$. 


The agent's utility parameters $\bar{u}(0)$, $\bar{u}(1)$, $W_G$, and $W_B$ enter the principal's problem only through the function $U(A, \mathbf{T})$ in \eqref{eq:IC}. To simplify the expression for $U(A, \mathbf{T})$, let
\[
    u(a) = \bar{u}(a) + a \l_G W_G + (1 - a) \l_B W_B. 
\]
In $u(a)$, we sum the agent's flow payoff $\bar{u}(a)$ with the agent's equivalent flow benefit from breakthroughs and breakdowns. Hereafter, we call $u(a)$ the agent's \emph{annuitized} flow payoff. Let $u_1 = u(1)$ and $u_0 = u(0)$. The optimal inspection policy is determined by the six parameters: $\l_G, \l_B, \d, r, u_1, u_0$.

\section{Warm-up: No inspections} \label{sec:no_inspections}

If the principal does not conduct inspections, then the agent's expected payoff from an action path $a$ is 
\[
    \E \Brac{ \int_{0}^{\infty} D_t(a) u(a_t) \de t }.
\]
Without inspections, the agent's problem is stationary. We show that the agent has a stationary best response---either always working or always shirking is optimal. Working until the project ends ($a = \bm{1}$) and shirking until the project ends ($a = \mathbf{0}$) respectively yield expected payoffs 
\[
    U_1 \coloneqq \frac{u_1}{\l_G + r}
    \quad
    \text{and}
    \quad
    U_0 \coloneqq \frac{u_0}{\l_B + r}.
\]
In each expression, the denominator reflects discounting and the rate at which the project ends under the specified action path.

\begin{prop}[No inspections] \label{res:NoInspections} Without inspections, working until the project ends is a best response for the agent if and only if $U_1 \geq U_0$. 
\end{prop}

We make the following standing assumption to rule out cases in which the principal's problem is either trivial or infeasible.

\begin{as} \label{as:nontrivial}  $U_0 > U_1 > 0$.
\end{as}

By \cref{res:NoInspections}, the inequality $U_0 > U_1$ means that inspections are necessary to induce the agent to work. Otherwise, it would be optimal for the principal to never inspect the agent. The inequality $U_1 > 0$ ensures that the agent strictly prefers always working to termination. Thus, termination can serve as a punishment to motivate work.

\section{Recursive formulation} \label{sec:recursive}

We analyze the principal's problem recursively. If the agent passes an inspection at time $t$, then $\th_t = 0$, so the principal's time-$t$ continuation problem is identical to the time-$0$ problem, as we formally argue below.\footnote{One implication is that the principal would use the same dynamic policy if she could commit only to the timing of the next inspection (and to termination if the agent fails that inspection).} 


In the recursive formulation,  after each passed inspection, the principal chooses the random time $T$ until the next inspection, and the agent chooses the action path $a = (a_{t})_{t \geq 0}$ that he will follow until the next inspection.\footnote{Time is measured relative to the last inspection. The agent plans to take action $a_t$ at time $t$, provided that the principal does not inspect first. The planned action $a_t$ is executed only if $T > t$.} Let $\AA$ denote the set of right-continuous action paths. The principal's optimal cost, denoted $K^\ast$,  satisfies the Bellman equation
\begin{equation} \label{eq:P_Bellman}
    K^\ast = \inf_{T} \E \Brac{ e^{-(\l_G + r) T} (1 + K^\ast)},
\end{equation} 
where the infimum is taken over all $(0,\infty)$-valued random times $T$ satisfying
\begin{equation} \label{eq:Bellman_agent}
\bm{1} \in \argmax_{a \in \AA} \, \E \Brac{ \int_{0}^{T} D_t(a) u(a_t) \de t + p_T(a) D_T(a)  U_1 }.
\end{equation}


First, consider the recursive obedience constraint \eqref{eq:Bellman_agent}. If the agent passes the next inspection, then his optimal value in the continuation problem is $U_1$ since the continuation policy must also induce work, and the agent's continuation value from always working is $U_1$. If the agent fails the next inspection, then the project is terminated. The associated term $(1 - p_T(a)) D_T(a) 0$ equals $0$, so it is omitted. Condition \eqref{eq:Bellman_agent} requires that if the agent behaves optimally \emph{after} the next inspection, then \emph{until} the next inspection the agent weakly prefers always working to any other action path. 


Next, consider the principal's Bellman equation \eqref{eq:P_Bellman}. The principal's expected cost is computed assuming that the agent chooses the action path $a = \bm{1}$. The next inspection is conducted at time $T$, provided that the project has not already ended. Thus, the principal uses the discount factor  $D_T ( \mathbf{1}) = e^{-(\l_G + r) T}$. The principal pays cost $1$ to conduct the inspection. On path, the agent passes the inspection, so the principal's cost in the continuation problem is $K^\ast$.

The principal's problem in \eqref{eq:P_Bellman}--\eqref{eq:Bellman_agent} can be expressed in the following more convenient form. The principal chooses a positive random time $T$ to solve
\begin{equation} \label{eq:next_inspection_A}
\begin{aligned}
    &\text{minimize} &&\E e^{- (\l_G + r) T} \\
    &\text{subject to} && \E \Brac{ \int_{0}^{T} D_t(a) u(a_t) \de t + p_T(a) D_T(a)  U_1 } \leq U_1, \quad a \in \AA.
\end{aligned}
\end{equation}
The constraint here is equivalent to \eqref{eq:Bellman_agent} since $U_1$ is the value of the expectation in \eqref{eq:Bellman_agent} with $a = \mathbf{1}$. We have divided the objective in \eqref{eq:P_Bellman} by $1 + K^\ast$, so the value of \eqref{eq:next_inspection_A} is $K^\ast / ( 1 + K^\ast)$; this scaling does not change the minimizer. An inspection policy $\mathbf{T}$ is optimal if and only if, conditional on almost every inspection history $(T_1, \ldots, T_{n-1})$, the increment  $T_n  - T_{n-1}$ follows a distribution that solves \eqref{eq:next_inspection_A}. 

In \eqref{eq:next_inspection_A}, the principal minimizes the expected cost of the next inspection subject to the constraint that the agent cannot profit by deviating before the next inspection. This problem is still complex because the set $\AA$ of action paths is large. The agent's best deviation changes with the principal's choice of random time $T$, and it could involve many alternating periods of working and shirking. To solve \eqref{eq:next_inspection_A}, we set up various relaxed problems, each of which imposes the inequality only for action paths $a$ in some subset $\AA'$ of $\AA$. The suitable set $\AA'$ of binding deviations depends on the parameter values. Each binding deviation takes one of three forms: always shirk; shirk and then work; or work and then shirk. We solve each relaxed problem by constructing Lagrange multipliers or by direct concavity arguments. Then we check that our relaxed solution is feasible in the original problem \eqref{eq:next_inspection_A}. To do so, we consider the dynamic optimization problem for the agent that is induced by the candidate inspection policy.  By solving the associated HJB equation or by direct comparison, we confirm that always working is optimal for the agent.

\section{Optimal timing of perfect inspections} \label{sec:perfect_inspections}

To highlight the main force in the model, we first solve the principal's problem in the special case of \emph{perfect} inspections. Formally, the passage probability is given by 
\[
    p_t(a) =  \begin{cases}
    1 &\text{if}~\int_{0}^{t} (1 - a_s) \de s = 0, \\
    0 &\text{if}~\int_{0}^{t} (1 - a_s) \de s > 0.
    \end{cases}
\]
That is, the agent passes the time-$t$ inspection if and only if he has not shirked for a positive duration before time $t$. This passage probability is the limit of the passage probability in the main model as the detectability parameter $\d$ tends to $\infty$. 

We separate the analysis into two regimes---innovation and maintenance---according to the relative sensitivities of breakthroughs and breakdowns to the agent's action.

\subsection{Innovation\texorpdfstring{: $\l_G > \l_B$}{}} \label{sec:innovation_perfect}

If $\l_G > \l_B$, then working increases the arrival rate of breakthroughs by more than it decreases the arrival rate of breakdowns. Consequently, working shortens the project in expectation. In particular, this case obtains if there are breakthroughs but no breakdowns ($\l_B=\nolinebreak0$).

\begin{thm}[Periodic perfect inspections] \label{res:speeding_up_perfect} Suppose that inspections are perfect and $\l_G > \l_B$. Then it is optimal to inspect periodically with some period $\tau^\ast$, i.e., $T_n = n \tau^\ast$ for all $n$.  If $u_0 \geq u_1$, then this policy is uniquely optimal and the period $\tau^\ast$ is given by
\begin{equation} \label{eq:tau_star}
     e^{- (\l_B + r) \tau^\ast} U_0 
       = U_0 - U_1.
\end{equation}
\end{thm}


If the agent's primary task is innovation---think of a start-up entrepreneur or a researcher working toward a new discovery---then it is optimal to conduct inspections at regular intervals.  When the agent shirks, the project has a lower hazard rate, so its survival probability is \emph{less} convex as a function of time.  Therefore, as we argue below, conducting the next inspection at a deterministic time is the cheapest way to deter shirking.

Before sketching the proof, we first consider how the optimal inspection policy depends on the model primitives. Fix the discount rate $r$ and arrival rates $\l_G$ and $\l_B$ with $\l_G > \l_B$. The ratio $U_1/U_0$ captures the relative attractiveness of working in the absence of inspections. This ratio is increasing in $\bar{u}(1)$ and $W_G$ and decreasing in $\bar{u}(0)$ and $W_B$. As $U_1/U_0$ increases, working becomes relatively more attractive, so fewer inspections are needed:  the period $\tau^\ast$ increases and the principal's associated expected inspection cost decreases. As $U_1/U_0$ tends to $0$, termination loses its punishment power relative to the benefit of shirking. Thus, the inspection period $\tau^\ast$ tends to $0$ and the principal's inspection cost tends to $\infty$. Conversely, as $U_1/U_0$ tends to $1$, the agent's relative preference for shirking in the absence of inspections vanishes. Thus, the inspection period $\tau^\ast$ tends to $\infty$ and the principal's inspection cost tends to $0$.

The comparative statics in the arrival rates are more subtle. Consider the case $u_0 \geq u_1$. Thus, $\tau^\ast$ is given by \eqref{eq:tau_star}. The continuation value $U_1$ is strictly increasing in $\l_G$ if and only if $r W_G > \bar{u}(1)$, i.e., the annuitized value from a breakthrough exceeds the flow utility from working. Similarly, $U_0$ is strictly increasing in $\l_B$ if and only if $r W_B > \bar{u}(0)$. As $\l_G$ increases, the inspection period $\tau^\ast$ changes only through the effect on $U_1$. As $\l_B$ increases, the inspection period $\tau^\ast$ changes through the effect on $U_0$, and there is also a separate reduction in $\tau^\ast$: holding $U_0$ fixed, higher $\l_B$ makes the project more likely to end before the inspection when the agent deviates, reducing the inspection's incentive power.

\label{page:periodic}
Consistent with \cref{res:speeding_up_perfect}, predictable inspections are standard in venture capital financing and research funding. \cite{gompers1999venture} observe that ``venture capitalists use staged investment to periodically evaluate'' firms they invest in, and they ``discontinue funding the project if they learn negative information'' (pp.~141--142).\footnote{In a sample of 794 firms backed by venture capital, \cite{gompers1999venture} find that firms subject to greater agency costs (as proxied by R\&D intensity or the industry ratio of tangible assets to total assets) have shorter financing rounds and thus greater monitoring frequency.} For research grants, the European Research Council (ERC) periodically reviews recipients' spending, according to a pre-announced schedule.\footnote{See \texttt{https://erc.europa.eu/manage-your-project/financial-reporting}.}

We illustrate the proof of \cref{res:speeding_up_perfect} in the case $u_0 \geq u_1$, i.e., shirking yields a weakly higher annuitized flow payoff than working. In this case, the binding deviation is for the agent to shirk immediately and continue shirking until the next inspection, as we check below. We consider the relaxation of \eqref{eq:next_inspection_A} that requires only this particular deviation ($a = \mathbf{0}$) to be unprofitable for the agent. After some algebra, this relaxed problem can be expressed as follows. The principal chooses a positive random time $T$ to solve
\begin{equation} \label{eq:perfect_delaying_problem}
\begin{aligned}
    &\text{minimize} && \E e^{- (\l_G + r) T} \\
    &\text{subject to} &&  \E  U_0 e^{-(\l_B + r) T} \geq U_0 - U_1. 
\end{aligned}
\end{equation}
The constraint requires that the payoff from this always-shirk deviation is weakly less than the payoff from always working. We have restated this inequality in terms of losses relative to $U_0$: the agent's loss from the always-shirk deviation is weakly greater than his loss from always working.

Crucially, $\l_B$ appears in the exponent in the constraint but $\l_G$ appears in the exponent in the objective. 
Under the always-shirk deviation, the agent's effective discount factor is $D_T ( \mathbf{0}) = e^{-(\l_B + r) T}$ because breakdowns arrive at Poisson rate $\l_B$ while the agent is shirking. On the other hand, the principal considers the expected cost of the next inspection on path. When the agent works, breakthroughs arrive at Poisson rate $\l_G$, so the principal's objective uses the effective discount factor $D_T ( \mathbf{1}) = e^{- (\l_G + r) T}$.

The solution of \eqref{eq:perfect_delaying_problem} becomes clear once we change variables. Instead of choosing the random time $T$ of the next inspection, the principal can equivalently choose the random variable $X = e^{-(\l_G + r)T}$, which is the on-path cost of conducting an inspection at time $T$. In terms of $X$, \eqref{eq:perfect_delaying_problem} becomes
\begin{equation} \label{eq:perfect_delaying_problem_cov}
\begin{aligned}
    &\text{minimize} && \E X \\
    &\text{subject to} &&  \E  U_0 X^{(\l_B + r)/(\l_G + r)} \geq U_0 - U_1.
\end{aligned}
\end{equation}

\begin{figure}
    \centering
\begin{tikzpicture}
	\begin{axis}[
				axis lines = center,
				scale = 0.85,
				xtick = {0.001}, 
				xticklabels = {$0$},
				ytick = \empty,
				xmin = 0,
				xmax = 2.55,
				ymin = 0,
				ymax = 2.2, 
				xlabel = {$t$},
				x label style={at={(current axis.right of origin)},anchor=west},	
				y label style={at={(current axis.above origin)},anchor=south},		
				]
				\addplot [Blue, thick, dashed,domain=0:2.5, samples=100]{1.25} node [pos = 0.8, above]{$U_0 - U_1$};
				\addplot [thick, domain=0:2.5, samples=100]{exp(-2 *x)} node [pos=0.3, above right, yshift = -4pt] {$e^{-(\l_G + r)t}$};
				\addplot [Orange,thick, domain=0:2.5,samples=100]{2*exp(-x)} node [pos = 0.16, above right] {$U_0 e^{-(\l_B + r)t}$};
			\end{axis}
		\end{tikzpicture}   
    \hspace{0.5cm}
		\begin{tikzpicture}
			\begin{axis}[
				axis lines = center,
				scale = 0.85,
				xtick = {0.001,39/64,1},
				xticklabels = {$1$,,$0$},
				ytick = \empty,
				xmin = 0,
				xmax = 1.05,
				ymin = 0,
				ymax = 2.2, 
				xlabel = {$x$},
				x label style={at={(current axis.right of origin)},anchor=west},	
				y label style={at={(current axis.above origin)},anchor=south},		
				clip = false
				]
				
				\addplot [Blue,dashed,thick,domain=0:1, samples=200]{1.25} 		node [pos = 0.2, above]{$U_0 - U_1$};
				\addplot [Orange,thick, domain=0:1,samples=200]{2*(1-x)^(1/2)} node [pos = 0.1, above right] {$U_0 x^{ (\l_B + r)/(\l_G+ r)}$};
				\addplot [thick, dotted, domain=0:1,samples=200] coordinates {(39/64, 0) (39/64, 1.25)} node [pos = 0, below] {$x^\ast$};
                \filldraw (39/64, 1.25) circle (2pt); 
			     
			\end{axis}
		\end{tikzpicture}

    \caption{Shirking agent's loss (orange) from a perfect inspection with $\l_G > \l_B$. In this example, $\l_G + r = 2$; $\l_B + r = 1$; $U_1 = 0.75$; and $U_0 = 2$.}
    \label{fig:speeding_up_perfect}
\end{figure}

\cref{fig:speeding_up_perfect} depicts the principal's problem before (left) and after (right) the change of variables, in an example with $\l_G > \l_B$. 
 The left panel plots, as a function of the inspection time $t$, the principal's on-path inspection cost (black) and the agent's loss from the inspection under the always-shirk deviation (orange). The principal chooses a distribution over the horizontal axis to minimize her expected on-path inspection cost, subject to the constraint that the agent's expected loss from the inspection under the always-shirk deviation is at least $U_0 - U_1$. As a function of the inspection time, the agent's shirking loss is \emph{less convex} than the principal's on-path cost because $\l_B < \l_G$. 

The right panel of \cref{fig:speeding_up_perfect} puts the principal's on-path inspection cost $x = e^{-(\l_G + r)t}$ on the horizontal axis $[0,1]$. The direction of this axis has been reversed so that time still moves from left to right. The principal chooses a distribution whose expectation is minimal (i.e., furthest right), subject to the constraint that the agent's expected loss from the inspection under the always-shirk deviation is at least $U_0 - U_1$. This loss is a strictly concave function of $X$, so replacing any nondegenerate random variable $X$ with the constant $\E X$ strictly slackens the constraint, without changing the principal's objective. Therefore, the unique solution of  \eqref{eq:perfect_delaying_problem_cov} is the constant $x^\ast = e^{-(\l_G + r) \tau^\ast}$ for which the constraint holds with equality. The constant $\tau^\ast$ is given by \eqref{eq:tau_star}. Returning to the original variables, we conclude that the constant $\tau^\ast$ is the unique solution of \eqref{eq:perfect_delaying_problem}.

We have argued that conducting the next inspection at the nonrandom time $\tau^\ast$  is the cheapest way to deter the agent from shirking continuously until the next inspection. It remains to check that if the next inspection is conducted at time $\tau^\ast$, then no other deviations are profitable for the agent. If the agent shirks for a positive duration before time $\tau^\ast$, then he is certain to fail the inspection. Given that he will fail the inspection, the agent's payoff is highest if he shirks continuously over the interval $[0, \tau^\ast]$---shirking maximizes both the discount factor $D_t(a)$ (because $\l_G > \l_B$) and the annuitized flow payoff $u(a)$ (because $u_0 \geq u_1$).

The argument above assumes that $u_0 \geq u_1$. Suppose instead that $u_0 < u_1$.\footnote{This is consistent with \cref{as:nontrivial} because $\l_B < \l_G$.} In this case,  periodic inspections are still optimal, but the argument is more subtle. If the agent has already shirked, then he knows that he will fail the next inspection.  As the next inspection (and hence the end of the game) nears, the agent becomes increasingly myopic. Since $u_0 < u_1$, the agent will find it optimal to work once the next inspection is sufficiently close. In the proof, we identify the binding shirk-before-work deviation and we give an explicit formula for the optimal period $\tau^\ast$.\footnote{If $u_0 < u_1$, the periodic policy is not the only solution of the relaxed problem. In the proof, we identify a time $\bar{t} > 0$ such that in every optimal policy, the time between any consecutive inspections is at least $\bar{t}$.} We analyze such shirk-before-work deviations in more detail below in the case of the imperfect inspection technology (\cref{sec:speeding_up}).

\begin{rem}[Comparison with static moral hazard]
In problem \eqref{eq:perfect_delaying_problem}, the principal's tradeoff is similar to that in the classic static moral hazard problem with binary actions (work and shirk) and a risk-averse agent. To induce work in that setting, the principal minimizes the on-path expected wage subject to the constraint that the agent's expected loss in wage utility from shirking relative to working is at least as large as the effort cost of working. In our model, the principal minimizes the on-path expected discounted inspection cost subject to the constraint that, under the deviation, the agent's expected loss from termination (upon failing an inspection) is at least as large as $U_0 - U_1$, the benefit from shirking in the absence of inspections. In the static problem, the principal shifts wages (the carrot) to states that are relatively unlikely under shirking. In our problem, the principal shifts the distribution of the inspection time (the stick) so that the inspection is relatively more likely to be carried out when the agent shirks.
\end{rem}

\subsection{Maintenance\texorpdfstring{: $\l_B >\l_G$}{}} \label{sec:maintentance_perfect}

If $\l_B > \l_G$, then working decreases the arrival rate of breakdowns by more than it increases the arrival rate of breakthroughs.  Consequently, working lengthens the project in expectation. In particular, this case obtains if there are breakdowns but no breakthroughs ($\l_G=0$). 

\begin{thm}[Random perfect inspections] \label{res:delaying_perfect} Suppose that inspections are perfect and $\l_B > \l_G$.  Then the following policy is uniquely optimal. The gaps $(T_n - T_{n-1})_{n \geq 1}$ are independently and identically distributed according to an exponential distribution with hazard rate $\g^\ast$, where 
\begin{equation} \label{eq:gamma_star}
    \frac{\g^\ast}{\l_B + r+\g^\ast} U_0=  
    U_0 - U_1.  
\end{equation}
\end{thm}

If the agent's primary task is maintenance---think of a worker following safety protocols to prevent an accident---then it is optimal to conduct inspections at random times. When the agent shirks, the project has a higher hazard rate, so its survival probability is \emph{more} convex as a function of time. As a result, conducting the next inspection at a random time is the cheapest way to deter shirking. 

The comparative statics parallel those in the innovation case. Fix the discount rate $r$ and arrival rates $\l_G$ and $\l_B$ with $\l_B > \l_G$. By \eqref{eq:gamma_star}, as the ratio $U_1/U_0$ increases, working becomes relatively more attractive, so fewer inspections are needed: the inspection hazard rate $\gamma^\ast$ decreases and the principal's associated expected inspection cost decreases. As $U_1/U_0$ tends to $0$, the inspection rate $\gamma^\ast$ tends to $\infty$ and the principal's inspection cost tends to $\infty$. Conversely, as $U_1/U_0$ tends to $1$, the inspection rate $\gamma^\ast$ tends to $0$ and the principal's inspection cost tends to $0$. The comparative statics in the arrival rates are more subtle. By \eqref{eq:gamma_star}, as $\l_G$ increases, the inspection hazard rate $\gamma^\ast$ changes only through the resulting change in $U_1$.  As $\l_B$ increases, the inspection hazard rate $\gamma^\ast$ changes through the resulting change in $U_0$, but there is also a separate increase in $\gamma^\ast$: holding $U_0$ fixed, higher $\l_B$ makes the project more likely to end before the inspection when the agent deviates, reducing the inspection's incentive power.

\label{page:random}
Consistent with \cref{res:delaying_perfect}, safety inspections are generally not announced in advance. In fact, the US Occupational Safety and Health
Administration (OSHA) explicitly prohibits advance notice of workplace inspections, outside of certain special circumstances \defcitealias{CFR1903}{OSHA, 1971}\citepalias[]{CFR1903}.\footnote{For example, advance notice is allowed if workplaces must make special preparations for an inspection. \cite{johnson2023improving} suggest an improved targeting approach for OSHA inspections. They acknowledge, however, that such targeting would make inspections more predictable, potentially reducing their general deterrence effect \citep[p.\ 33]{johnson2023improving}.}  A similar logic applies to politics since elected representatives understand that shirking (e.g., engaging in corruption) tends to shorten their time in office, by increasing their chances of being ousted. Brazil has  a federal anti-corruption initiative that randomly audits the finances of municipal governments.  Municipalities are selected for audit by public lotteries \citep{AvisEtal2018}.\footnote{The details of the randomization resemble our solution with the imperfect inspection technology, as we discuss in \cref{sec:imperfect_inspections}.}

\begin{figure}
    \centering
\begin{tikzpicture}
	\begin{axis}[
				axis lines = center,
				scale = 0.85,
				xtick = {0.001}, 
				xticklabels = {$0$},
				ytick = \empty,
				xmin = 0,
				xmax = 2.55,
				ymin = 0,
				ymax = 2.2, 
				xlabel = {$t$},
				x label style={at={(current axis.right of origin)},anchor=west},	
				y label style={at={(current axis.above origin)},anchor=south},		
				]
				\addplot [Blue,dashed, thick,domain=0:2.5, samples=200]{1.25} node [pos = 0.8, above]{$U_0 - U_1$};
				\addplot [thick, domain=0:2.5, samples=200]{exp(-x)} node [pos=0.65, above right] {$e^{-(\l_G + r)t}$};
				\addplot [Orange,thick, domain=0:2.5,samples=200]{2*exp(-2 *x)} node [pos = 0.15, above right] {$U_0 e^{-(\l_B + r)t}$};
			\end{axis}
		\end{tikzpicture}   
    \hspace{0.5cm}
\begin{tikzpicture}
		\begin{axis}[
			axis lines = center,
			scale = 0.85,
			xtick = {0.001,1},
			xticklabels = {$1$, $0$},
			ytick = \empty,
			xmin = 0,
			xmax = 1.05,
			ymin = 0,
			ymax = 2.2, 
			xlabel = {$x$},
			x label style={at={(current axis.right of origin)},anchor=west},	
			y label style={at={(current axis.above origin)},anchor=south},		
			]
			\addplot [Blue,dashed, thick,domain=0:1, samples=200]{1.25} node [pos = 0.8, above]{$U_0 - U_1$};
			\addplot [Orange,thick, domain=0:1,samples=200]{2*(1-x)^2} node [pos = 0.1, right, xshift = 6pt] {$U_0 x^{ (\l_B + r)/(\l_G+ r)}$};
		\end{axis}
\end{tikzpicture}
\caption{Shirking agent's loss (orange) from a perfect inspection with $\l_B > \l_G$. In this example, $\l_G + r = 1$; $\l_B + r = 2$; $U_1 = 0.75$; and $U_0 = 2$.}
\label{fig:delaying_perfect}
\end{figure}

To build intuition for \cref{res:delaying_perfect}, first consider the same relaxed problem  \eqref{eq:perfect_delaying_problem} as in the innovation regime. This problem requires only that it is unprofitable for the agent to shirk immediately and continue shirking until the
next inspection. As before, we can change variables to get \eqref{eq:perfect_delaying_problem_cov}. \cref{fig:delaying_perfect} plots the same functions as \cref{fig:speeding_up_perfect}, before and after the change of variables, in an example with $\l_B > \l_G$. 
 As a function of the inspection time, the agent's loss from the inspection under the always-shirk deviation is \emph{more convex} than the principal's on-path inspection cost. In the right panel, we express this loss as a convex function of the principal's on-path inspection cost $x = e^{-(\l_G 
+ r)t}$. In the relaxed problem, the principal would like to spread out the distribution of time until the next inspection by inspecting either very early or very late.  But such a policy is infeasible in the original problem. If the agent is not inspected early on, then he can infer that he will not be inspected for a very long time. Instead of working continuously, the agent can profitably deviate by working briefly and then, if he is not inspected, shirking thereafter. Deviations of this work-before-shirk form will indeed bind.

In the maintenance regime, the binding deviations take the following form: work until time $s$, and then shirk until the next inspection, for each time $s \geq 0$. We consider the relaxation of \eqref{eq:next_inspection_A} requiring that none of these work-before-shirk deviations is profitable. After some algebra, this relaxed problem can be expressed as follows. Let $\E_s = \E [ \cdot | T > s]$. The principal chooses a positive random time $T$ to solve\footnote{If $\P (T > s) = 0$, then the agent does not actually shirk under the specified deviation, so we consider the inequality to be satisfied, even though the conditional expectation is not well-defined.} 
\begin{equation} \label{eq:perfect_speeding_problem}
\begin{aligned}
    &\text{minimize} && \E e^{- (\l_G + r) T} \\
    &\text{subject to} &&  \E_s U_0 e^{-(\l_B + r) (T- s)}  \geq U_0 - U_1, \quad s \geq 0.
\end{aligned}
\end{equation}
For each fixed $s$, suppose time $s$ has passed since the last inspection and the agent has worked continuously since the last inspection. The inequality requires that it is unprofitable for the agent to begin shirking and continue shirking until the next inspection.

Since $\l_B > \l_G$, the loss function in the constraint is more convex than the objective, as illustrated in \cref{fig:delaying_perfect}. Therefore, the solution of the relaxed problem \eqref{eq:perfect_speeding_problem} is to conduct the next inspection at a constant hazard rate (see \cref{res:exponential_sol}, \cref{sec:prelim}). This policy is memoryless---the conditional distribution of time until the next inspection is the same, no matter how much time has passed since the last inspection. The hazard rate $\g^\ast$ in \eqref{eq:gamma_star} ensures that each constraint in \eqref{eq:perfect_speeding_problem} holds with equality. 

We have argued that inspecting with the constant hazard rate $\g^\ast$ is the cheapest way to deter all work-before-shirk deviations. It remains to check that if the next inspection is conducted with hazard rate $\g^\ast$, then no other deviations are profitable. In the proof, we show that under this policy, once the agent begins shirking, he finds it optimal to continue shirking until the next inspection. To see why, note that at each time, if the agent has not previously shirked, then he is indifferent between working continuously and shirking continuously until the next inspection. If the agent has previously shirked, then he is certain to fail the next inspection, so his incentives to work are weaker. Hence, he \emph{strictly} prefers to shirk until the next inspection. 



\section{Optimal timing of imperfect inspections} \label{sec:imperfect_inspections} 

Next, we consider the imperfect inspection technology, with finite detectability parameter $\d$,  introduced in \cref{sec:setting}.  Under this technology, shirking generates evidence stochastically, but each inspection perfectly reveals whether evidence is present. To ensure that the principal's problem is feasible, we impose the additional standing assumption that the rate at which shirking generates evidence is sufficiently high.

\begin{as}
\label{as:sufficient} $\d > (\l_B + r) (U_0 - U_1)/U_1$.
\end{as}

The imperfect inspection technology creates a new motive for the principal to space apart inspections. To deter \emph{global} deviations, it is wasteful for the principal to conduct imperfect inspections in short succession. Suppose that after passing an inspection, the agent begins to shirk
for a positive duration. If the principal conducts another inspection soon after the last inspection, then the agent is very likely to pass because his shirking is unlikely to leave behind new evidence in a short time interval.\footnote{Under our imperfect inspection technology, the results of inspections at time $t$ and time $t + \D$ become perfectly correlated in the limit as $\D \to 0$. This would no longer be true in an alternative model in which an inspection at time $t$ reveals a noisy signal of $\th_t$, where the noise is independent across inspections. With such noise, closely spaced inspection results are still correlated, but the correlation is bounded away from $1$. Thus, the motive to space apart inspections is weakened but not eliminated. Fully analyzing this alternative model is challenging because the principal's problem would no longer renew after each inspection.} 

The optimal policy reflects this new motive to space apart inspections. Periodic inspections are still optimal in the innovation regime ($\l_B < \l_G$), but they are also optimal if $\l_B$ is below a higher threshold $\bar{\l}_B > \l_G$. On the other hand, if $\l_B$ is above the threshold $\bar{\l}_B$, then the optimal policy leverages the benefits of randomization while also spacing apart inspections. Inspections are conducted randomly, but only after a fixed period has elapsed since the last inspection.

To formally state the optimal inspection policy, let $\operatorname{Exp} (\g)$ denote an (independent) exponentially distributed random variable with hazard rate $\g$. 

\begin{thm}[Imperfect inspections] \label{res:imperfect_inspections}
Assume $\d > 2 \l_G - \l_B + r$. There exists a threshold $\bar{\l}_B = \bar{\l}_B (\l_G, \d, r, u_0, u_1)$, with $\bar{\l}_B  > \l_G$, such that the following hold.  
\begin{enumerate}[label = (\roman*), ref = \roman*]
    \item \label{res:speeding_up_imperfect} If $\l_B \le \bar \l_B$, then it is optimal to inspect periodically with some period $\tau^\ast$, i.e., $T_n = n \tau^\ast$ for all $n$. Moreover, there exists a time $\bar{t} = \bar{t} ( \l_B, \l_G, \d, r, u_0, u_1) > 0$ such that under every optimal policy,  $T_n - T_{n-1} \geq \bar{t}$ for each $n$.
    \item \label{res:delaying_imperfect} If $\l_B > \bar \l_B$, then the following policy is uniquely optimal. The gaps $(T_n - T_{n-1})_{n \geq 1}$ are independently and identically distributed. For each $n$, the gap $T_n - T_{n-1}$ equals $\hat{\tau}$ with some probability $\pi^\ast$ in $(0,1)$, defined in the proof. With probability $1 - \pi^\ast$, the gap $T_n - T_{n-1}$ has the distribution of $\hat{\tau} + \operatorname{Exp}(\g^\ast)$, where 
   \begin{equation} \label{eq:gamma_ast_0}
    e^{-\d \hat{\tau}} = \frac{\l_B - \l_G}{\l_B - \l_G + \d}, 
    \qquad
    \frac{\g^\ast \, U_0}{\l_B +r + \g^\ast}  - \frac{\g^\ast\, U_1}{\l_B + r + \d + \g^\ast}  = U_0 - U_1.
    \end{equation}
\end{enumerate}
\end{thm}

The additional assumption $\d > 2 \l_G - \l_B + r$ ensures that the passage probability is sufficiently convex (as a function of the duration of shirking) that the binding deviations are not local; 
\cref{sec:recovery} analyzes a setting in which local deviations bind. We separately discuss the two cases of \cref{res:imperfect_inspections} below.

\begin{rem}[Threshold $\bar{\l}_B$] \label{rem:threshold_B} In the proof of \cref{res:imperfect_inspections} in \cref{sec:main_proof}, we show that the condition $\l_B > \bar{\l}_B$ can be expressed as 
\begin{equation} \label{eq:threshold_B}
    \frac{ \l_B  + r }{\l_G + r} > 1 + g \Paren{  \frac{\d}{\l_G + r}, \frac{ u_0}{u_1}},
\end{equation}
for some strictly positive function $g$ that is strictly decreasing in its first argument and strictly increasing in its second argument. For any fixed ratio $u_0/u_1 > 1$, we have $g (d, u_0/u_1) \downarrow 0$ as $d \uparrow \infty$, consistent with the perfect-inspection limit.\footnote{If $u_0 / u_1 \leq 1$, then \cref{as:nontrivial} implies that $\l_B < \l_G$, so the threshold $\bar{\l}_B$ is arbitrary, as long as $\bar{\l}_B > \l_G$.} The condition in \eqref{eq:threshold_B} is expressed in terms of the annuitized flow payoff $u_0 = \bar{u} (0) + \l_B W_B$, which itself depends on $\l_B$. But as long as the continuation payoff $W_B$ after a breakdown is weakly below the agent's outside option payoff $0$, we can equivalently express the condition $\l_B > \bar{\l}_B( \l_G, \d, r, u_0, u_1)$ as $\l_B > \hat{\l}_B ( \l_G, \d, r, \bar{u}(0), \bar{u}(1), W_G, W_B)$ for some alternative threshold function $\hat{\l}_B$. 
\end{rem}

\subsection{Periodic imperfect inspections} \label{sec:speeding_up}

Within the case $\l_B \leq \bar{\l}_B$, different deviations can bind, depending on the parameter values.  First consider the relaxation of \eqref{eq:next_inspection_A} requiring that it is unprofitable for the agent to shirk continuously until the next inspection. This relaxed problem can be expressed as follows. The principal chooses a positive random time $T$ to solve
\begin{equation} \label{eq:imperfect_maintenance}
\begin{aligned}
    &\text{minimize} && \E  e^{- (\l_G + r) T} \\
    &\text{subject to} &&  \E  L_S (T) \geq U_0 - U_1,
\end{aligned}
\end{equation}
where $L_S (t) = e^{-(\l_B + r) t} (U_0 - U_1 e^{-\d t})$. 

\begin{figure}
    \centering
\begin{tikzpicture}
	\begin{axis}[
		axis lines = center,
			scale = 0.85,
			xtick = {0.001},
			xticklabels = {$0$},
			ytick = \empty,
			xmin = 0,
			xmax = 2.55,
			ymin = 0,
			ymax = 1.3, 
			xlabel = {$t$},
			x label style={at={(current axis.right of origin)},anchor=west},	
			y label style={at={(current axis.above origin)},anchor=south},	
                ]
				\addplot [Blue,dashed, thick,domain=0:2.5, samples=200]{0.75} node [pos = 0.8, above]{$U_0 - U_1$};
				\addplot [thick, domain=0:2.5, samples=200]{exp(-2*x)} node [pos=0.3, right, xshift = 6pt] {$e^{-(\l_G + r)t}$};
				\addplot [Orange,thick, domain=0:2.5,samples=200]{2*exp(-x)-(5/4)*exp(-(1+3.5)*x)} node [pos = 0.3, right, xshift = 2pt] {$L_S (t)$}; 
			\end{axis}
		\end{tikzpicture}   
    \hspace{0.5cm}
\begin{tikzpicture}
		\begin{axis}[
			axis lines = center,
			scale = 0.85,
			xtick = {0.001,0.418, 0.823,1},
			xticklabels = {$1$, $\bar{x}$, $x^\ast$, $0$},
			ytick = \empty,
			xmin = 0,
			xmax = 1.05,
			ymin = 0,
			ymax = 1.3, 
			xlabel = {$x$},
			x label style={at={(current axis.right of origin)},anchor=west},	
			y label style={at={(current axis.above origin)},anchor=south},		
			]
			\addplot [Blue,dashed, thick,domain=0:1, samples=100]{0.75} node [pos = 0.25, below]{$U_0 - U_1$};
			\addplot [Orange,thick, domain=0:1,samples=100]{2*(1 - x)^(1/2) - (5/4)* (1 - x)^(4.5/2)} node [pos = 0.42, above right] {$\bar{L}_S(x)$}; 
            \addplot [Green, thick, dashed, domain = 0:0.418, samples = 200] { 2*(1-x)^(1/2) - (5/4)*(1 - x)^(4.5/2)};
            \addplot [Green, thick, domain = 0.418:1, samples = 200] { 2*0.582^(1/2) - (5/4)*0.582^(1/2)*(1-(1-x)/0.582*(1-0.582^(3.5/2))) } node [pos = 0.3, left, yshift = -4pt] {$\bar{L}(x)$};
            \filldraw (0.823, 0.75) circle (2pt);
            \addplot[ dotted, thick] coordinates {(0.823,0) (0.823, 0.75)};
		\end{axis}
\end{tikzpicture}
\caption{Agent's loss from an imperfect inspection (orange for shirk-always deviation, green for shirk-before-work deviation) with $\l_B \leq \bar{\l}_B$. In this example, $\l_G + r = 2$; $\l_B + r = 1$; $U_1 = 1.25$; $U_0 = 2$; and $\d = 3.5$. Here, $\bar{L}_S$ is globally concave because $\d + \l_B  \geq \l_G  \geq \l_B$.}
\label{fig:shirk_work}
\end{figure}

The function $L_S$ is plotted in the left panel of \cref{fig:shirk_work}, in an example with $\l_B \leq \bar{\l}_B$. As a function of the time $t$ until the next inspection, $L_S$ represents the agent's loss (relative to $U_0$) from shirking immediately and continuing to shirk until the next inspection. Naturally, this loss is smaller than in the case of perfect inspections. Here, the agent passes the time-$t$ inspection with probability $e^{-\d t}$, in which case he gets his continuation payoff $U_1$. As $t$ tends to $0$, this passage probability converges to $1$, so the loss $L_S (t)$ converges to $U_0 - U_1$.

We apply the same change of variables as in the case of perfect inspections. In terms of the principal's on-path inspection cost $X = e^{-(\l_G + r) T}$,  problem \eqref{eq:imperfect_maintenance} becomes
\begin{equation} \label{eq:imperfect_maintenance_cov}
\begin{aligned}
    &\text{minimize} && \E X \\
    &\text{subject to} &&  \E \bar{L}_S (X) \geq U_0 - U_1,
\end{aligned}
\end{equation}
where $\bar{L}_S (x) = x^{(\l_B + r)/(\l_G + r)} (U_0 - U_1 x^{\d/(\l_G + r)} )$. The right panel of \cref{fig:shirk_work} plots the loss $\bar{L}_S$ as a function of the principal's on-path inspection cost, with $\l_B \leq \bar{\l}_B$. 


For some parameter values within the case $\l_B \leq \bar{\l}_B$, shirking all the way until the next inspection is the binding deviation. In this case, we prove that $\bar{L}_S$ is concave on a suitable region (\cref{claim:LS} in \cref{sec:main_proof}); therefore, the unique solution of \eqref{eq:imperfect_maintenance_cov} is the constant at which $\bar{L}_S$ intersects the horizontal line $U_0 - U_1$. Returning to the original variables, we conclude that the associated constant time is the unique solution of the relaxed problem \eqref{eq:imperfect_maintenance}. In the proof, we verify that this solution is feasible in the original problem \eqref{eq:next_inspection_A}.

For other parameter values within the case $\l_B \leq \bar{\l}_B$, the binding deviation takes the following form: shirk until time $\bar{t}$ and then work until the next inspection, for some fixed time $\bar{t} > 0$.\footnote{In the proof, we must consider a third case. For $\l_B$ sufficiently close to (but below) $\bar{\l}_B$, the binding deviations are the same as in the case $\l_B > \bar{\l}_B$ discussed below. We show that inspecting periodically is then the unique solution.}  In this case, we consider a variant of \eqref{eq:imperfect_maintenance_cov}
requiring that the loss from this shirk-before-work deviation (relative to $U_0)$ is at least $U_0 - U_1$. This loss, expressed as a function of the principal's on-path inspection cost, is denoted by $\bar{L}$ and is plotted in the right panel of \cref{fig:shirk_work}. We label the point $\bar{x} = e^{-(\l_G + r) \bar{t}}$. If the next inspection is conducted before time $\bar{t}$, then the agent does not begin working before the inspection. So, for $x \geq \bar{x}$, the agent's loss $\bar{L}$ coincides with $\bar{L}_S$. At time $\bar{t}$, the agent begins working, and thereafter the hazard rate of the project is the same as it is on path. So, for $x \leq \bar{x}$, the agent's loss $\bar{L}$ is affine in the principal's on-path inspection cost. 


Let $x^\ast = e^{-(\l_G + r) \tau^\ast}$ be the point at which the function $\bar{L}$ intersects the horizontal line $U_0 - U_1$, as shown on the plot. We prove that $\bar{L}$ is concave on a suitable region (\cref{claim:U} in \cref{sec:main_proof});  therefore, the solutions of this relaxed problem (the variant of \eqref{eq:imperfect_maintenance_cov} with $\bar{L}$ in place of $\bar{L}_S$) are precisely the random variables $X$ with expectation $x^\ast$ that concentrate on $(0, \bar{x}]$, the affine segment of $\bar{L}$. In terms of calendar time, the solutions are precisely the random variables $T$ that concentrate on $[\bar{t}, \infty)$ and satisfy $\E [ e^{-(\l_G + r)T}] = e^{-(\l_G + r) \tau^\ast}$.\footnote{Since the transformation $t  \mapsto x(t)$ is strictly decreasing, we have $x(t) \leq \bar{x}$ if and only if $t \geq \bar{t}$.}  We show in the proof that the constant $T = \tau^\ast$ is feasible in the original problem \eqref{eq:next_inspection_A}. Some other relaxed solutions may also be feasible in the original problem, so the solution $T = \tau^\ast$ is not necessarily unique, but every solution concentrates on $[\bar{t}, \infty)$ and hence has a no-inspection period of at least $\bar{t}$.

\subsection{Random imperfect inspections} \label{sec:imperfect_random}

Suppose $\l_B > \bar{\l}_B$. To build intuition for part \ref{res:delaying_imperfect} of \cref{res:imperfect_inspections}, first consider the same relaxed problem \eqref{eq:imperfect_maintenance} as in the case $\l_B \leq \bar{\l}_B$. 
\cref{fig:delaying_breakdown_imperfect} plots the same loss functions from \cref{fig:shirk_work} in an example with $\l_B > \bar \l_B$. 
Since $\bar{\l}_B > \l_G$, we have $\l_B > \l_G$. Thus, the loss function $\bar{L}_S$, as a function of the on-path inspection cost, is concave for high $x$ and then convex for low $x$. We can solve the relaxed problem \eqref{eq:imperfect_maintenance} by computing the concavification of the function $\bar{L}_S$ and finding the point at which this concavification intersects $U_0 - U_1$. The solution $X$ to this relaxed problem concentrates on two points---the point $x = 0$ and the smallest strictly positive point of intersection between $\bar{L}_S$ and its concavification $\cav \bar{L}_S$.\footnote{Technically, this problem does not allow $X = 0$. The distribution we are describing is the limit of every minimizing sequence.} Returning to the original variables, the associated policy either inspects at a fixed time or never inspects. But such a policy is infeasible in the original problem. If the agent is not inspected at the fixed time, then he can infer that he will never be inspected. The agent can profitably deviate by working until the fixed time and then, if he is not inspected, shirking forever after. We conclude that additional deviations must bind.

\begin{figure}
\centering
\begin{tikzpicture}
	\begin{axis}[
		axis lines = center,
			scale = 0.85,
			xtick = {0.001},
			xticklabels = {$0$},
			ytick = \empty,
			xmin = 0,
			xmax = 2.55,
			ymin = 0,
			ymax = 1.3, 
			xlabel = {$t$},
			x label style={at={(current axis.right of origin)},anchor=west},	
			y label style={at={(current axis.above origin)},anchor=south},	
                ]
				\addplot [Blue,dashed, thick,domain=0:2.5, samples=200]{0.75} node [pos = 0.8, above]{$U_0 - U_1$};
				\addplot [thick, domain=0:2.5, samples=200]{exp(-x)} node [pos=0.7, above right] {$e^{-(\l_G + r)t}$};
				\addplot [Orange,thick, domain=0:2.5,samples=200]{2*exp(-2*x)-(5/4)*exp(-7*x)} node [pos = 0.15, above right] {$L_S (t)$}; 
			\end{axis}
		\end{tikzpicture}   
    \hspace{0.5cm}
	\begin{tikzpicture}
		\begin{axis}[
			axis lines = center,
                scale = 0.85,
			xtick = {0.001,0.301173, 1},
			xticklabels = {$1$,$\hat{x}$, $0$},
			ytick = \empty,
			xmin = 0,
			xmax = 1.05,
			ymin = 0,
			ymax = 1.3, 
			xlabel = {$x$},
			x label style={at={(current axis.right of origin)},anchor=west},	
			y label style={at={(current axis.above origin)},anchor=south},		
			]
			\addplot [Blue,dashed, thick,domain=0:1, samples=100]{0.75} node [pos = 0.8, above]{$U_0 - U_1$};
			\addplot [Orange,thick, domain=0:1,samples=100]{2*(1 - x)^2 - (5/4)*(1 - x)^(7)} node [pos = 0.2, above right] {$\bar{L}_S (x)$}; 
            \filldraw (.301173,0.875) circle (2pt);
		\end{axis}
	\end{tikzpicture}
	\caption{Shirking agent's loss (orange) from an imperfect inspection with $\l_B > \bar\l_B$. In this example, $\l_G + r = 1$; $\l_B + r = 2$; $U_0 = 2$; $U_1 = 1.25$; and $\d = 5$.}
	\label{fig:delaying_breakdown_imperfect}
\end{figure}


The binding deviations are of the same form as in the maintenance regime under perfect inspections: work until time $s$, then shirk until the next inspection. But now these constraints bind only for $s = 0$ and for $s \geq \hat{\tau}$, where $\hat{\tau}$ is given in \eqref{eq:gamma_ast_0}.
 The point $\hat{x} = e^{-(\l_G + r) \hat{\tau}}$ is shown in the right panel of \cref{fig:delaying_breakdown_imperfect}.
 The optimal policy proceeds as follows. First, there is an inspection-free period of length $\hat{\tau}$. With positive probability, the agent is inspected exactly at time $\hat{\tau}$. This atom deters the agent from  beginning to shirk immediately after the last inspection. If the agent is not inspected at time $\hat{\tau}$, then the next inspection is conducted with a constant hazard rate thereafter. Once the inspection is conducted, the cycle repeats, beginning with a fresh period without inspections. 
 
 This inspection format is observed in practice. Under the Brazilian anti-corruption initiative discussed in \cref{sec:maintentance_perfect}, municipalities are selected for audit in regular lotteries, but ``once audited, the municipality can be audited again only after several lotteries have elapsed'' \citep[p.\ 1920]{AvisEtal2018}. This rule is a convenient way to (approximately) implement an inspection-free period followed by inspections with a constant hazard rate.

 As the detectability parameter $\d$ increases, the inspection-free period $\hat{\tau}$ decreases (\cref{res:thresholds} in \cref{sec:main_proof}) since more precise inspections are more likely to detect even short periods of shirking. This does not mean that inspections occur more frequently on average: the mass $\pi^\ast$ and the subsequent hazard rate $\g^\ast$ are strictly decreasing in $\d$. As $\d$ tends to $\infty$, the optimal policy in \cref{res:imperfect_inspections}.\ref{res:delaying_imperfect} converges to the exponential policy in \cref{res:delaying_perfect}: the inspection-free period $\hat{\tau}$ and the probability $\pi^\ast$ of inspecting at time $\hat{\tau}$ each tend to $0$, and the threshold $\bar{\l}_B$ converges to $\l_G$.

The structure of the policy in \cref{res:imperfect_inspections}.\ref{res:delaying_imperfect} is similar to the optimal policy in \citet[Theorem 1, p.~2905]{Varas2020}, but their policy arises for different reasons. In our problem, the binding deviations are global, and the policy in \cref{res:imperfect_inspections}.\ref{res:delaying_imperfect} is the cheapest way to incentivize work. Under the inspection technology in \cite{Varas2020}, by contrast, binding deviations are local, and the \emph{cheapest} way to incentivize work is to inspect with a constant hazard rate; see our \cref{res:high_recovery_exponential} below. Their solution features an inspection-free period in order to provide the public with better information about the state, which is payoff-relevant. 

\section{Inspection technology with recovery} \label{sec:recovery}

In the main model, we assume that once the agent's shirking leaves behind evidence, the agent is certain to fail the next inspection. We now consider an alternative inspection technology that allows the agent to recover from past shirking. Formally, the state $\th_t$ evolves as follows. Transitions from state $0$ to state $1$ occur at Poisson rate $(1 - a_t) \d$, as before. Now transitions from state $1$ to state $0$ occur at Poisson rate $ a_t  \rho$.\footnote{Let $p_{t|t'} (a)$ denote the conditional probability of passing an inspection at time $t$, given that the agent passed an inspection at time $t'$. If $\rho = 0$, then $p_{t|t'} (a) = p_t (a) / p_{t'}(a)$. If $\rho > 0$, this equality no longer holds. In this case, the term $p_{T_{n-1}}(A)$  in \eqref{eq:IC} must be replaced with the full expression $\prod_{j=1}^{n-1} p_{T_j | T_{j-1} } (A)$.} Call $\rho$ the recovery rate. Our main model considers the case $\rho = 0$. The state process introduced in \cite{board2013reputation} considers the case $\rho = \d$. 


The policies in \cref{res:imperfect_inspections} remain optimal if $\rho$ is locally perturbed above $0$, provided that the detectability parameter $\d$ is large enough and $u_0 > u_1$, i.e., shirking yields a strictly higher annuitized flow payoff than working; for a formal statement, see \cref{res:low recovery robustness} in \cref{sec:robust_small_rho}. Perturbing $\rho$ affects the agent's passage probability only when he works after having previously shirked. Such shirk-before-work deviations do not bind under the above conditions. 

In the remainder of this section, we illustrate how a higher recovery rate affects the form of the optimal inspection policy. We first describe how the inspection technology affects the agent's dynamic work incentives. We show that if $\rho$ is large enough, then only local deviations bind, and it is optimal to inspect with a constant hazard rate (\cref{res:high_recovery_exponential}). 

\subsection{Work incentives under the inspection technology} \label{sec:inspection_role}

Consider the inspection technology with recovery rate $\rho \geq 0$. Fix $t > s > 0$. In state $\th_s$ (which is hidden), the marginal effect of working, rather than shirking, at time $s$ on the probability of passing a time-$t$ inspection can be shown to equal
\begin{equation} \label{eq:marginal_effect}
    \Brac{  \d (1 - \th_s) + \rho \th_s}  \exp \Set{  - \d \int_{s}^{t} (1 - a_\tau) \de \tau - \rho \int_{s}^{t} a_\tau \de \tau}.
\end{equation}
The expression in \eqref{eq:marginal_effect} is the product of two terms. The first term, in brackets, captures the effect of working on the state transition at time $s$. In state $0$, working prevents transitions to state $1$ at rate $\d$. In state $1$, working generates transitions to state $0$ at rate $\rho$.  The exponential term in \eqref{eq:marginal_effect} captures the effect of $\th_s$ on $\th_{t}$. Formally, this term is the difference in conditional probabilities, $\P (\th_{t} = 0 | \th_s= 0) - \P (\th_{t} = 0 | \th_s= 1)$, for a given action path $(a_\tau)_{s \leq \tau < t}$. 


From \eqref{eq:marginal_effect}, we see that
the relative values of $\rho$ and $\d$ determine whether the passage probability $p_t$ is a supermodular or submodular function of the action history $(a_s)_{0 \leq s < t}$. If $\rho <  \delta$, then the marginal effect in \eqref{eq:marginal_effect} is greatest if the agent works before time $s$ (making $\th_s = 0$ most likely) and after time $s$ (making  the exponential term largest). Thus, the passage probability is a \emph{supermodular} function of the action path---different periods of work are complements. In this case, \emph{global} deviations are tempting because once the agent shirks for a short period, additional shirking becomes more attractive. 



If $\rho > \delta$, then the marginal effect in \eqref{eq:marginal_effect} is greatest if the agent shirks before time $s$ (making $\th_s = 1$ most likely) and after time $s$ (making  the exponential term largest). The passage probability is a \emph{submodular} function of the action path---different periods of work are substitutes. This force makes \emph{local} deviations tempting; once the agent shirks for a short period, additional shirking becomes less attractive. 



In the special case $\rho = \d$, the expression in \eqref{eq:marginal_effect} simplifies dramatically to $\d e^{-\d (t - s)}$. The marginal effect of working at time $s$ on the probability of passing a time-$t$ inspection is independent of all other action choices; this is the case studied in \cite{Varas2020} and \cite{wagner2021relational,achim2024tension}.

\subsection{Optimal inspection policy with high recovery rate}

With a sufficiently high recovery rate $\rho$,  the binding deviations are local. Consider the relaxed problem requiring only that locally shirking at each time $s$ is unprofitable. Let $\E_s = \E [ \cdot | T > s]$.  The principal chooses a random inspection time $T$ to solve 
\begin{equation} \label{eq:local_relaxed}
\begin{aligned}
    &\text{minimize} && \E  e^{- (\l_G + r) T} \\
    &\text{subject to} &&       \E_s \d e^{-(\l_G + r + \rho) ( T - s)}  U_1 \geq (u_0-u_1) +  (\l_G - \l_B)U_1, \quad s \geq 0.
\end{aligned}
\end{equation}
The right side of the inequality captures the marginal benefit from locally shirking at time $s$, in the absence of inspections. This benefit reflects the change in the annuitized flow payoff and the marginal effect on the probability of the project ending. The left side captures the marginal loss, due to the upcoming inspection, from locally shirking at time $s$. From \eqref{eq:marginal_effect}, the marginal effect of locally shirking at time $s$ on the probability of failing a time-$T$ inspection is $\d e^{-\rho (T - s)}$. This term  is multiplied by the discounted utility $e^{-(\l_G + r)(T-s)} U_1$ from passing a time-$T$ inspection. Here, $\l_G$ appears in the exponent because the agent works after locally shirking.

Whenever $\rho > 0$, the loss function in the local constraint is more convex, as a function of $T$, than the objective.  Therefore, inspecting with a constant hazard rate is the cheapest way to deter \emph{local} deviations, regardless of the values of $\d$, $\l_G$, and $\l_B$; see \cref{res:exponential_sol} (\cref{sec:prelim}). On the other hand, these parameters determine whether deterring local deviations is sufficient to deter all deviations.

We next identify a range of parameters for which deterring local deviations is sufficient. In this case, it is uniquely optimal to inspect with a constant hazard rate.

\begin{thm}[Random inspections with recovery] \label{res:high_recovery_exponential}
   Consider the inspection technology with recovery rate $\rho > 0$. 
   If $\rho + \l_G \geq \d+\l_B$, then the following policy is uniquely optimal. The gaps $(T_n - T_{n-1})_{n \geq 1}$ are independently and identically distributed according to an exponential distribution with hazard rate
    \begin{equation} \label{eq:gamma_ast}
    \frac{\g^\ast \,  \d \, U_1}{\l_G + r + \rho + \g^\ast}
    =
    (u_0 - u_1) + (\l_G - \l_B) U_1. 
    \end{equation}
\end{thm}

The condition $\rho + \l_G \geq \d + \l_B$ ensures that the binding constraints are local. If $\l_G = \l_B$, then this inequality reduces to $\rho \geq \d$. This is precisely the condition under which the passage probability is submodular in the action path.\footnote{If $\l_G \neq \l_B$, then there is an additional effect. When the agent shirks, his continuation value decreases. If shirking lengthens the project ($\l_G > \l_B$), then a lower continuation value makes working more attractive, which encourages local deviations. If shirking shortens the project $(\l_G < \l_B$), then a lower continuation value makes shirking more attractive, which encourages global deviations.} 
If $\rho + \l_G > \d + \l_B$, then the hazard rate $\g^\ast$ in \cref{res:high_recovery_exponential} is strictly higher than the hazard rate in \cref{res:imperfect_inspections}.\ref{res:delaying_imperfect}.\footnote{In the limit as $\rho$ tends to $0$, the condition $\rho + \l_G \geq \d + \l_B$ reduces to $\d \leq \l_G - \l_B$, which can be satisfied only in the innovation regime ($\l_G > \l_B$).  Note that the inequality $\d \leq \l_G - \l_B$ is inconsistent with the condition $\d > 2 \l_G - \l_B + r$ imposed in \cref{res:imperfect_inspections}.} This higher rate is necessary to deter local deviations, which become more attractive with quicker recovery.

A striking feature of the optimal inspection policy in \cref{res:high_recovery_exponential} is that inspections are conducted in arbitrarily short succession with positive probability, even though the probability that the evidence state changes between such inspections becomes arbitrarily small. To be sure, this inspection timing is not a cost-effective way to deter \textit{global} deviations, as discussed in \cref{sec:imperfect_inspections}.  Nevertheless, the policy in \cref{res:high_recovery_exponential} is the cheapest way to deter all \textit{local} deviations, and the local deviations are binding in this setting.  

In a setting without breakthroughs or breakdowns, \cite{Varas2020} consider an inspection technology with $\rho = \d$. They show that conducting inspections with a constant hazard rate is the cheapest way to induce full effort (Proposition 4, p.\ 2913). This result is essentially a special case of our \cref{res:high_recovery_exponential} with $\l_G = \l_B = 0$ and $\rho = \d$.\footnote{Their result is not exactly a special case of ours because their payoff structure is different and they restrict the agent's (continuous) action to an interval $[0, \bar{a}]$, where $\bar{a} < 1$. So even under maximal effort, there are random state transitions.}  By considering different inspection technologies, our paper highlights how the inspection technology determines the structure of the binding deviations, which in turn shapes the form of the optimal inspection policy.

\section{Conclusion} \label{sec:conclusion}
We study the optimal timing of inspections in a dynamic moral hazard setting. Under the main specification, we find that different forms of inspection policies are optimal for encouraging different kinds of tasks. Predictable inspections are better for motivating an agent to work toward a breakthrough, such as a technological innovation. Random inspections are typically better for motivating an agent to work to prevent a breakdown, such as a workplace accident. Under our inspection technology, global deviations bind. The optimal inspection policy is designed to most cheaply deter such deviations. As a result, the optimal timing of inspections is determined by the relative curvature of the project's survival probability when the agent works and when the agent shirks for an extended duration.

\appendix

\newpage
\section{Main proofs} \label{sec:proofs}

\subsection{Preliminaries} \label{sec:prelim}

\paragraph{Notation} Throughout the proofs we use the notation $\l_0 \coloneqq \l_B + r$ and $\l_1 \coloneqq \l_G + r$. With this notation, $U_i = u_i/ \l_i$ for $i = 0,1$. We express the solutions in terms of the five (strictly positive) parameters $\l_0$, $\l_1$, $\d$, $U_0$, $U_1$. To be sure, comparative statics are evaluated with respect to the primitive parameters $\l_B$, $\l_G$, $\d$, $r$, $u_0$, $u_1$.


\paragraph{Constrained optimization} In the proofs below, we solve two constrained optimization problems of the following general form. Given parameters $A$, $\a$, and $\b$, choose a distribution $F$ over $(0, \infty)$ to solve
\begin{equation} \label{eq:program_lemma}
\begin{aligned}
&\text{minimize} && \int_{(0, \infty)} e^{-\b t} \de F (t) \\
    &\text{subject to} && \int_{(s, \infty)} (e^{-\a ( t- s)} - A) \de F (t) \geq 0, \qquad s \geq 0. 
\end{aligned}
\end{equation}

The next lemma states the solution for a range of parameter values. 

\begin{lem}[Exponential solution] \label{res:exponential_sol}
Given $0 < A < 1$ and $\a > \b > 0$, the unique solution of \eqref{eq:program_lemma} is the exponential distribution with hazard rate $\g^\ast  = \a A / (1 - A)$. 
\end{lem}
The proof of \Cref{res:exponential_sol} is in \cref{sec:proof_exponential_lemma}.

\paragraph{Deviations} We introduce notation for the agent's utility from deviations. For $a \in \AA$ and $\tau \geq 0$, let
\begin{equation} \label{eq:deviation_notation}
    U(a ; \tau) = 
\int_{0}^{\tau} D_t(a) u(a_t) \de t + p_\tau(a) D_\tau(a)  U_1.
\end{equation}
With this notation, the constraints in \eqref{eq:next_inspection_A} take the form $\E U(a ; T) \leq U_1$, for each $a \in \AA$. We consider a few special action paths: always shirk, shirk-before-work, and work-before-shirk. For $s, \tau \geq 0$, let
\begin{equation} \label{eq:general_def}
    U_S(\tau) = U ( \mathbf{0}; \tau), \qquad U_{\SW}(s;\tau) = U ( 1_{[s, \infty)} ; \tau), \qquad U_{\WS} ( s; \tau) = U ( 1_{[0,s)} ; \tau).
\end{equation}
For $s \leq \tau$, evaluating these integrals and simplifying gives
\begin{equation} \label{eq:deviation_expressions}
\begin{aligned}
    U_{S} (\tau) &=  U_0( 1 - e^{-\l_0 \tau}) + U_1 e^{-(\d + \l_0) \tau},  \\
    U_{\SW} (s; \tau) &= U_0( 1 - e^{-\l_0 s}) + U_1 e^{-\l_0 s} \Brac{1 - e^{-\l_1 ( \tau - s)} (1 - e^{-\d s}) }, \\
    U_{\WS} (s; \tau) &= U_1 + e^{-\l_1 s} \Brac{ U_0 (1 - e^{-\l_0 (\tau- s)}) - U_1 ( 1 - e^{- (\d + \l_0) (\tau - s)})}.
\end{aligned}
\end{equation}
For $s > \tau$, we have $U_{\SW} (s;\tau) = U_{\SW} ( \tau; \tau)$ and $U_{\WS} (s ;\tau) = U_{\WS} (\tau ; \tau) = U_1$. Note that $U_{S} (\tau) = U_{\SW}(\tau;\tau)  = U_{\WS} (0;\tau)$. 

We often work with losses rather than gains. Setting $L ( a;\tau) = U_0 - U(a;\tau)$, the constraints in \eqref{eq:next_inspection_A} can be expressed as $\E L (a; T) \geq U_0 - U_1$, for $a$ in $\AA$.

\paragraph{Bounding number of zeros}
In the proofs below, we use the following bound on the number of zeros of certain sums of exponentials. 

\begin{lem}[Zeros]  \label{res:zeros} Define $g \colon \R \to \R$ by $g(t) = \sum_{i=1}^{n} A_i e^{-\a_i t}$, for some integer $n\geq 1$,  distinct real exponents $\a_1, \ldots, \a_n$, and nonzero coefficients $A_1, \ldots, A_n$. The function $g$ has at most $n - 1$ zeros.
\end{lem}

Through a change of variables $x = e^{-t}$, we can equivalently bound the number of zeros of the function $\bar{g}(x) = \sum_{i=1}^{n} A_i x^{\a_i}$ on the domain $(0, \infty)$.
This bound can be proven by induction, using the following observations. If $\a_i \neq 0$, then the function $\bar{h}(x) = x^{-\a_i} \bar{g}(x)$ has the same strictly positive zeros as $\bar{g}$. The function $\bar{h}$ has a constant term, so its derivative $\bar{h}'$ has at most $n-1$ nonzero terms. The function $\bar{h}'$ has a zero between any two zeros of $\bar{h}$, so $\bar{h}$ has at most one more strictly positive zero than $\bar{h}'$.

\subsection{Proof of \texorpdfstring{\cref{res:NoInspections}}{Proposition~\ref{res:NoInspections}}}   \label{sec:proof_no_inspections}

We discretize the agent's problem. Fix $\D > 0$. In the $\D$-discretized problem, the agent can change his action only at times $k \D$ for $k =0,1, \ldots$. Let $V_\D$ denote the agent's supremal utility in the $\D$-discretized problem. The Bellman equation reads
\[
    V_\D = \max_{i=0, 1} \Set{ U_i (1 - e^{-\l_i \D}) + e^{-\l_i \D} V_\D}.
\]
The unique solution is $V_\D = \max \{ U_0, U_1\}$. By a limiting argument,\footnote{\label{ft:discretization}Any right-continuous function $a \colon [0, \infty) \to \{0,1\}$ can be expressed as the pointwise limit of a sequence of step functions $a_n \colon [0, \infty) \to \{0, 1\}$ defined by $a_n(t) = a (k/2^n)$ if $(k-1)/2^n \leq t < k /2^n$, for $k = 1, 2, \ldots$. By dominated convergence, as $n$ tends to $\infty$, the agent's expected utility from $a_n$ converges to the agent's expected utility from $a$.} it follows that the agent's value in the continuous-time problem is also $\max \{U_0, U_1\}$. Thus, working forever is optimal if and only if $U_1 \geq U_0$.

\subsection{Proof of \texorpdfstring{\cref{res:speeding_up_perfect}}{Theorem \ref{res:speeding_up_perfect}}}

We separate into two cases.
\paragraph{Case 1} Suppose $u_0 \geq u_1$. The period $\tau^\ast$ in  \eqref{eq:tau_star} is well-defined because $0 < (U_0 - U_1)/U_0 < 1$. 

First we check that the constant $x^\ast \coloneqq e^{-\l_1 \tau^\ast}$ is the unique solution of the relaxed problem \eqref{eq:perfect_delaying_problem_cov}. Define $\bar{L}_S \colon [0,1] \to \R$ by $\bar{L}_S (x) = U_0 x^{\l_0 /\l_1}$. Since $\l_1 > \l_0$, the function $\bar{L}_S$ is strictly concave and strictly increasing. From \eqref{eq:tau_star}, we have $\bar{L}_S(x^\ast) = U_0 - U_1$. If a $(0,1)$-valued 
random variable $X$ satisfies $\E X \leq x^\ast$, then
\[
    \E \bar{L}_S(X) \leq \bar{L}_S ( \E X) \leq \bar{L}_S (x^\ast )= U_0 - U_1,
\]
and equality holds in both inequalities if and only if $X$ is the constant $x^\ast$. Thus, the constant $x^\ast$ is the unique solution of \eqref{eq:perfect_delaying_problem_cov}. In terms of the original variables, the constant $\tau^\ast$ is the unique solution of \eqref{eq:perfect_delaying_problem}.

It remains to check that $\tau^\ast$ is feasible in the original problem \eqref{eq:next_inspection_A}. Recall the notation in  \eqref{eq:deviation_notation}. From the definition of $\tau^\ast$ in \eqref{eq:tau_star}, we have $U( \mathbf{0}; \tau^\ast) = U_1$. Consider an arbitrary action path $a$ in $\AA$. If $\int_{0}^{\tau^\ast} a_t \de t = \tau^\ast$, then $U(a;\tau^\ast) = U_1$. If $\int_{0}^{\tau^\ast} a_t \de t  < \tau^\ast$, then $p_{\tau^\ast}(a) =  0$, so
\[
   U(a;\tau^\ast) =  \int_{0}^{\tau^\ast} D_t (a) u (a_t) \de t
    \leq    
     \int_{0}^{\tau^\ast} D_t (\mathbf{0}) u_0 \de t
    = U(\mathbf{0}; \tau^\ast),
\]
where the inequality holds because for each time $t$, we have $D_t(a) \leq D_t( \mathbf{0})$  (because $\l_G > \l_B$) and $u (a_t) \leq u_0$ (because $u_1 \leq u_0$).

\paragraph{Case 2} Suppose $u_1 > u_0$. In this case, we consider shirk-before-work deviations. Recall the notation from \eqref{eq:general_def}. With perfect inspections, for $0 < s < t$ we have
\begin{equation}
\begin{aligned}
    U_{\SW} ( s; t) &= U_0 (1 - e^{-\l_0 s}) + U_1 e^{-\l_0 s} (1 - e^{-\l_1 (t -s)}), \\
    U_{\SW}' (s;t) &= e^{-\l_0 s} \Brac{  (U_0 - U_1) \l_0 - U_1 ( \l_1 - \l_0)e^{-\l_1 (t -s)} },
\end{aligned}
\end{equation}
where $U_{\SW}'$ denotes the derivative of $U_{\SW}$ with respect to its \emph{first} argument. By assumption, $\l_G > \l_B$, so $\l_1 > \l_0$. Therefore, for each fixed $t > 0$, the derivative $U_{\SW}'( \cdot ; t)$ is strictly single-crossing from above over $(0,t)$.

We identify the binding shirk-before-work deviation. Define $\bar{s}$ and $\tau^\ast$ by
    \begin{equation} \label{eq:inspection_time_perfect_periodic_case2}
    e^{-\l_0 \bar{s}} =  \frac{\l_1 - \l_0}{\l_1}, \qquad
 e^{-\l_0 \tau^\ast} =\frac{\l_1 - \l_0}{\l_1} \Paren{ \frac{ (U_0 - U_1)\l_0}{U_1 ( \l_1 - \l_0)} }^{\l_0 /\l_1}.
\end{equation} 
Since $\l_1 > \l_0$ and $\l_1 U_1 = u_1 > u_0 = \l_0 U_0$, these values $\bar{s}$ and $\tau^\ast$ are well-defined and satisfy $0 < \bar{s} < \tau^\ast$. It can be checked that $U_{\SW}' ( \bar{s}; \tau^\ast) = 0$ and $U_{\SW} ( \bar{s}; \tau^\ast) = U_1$. 

Consider the relaxed problem of choosing a $(0, \infty)$-valued random variable $T$ to solve
\begin{equation} \label{eq:relaxed_problem_t_case2}
    \begin{aligned}
    &\text{minimize} && \E e^{- \l_1 T} \\
    &\text{subject to} &&  \E U_{\SW} ( \bar{s}  ; T) \leq U_1. 
\end{aligned}
\end{equation}
We change variables. Let $\bar{x} = e^{-\l_1 \bar{s}}$ and $x^\ast = e^{- \l_1 \tau^\ast}$. Note that $\bar{x} > x^\ast$. Upon setting $X = e^{-\l_1 T}$, we obtain the equivalent problem of choosing a $(0,1)$-valued random variable $X$ to solve
\begin{equation} \label{eq:relaxed_problem_x_case2}
\begin{aligned}
    &\text{minimize} && \E X  \\
    &\text{subject to} &&  \E \bar{L}(X) \geq U_0 - U_1,
\end{aligned}
\end{equation}
where $\bar{L} \colon [0,1] \to \R$ is defined by
\[
    \bar{L} (x) 
    = \begin{cases} 
        U_0 x^{\l_0 /\l_1} &\text{if}~ x \geq \bar{x}, \\
        U_0 \bar{x}^{\l_0 /\l_1}  - U_1 \bar{x}^{\l_0/\l_1} (1  - x/\bar{x})  &\text{if}~ x < \bar{x}.
    \end{cases}
\]

Since $\l_1 > \l_0$, the function $\bar{L}$ is concave and strictly increasing.\footnote{To see that the kink preserves concavity, note that the derivative of $\bar{L}$ jumps at $\bar{x}$ by 
\[
   (\l_0/\l_1) U_0 \bar{x}^{(\l_0-\l_1)/\l_1} - U_1 \bar{x}^{(\l_0-\l_1)/\l_1}  = \bar{x}^{(\l_0 - \l_1)/\l_1} (U_0 \l_0 - U_1 \l_1)/\l_1,
\]
which is strictly negative because $u_1 > u_0$.} Moreover, $\bar{L}$ is strictly concave over $[\bar{x},1]$. From the definition of $\tau^\ast$ in \eqref{eq:inspection_time_perfect_periodic_case2}, we have $\bar{L}(x^\ast) = U_0 - U_1$. If a $(0,1)$-valued random variable $X$ satisfies $\E X \leq x^\ast$, then 
\[
    \E \bar{L}(X) \leq \bar{L} ( \E X) \leq \bar{L} (x^\ast )= U_0 - U_1,
\]
and equality holds in both inequalities if and only if $\E X = x^\ast$ and $X$ concentrates on $(0, \bar{x}]$, the affine segment of $\bar{L}$. Therefore, the constant $x^\ast$ solves \eqref{eq:relaxed_problem_x_case2}, and all solutions of \eqref{eq:relaxed_problem_x_case2} must concentrate on $(0, \bar{x}]$. In terms of the original variables, the constant $\tau^\ast$ solves \eqref{eq:relaxed_problem_t_case2}, and all solutions of \eqref{eq:relaxed_problem_t_case2} must concentrate on $[\bar{s}, \infty)$. 

It remains to check that $\tau^\ast$ is feasible in the original problem \eqref{eq:next_inspection_A}. We claim that over any inspection-free interval, the agent strictly prefers shirking for length $s$ and  then working for length $w$ to working for length $w$ and then shirking for length $s$. The passage probability is the same under both action paths, so it suffices to check that
\begin{equation} \label{eq:shirk_first_binding}
    U_0 ( 1 - e^{-\l_0 s}) + e^{-\l_0 s} U_1 ( 1 - e^{-\l_1 w}) 
    >
    U_1 ( 1 - e^{-\l_1 w}) + e^{-\l_1 w} U_0 ( 1 - e^{-\l_0 s}).
\end{equation}
To see this, observe that the difference between the left side and the right side is
\[
    (U_0 - U_1) (1 - e^{- \l_0 s})(1 - e^{-\l_1 w}),
\]
which is strictly positive since $U_0 > U_1$. 

Since every right-continuous action path is the pointwise limit of some sequence of step functions, we conclude that for any $a$ in $\AA$,\footnote{This inequality is immediate if $\int_{0}^{\tau^\ast} (1 - a_s) \de s = 0$, so we may assume $\int_{0}^{\tau^\ast} (1 - a_s) \de s > 0$. Thus, $p_{\tau^\ast} (a)$ is the limit of the passage probabilities along the sequence.} 
\[
    U ( a ; \tau^\ast) \leq \sup_{0 \leq s \leq \tau^\ast} U_{\SW}(s;\tau^\ast) = U_{\SW} ( \bar{s}; \tau^\ast ) = U_1.
\]

\subsection{Proof of \texorpdfstring{\cref{res:delaying_perfect}}{Theorem \ref{res:delaying_perfect}}}

The relaxed problem \eqref{eq:perfect_speeding_problem}, expressed in terms of the distribution $F$ over $(0,\infty)$ of the random variable $T$, takes the form 
\begin{equation*}
\begin{aligned}
    &\text{minimize} && \int_{(0,\infty)} e^{- (\l_G + r) t} \de F(t) \\
    &\text{subject to} &&  \int_{(s,\infty)} \Brac{  U_0 e^{-(\l_B + r) (t-s)} -  (U_0 - U_1)}  \de F(t) \geq 0,  \quad s \geq 0.
\end{aligned}
\end{equation*}
This problem can be expressed in the form of \eqref{eq:program_lemma} with $A = (U_0 - U_1)/U_0$; $\a = \l_0$; and $\b = \l_1$. Since $0 < (U_0 - U_1)/U_0 < 1$ and $\l_0 > \l_1 > 0$, we can apply \cref{res:exponential_sol} to conclude that the unique solution of \eqref{eq:perfect_speeding_problem} is the 
exponential distribution with hazard rate $\g^\ast = \l_0 (U_0 - U_1)/U_1$. 

It remains to check that this exponential distribution is feasible in the original problem \eqref{eq:next_inspection_A}. Suppose that the time until the next inspection is exponentially distributed with hazard rate $\g^\ast$. By construction, no work-before-shirk deviation is profitable. To show that no other deviations are profitable, it suffices to show that once the agent has shirked (for a positive duration), he finds it optimal to shirk until the next inspection. Once the agent has shirked, he is certain to fail the next inspection, so his continuation problem is equivalent to the no-inspection problem with discount rate $r + \g^\ast$. By the proof of \cref{res:NoInspections} (\cref{sec:proof_no_inspections}), shirking is optimal if
\[  
   \frac{ U_1 \l_1}{\l_1 + \g^\ast} \leq    \frac{U_0  \l_0}{\l_0 + \g^\ast},
\]
which holds (strictly) because $U_1 < U_0$ and $\l_1 < \l_0$.

\subsection{Proof of \texorpdfstring{\cref{res:imperfect_inspections}}{Theorem \ref{res:imperfect_inspections}}} \label{sec:main_proof}

We separate into cases and solve a different relaxed problem in each case. Then we use these relaxed solutions to prove the result. Recall the expressions in \eqref{eq:deviation_expressions} for the agent's payoffs from the shirk-before-work and work-before-shirk deviations. In particular, for $s \leq \tau$, we have
\[
 U_{\SW} (s; \tau) = U_0( 1 - e^{-\l_0 s}) + U_1 e^{-\l_0 s} \Brac{1 - e^{-\l_1 ( \tau - s)} (1 - e^{-\d s}) }.
\]

Define the period $\tau^\ast$ to be the largest time $t$ such that
\begin{equation} \label{eq:SW_Bellman_imperfect}
     \max_{s \in [0, t]}\, U_{\SW} (s; t) \leq U_1.
\end{equation}
It can be checked that $\tau^\ast$ is well-defined and strictly positive; moreover, \eqref{eq:SW_Bellman_imperfect} holds with equality at $t = \tau^\ast$.%
\footnote{\label{ft:well_defined}For $0 \leq s < t$, as $(s, t)$ tends to $(0,0)$, the derivative $U_{\SW}'(s ; t)$ tends to $\l_0 ( U_0 - U_1) - \d U_1$, which is strictly negative by \cref{as:sufficient}. So, for $t$ sufficiently small, the maximum on the left side of \eqref{eq:SW_Bellman_imperfect} is achieved at $s = 0$ and hence \eqref{eq:SW_Bellman_imperfect} holds. On the other hand, $U_{\SW} ( t ; t) \to U_0$ as $t \to \infty$, so \eqref{eq:SW_Bellman_imperfect} is violated for $t$ sufficiently large. By Berge's maximum theorem, the left side of \eqref{eq:SW_Bellman_imperfect} is continuous in $t$. Thus, $\tau^\ast$ is well-defined, and  the inequality in \eqref{eq:SW_Bellman_imperfect} must hold with equality at $t = \tau^\ast$.} 
For all $s$ in $[0, \tau^\ast]$, we have
\[
U_{\SW}( s ; \tau^\ast) \leq U_1 = U_{\SW} (0 ; \tau^\ast),
\]
so $U_{\SW}' (0 ; \tau^\ast ) \leq 0$. Here and below, we add a prime to denote
the derivative of a function with respect to its \emph{first} argument. 
We have
\begin{equation*}
\begin{aligned}
    U_{\SW}'(0 ; \tau^\ast)
   &= (U_0 - U_1) \l_0 - e^{- \l_1 \tau^\ast} U_1 \d, \\
   U_{\SW}''(0 ; \tau^\ast) 
    &= - (U_0 - U_1) \l_0^2 + e^{-\l_1 \tau^\ast} U_1 \d (\d +2 \l_0 - 2\l_1).
\end{aligned}
\end{equation*}
We check that $U_{\SW}'(0;\tau^\ast) < 0$. Suppose for a contradiction that $U_{\SW}'(0; \tau^\ast )  = 0$. Then 
\begin{equation} \label{eq:suff_condition}
    \begin{aligned}
    U_{\SW}'' (0;\tau^\ast) &=  U_{\SW}'' (0;\tau^\ast) + U_{\SW}'(0;\tau^\ast) ( \d + 2\l_0 - 2 \l_1) \\
    &= (U_0 - U_1) \l_0 (\d + \l_0 - 2 \l_1) \\
    &> 0,
    \end{aligned}
\end{equation}
where the last inequality holds because $\d > 2\l_1 - \l_0$ by assumption. Thus, $U_{\SW} (s ; \tau^\ast) > U_{\SW} (0 ; \tau^\ast) = U_1$ for $s$ sufficiently small, contrary to the definition of $\tau^\ast$.

We separate into cases according to the condition 
\begin{equation} \label{eq:cutoff}
    U_1 ( \d + \l_0 - \l_1) e^{- \d \tau^\ast} \geq U_0 (\l_0 - \l_1).
\end{equation}
If $\l_1 \geq \l_0$, then \eqref{eq:cutoff} holds: by assumption, $\d + \l_0 > 2 \l_1 > \l_1$, so the left side is strictly positive and the right side is nonpositive.

The rest of the proof proceeds as follows. We consider three different relaxed problems. We use the solutions of these relaxed problems to establish the two forms of the optimal inspection policy. Then we characterize the threshold $\bar{\l}_B$, as discussed in \cref{rem:threshold_B}. The four claims are proved in \cref{sec:proofs_of_claims}.  

\paragraph{Relaxed problem: shirk} Suppose that \eqref{eq:cutoff} holds and $U_S ( \tau^\ast) = U_1$.  We check that the constant $x^\ast \coloneqq e^{-\l_1 \tau^\ast}$ is the unique solution of the relaxed problem \eqref{eq:imperfect_maintenance_cov}. Recall the loss function $\bar{L}_S \colon [0,1] \to \R$ given by
\[
    \bar{L}_S (x) = x^{\l_0/\l_1} (U_0 - U_1 x^{\d/\l_1} ).
\]
If $\l_0 > \l_1$, define $x_c$ by $U_1 (\d + \l_0 - \l_1) x_c^{\d/ \l_1} = U_0 ( \l_0 - \l_1)$. In the case $\l_0 > \l_1$, the condition \eqref{eq:cutoff} holds if and only if $x^\ast \geq x_c$. 

\begin{claim} \label{claim:LS} The function $\bar{L}_S$ is strictly quasiconcave and has an interior maximizer, denoted $x_{0,S}$. If $\l_1 \geq \l_0$, then $\bar{L}_S$ is strictly concave over $[0, 1]$. If $\l_1 < \l_0$, then over the interval $[x_c, 1]$, the function $\bar{L}_S$ is strictly concave and coincides with its concavification $\cav \bar{L}_S$.
\end{claim}

\cref{claim:LS} and the subsequent claims are proven in \cref{sec:proofs_of_claims}. Using \cref{claim:LS}, we show that the constant $x^\ast$ is the unique solution of \eqref{eq:imperfect_maintenance_cov}. By assumption, $U_S (\tau^\ast) = U_1$, so $\bar{L}_S(x^\ast) = U_0 - U_1$. Since $\bar{L}_S (1) = U_0 - U_1$, it follows that $x^\ast < x_{0,S}$. There are two cases. 

\begin{enumerate}
    \item $\l_1 \geq \l_0$. Suppose that a $(0,1)$-valued random variable $X$ satisfies $\E X \leq x^\ast$. By \cref{claim:LS},
\[
    \E \bar{L}_S ( X) \leq \bar{L}_S ( \E X) \leq \bar{L}_S ( x^\ast) = U_0 - U_1,
\]
with equality only if $X$ equals the constant $x^\ast$.

\item  $\l_1 < \l_0$. Suppose that a $(0,1)$-valued random variable $X$ satisfies $\E X \leq x^\ast$. By the definition of concavification and \cref{claim:LS},
\begin{equation} \label{eq:three_ineq}
    \E  \bar{L}_S (X) \leq \cav \bar{L}_S ( \E X) \leq \cav \bar{L}_S(x^\ast) = \bar{L}_S ( x^\ast) = U_0 - U_1,
\end{equation}
where the second inequality holds because $\cav \bar{L}_S$ is strictly increasing\footnote{This holds because $\cav \bar{L}_S$ is concave and is uniquely maximized at the point $x_{0,S}$.} over $[0,x_{0,S}]$ and $\E X \leq x^\ast < x_{0,S}$; the first equality holds because $x^\ast \geq x_c$ by \eqref{eq:cutoff}. Moreover, since $x^\ast \geq x_c$, both inequalities in \eqref{eq:three_ineq} hold with equality only if $X$ equals the constant $x^\ast$.
\end{enumerate}

\paragraph{Relaxed problem: shirk-before-work} Suppose that \eqref{eq:cutoff} holds and $U_S ( \tau^\ast) < U_1$. Let $\bar{s}$ be the largest maximizer of $U_{\SW} ( \cdot ; \tau^\ast)$ over $[0, \tau^\ast]$. Since $U_{S} (\tau^\ast) < U_1$, we have $\bar{s} < \tau^\ast$. Since $U_{\SW}'(0 ; \tau^\ast) < 0$, it can be checked that $\bar{s} > 0$.\footnote{\label{ft:deriv} Suppose for a contradiction that $\bar{s} = 0$. Then $U_{\SW} ( s ; \tau^\ast) < U_1$ for all $s$ in $(0, \tau^\ast]$. Since $U_{\SW}' ( 0 ; \tau^\ast) < 0$, it follows that \eqref{eq:SW_Bellman_imperfect} holds for some $t$ strictly greater than $\tau^\ast$, contrary to the definition of $\tau^\ast$. (We must consider the derivative because $U_{\SW} ( \cdot ; \tau^\ast)$ is not bounded away from $U_1$ over $(0, \tau^\ast]$.)}

Consider the relaxed problem of choosing a $(0,\infty)$-valued random variable $T$ to solve
\begin{equation} \label{eq:shirk_work_original}
\begin{aligned}
    &\text{minimize} &&\E e^{-\l_1 T} \\
    & \text{subject to} && \E U_{\SW} ( \bar{s} ; T) \leq U_1.
\end{aligned}
\end{equation}
We change variables.  Let $\bar{x} = e^{-\l_1 \bar{s}}$ and $x^\ast = e^{- \l_1 \tau^\ast}$. Note that $\bar{x} > x^\ast$. Define $\bar{L} \colon [0,1] \to \R$ by\footnote{We extend the function $\bar{L}$ to the point $0$ by continuity.}
\[
    \bar{L} (x) = U_0 - U_{\SW} (\bar{s}; - \l_1^{-1} \log x).
\]
Consider the equivalent relaxed problem of choosing a $(0,1)$-valued random variable $X = e^{-\l_1 T}$ to solve
\begin{equation} \label{eq:relaxed_imperfect}
\begin{aligned}
    &\text{minimize} &&\E X\\
    & \text{subject to} && \E \bar{L}(X) \geq U_0 - U_1.
\end{aligned}
\end{equation}


\begin{claim} \label{claim:U} The function $\bar{L}$ is strictly quasiconcave, and its unique maximizer, $x_0$, satisfies $x_0 \geq \bar{x}$. The function $\bar{L}$ is concave over $[0, x_0]$ and affine over $[0, \bar{x}]$. If $x_0 > \bar{x}$, then $\bar{L}$ is strictly concave over $[\bar{x}, x_0]$. 
\end{claim}

Using \cref{claim:U}, we show that the constant $x^\ast$ solves \eqref{eq:relaxed_imperfect}. From the definition of $\tau^\ast$ in \eqref{eq:SW_Bellman_imperfect}, we have
$\bar{L}(x^\ast)= U_0 - U_1$. Suppose that a $(0,1)$-valued random variable $X$ satisfies $\E X \leq x^\ast$. Let $X' = \min \{ X, x_0\}$. We have
\begin{equation} \label{eq:prime_inequalities}
    \E  \bar{L} (X) \leq \E \bar{L}(X') \leq  \bar{L}( \E X') \leq \bar{L} (x^\ast) = U_0 - U_1,
\end{equation}
where the first inequality uses the definition of $X'$; the second inequality holds because $\bar{L}$ is concave over $[0, x_0]$; and the third inequality holds because $\bar{L}$ is strictly increasing over $[0, x_0]$, and $\E X' \leq \E X \leq x^\ast < \bar{x} \leq x_0$. Moreover, equality holds in all three inequalities in \eqref{eq:prime_inequalities} if and only if $\E X = x^\ast$ and $X$ concentrates on $(0, \bar{x}]$. Therefore, the constant $x^\ast$ solves \eqref{eq:relaxed_imperfect}, and all solutions of  \eqref{eq:relaxed_imperfect} must concentrate on $(0, \bar{x}]$. In terms of the original variables, the constant $\tau^\ast$ solves \eqref{eq:shirk_work_original}, and all solutions of \eqref{eq:shirk_work_original} must concentrate on $[\bar{s}, \infty)$. 

\paragraph{Relaxed problem: work-before-shirk} Suppose that \eqref{eq:cutoff} is violated. It follows that $\l_0 > \l_1$. 
Consider the relaxed problem of choosing a $(0,\infty)$-valued random variable $T$ to solve
\begin{equation} \label{eq:program_ws_1}
\begin{aligned}
&\text{minimize} && \E e^{-\l_1 T} \\
&\text{subject to} &&  \E U_{\WS} (s ; T ) \leq U_1, \quad s \in \{0\} \cup [\tau, \infty),
\end{aligned}
\end{equation}
where the value of $\tau$ will be specified below. From \eqref{eq:deviation_expressions}, we have 
\[
    U_{\WS} (s;T) = U_1 + e^{-\l_1 (s \wedge T)} h ( s \wedge T; T),
\]
where the function $h$ is defined by
\[
    h(s;t) =  U_0 (1 - e^{-\l_0 (t- s)}) - U_1 ( 1 - e^{- (\d + \l_0) (t - s)}). 
\]
Therefore, problem \eqref{eq:program_ws_1} is equivalent to the problem of choosing a distribution $F$ on $(0, \infty)$ to solve
\begin{equation} \label{eq:relaxed_full}
\begin{aligned}
&\text{minimize} && \int_{(0, \infty)} e^{-\l_1 t} \de F(t) \\
&\text{subject to} &&\int_{ (s, \infty)} h (s; t) \de F(t) \leq 0, \quad s \in  \{0\} \cup [\tau, \infty).
\end{aligned}
\end{equation}

To solve \eqref{eq:relaxed_full}, we construct Lagrange multipliers. The constraints are indexed by $s$ in 
$\{0\} \cup [\tau, \infty)$. Attach a nonnegative mass multiplier $\h_0$ to the $s= 0$ constraint and a nonnegative, integrable density multiplier $\h(s)$ to the time-$s$ constraint, for all $s \geq \tau$. The Lagrangian becomes
\begin{multline*}
    L(F;\h_0, \h)
    = 
    \int_{ (0,\infty)}  e^{-\l_1 t} \de F(t)
    + \h_0 \int_{(0,\infty)} h(0;t) \de F(t) \\
    + \int_{\tau}^{\infty} \Brac{\int_{(s,\infty)} h(s;t) \de F(t) } \h(s) \de s.
\end{multline*}
Change the order of integration in the double integral to get
\[
   L(F; \h_0, \h) = \int_{(0, \infty)} I(t) \de F(t),
\]
where
\begin{equation} \label{eq:I_general}
    I(t) = e^{-\l_1 t} + \h_0  h(0;t) + \int_{\tau \wedge t}^{t} \h(s) h(s;t) \de s.
\end{equation}

Now we define the multipliers. For some $\bar{\h} \geq 0$, let $\h(s) = \bar{\h} e^{-\l_1 s}$ for all $s \geq \tau$. Plug in this expression, integrate, and group like terms. For $t \geq \tau$, we get
\begin{equation} \label{eq:I_long}
\begin{aligned}
    I(t) &= e^{- \l_1 t} \Brac{1  - \frac{\bar{\h}}{\l_1} \Paren{ \frac{ U_0 \l_0}{\l_0 - \l_1} - \frac{U_1 ( \d + \l_0)}{\d + \l_0 - \l_1}}}\\
    &\quad + e^{-\l_0 t} U_0 \Brac{ -\h_0 + \frac{\bar{\h}}{\l_0 - \l_1} e^{ (\l_0 - \l_1 )\tau} }\\
    &\quad + e^{- (\d + \l_0) t} U_1 \Brac{ \h_0  - \frac{\bar{\h}}{\d + \l_0 - \l_1} e^{(\d + \l_0 - \l_1) \tau}} \\
    &\quad  +  (U_0 - U_1) \Brac{ \h_0 + \frac{\bar{\h}}{\l_1} e^{-\l_1 \tau}}.
\end{aligned}
\end{equation}
By assumption, \eqref{eq:cutoff} is violated, so $\l_0 > \l_1$. Therefore, $\hat{\tau}$ is well-defined by \eqref{eq:gamma_ast_0}:
$e^{-\d \hat{\tau}}  = (\l_0 - \l_1) / (\d + \l_0 - \l_1)$. Let $\tau = \tau^\ast \wedge \hat{\tau}$. Define $\bar{\h}$ and $\h_0$ by
\begin{equation} \label{eq:optimal_multipliers}
    \bar{\h} = \l_1 \Paren{ \frac{ U_0 \l_0}{\l_0 - \l_1} - \frac{U_1 ( \d + \l_0)}{\d + \l_0 - \l_1}}^{-1}, \qquad
    \h_0 = \frac {\l_1 e^{(\l_0 - \l_1) \tau}}{U_0 \l_0 - e^{-\d \tau} U_1 ( \d + \l_0)}.
\end{equation}
The multiplier $\bar{\h}$ is well-defined and positive because $U_0 > U_1$ and $\l_0 (\d + \l_0 - \l_1) > (\d + \l_0) (\l_0 - \l_1)$. The multiplier $\h_0$ is well-defined and positive because
\begin{equation} \label{eq:integrand_ineq}
    e^{-\d \tau} < \frac{ U_0 (\l_0 - \l_1)}{U_1 (\d + \l_0 - \l_1)} < \frac{ U_0 \l_0 - U_1 \l_1}{U_1 (\d + \l_0 - \l_1)} < \frac{ U_0 \l_0}{U_1 ( \d + \l_0)},
\end{equation}
as we now check. The first inequality can be checked by cases: if $\tau = \hat{\tau}$, use the inequality $U_0 > U_1$; if $\tau = \tau^\ast$, use the fact that \eqref{eq:cutoff} is violated. The second inequality uses $U_0 > U_1$, and the third inequality uses \cref{as:sufficient}. 

Hereafter, we consider the function $I \colon (0,\infty) \to \R$ defined in \eqref{eq:I_general}, with the multipliers defined in \eqref{eq:optimal_multipliers}. 

\begin{claim} \label{claim:I} If $\tau^\ast < \hat{\tau}$, then $\argmin_{t > 0} I(t) = \{ \tau^\ast \}$. If $\tau^\ast \geq \hat{\tau}$, then  $\argmin_{t > 0} I(t) = [\hat{\tau}, \infty)$. 
\end{claim}

We apply \cref{claim:I} to solve \eqref{eq:relaxed_full}. There are two cases. 

First suppose $\tau^\ast \leq \hat{\tau}$. Let $\bar{s}$ be the largest maximizer of $U_{\SW} ( \cdot ; \tau^\ast)$ over $[0, \tau^\ast]$. Thus, $\bar{s} > 0$. We claim that a point mass 
on $\tau^\ast$  is the unique solution of  \eqref{eq:relaxed_full}.  First we check that the point mass $\d_{\tau^\ast}$ is a solution of \eqref{eq:relaxed_full}. By \cref{claim:I},  the point mass $\d_{\tau^\ast}$ minimizes the Lagrangian. We check that $\d_{\tau^\ast}$ satisfies all the constraints in \eqref{eq:relaxed_full} with equality.  Clearly, $U_{\WS} (s ; \tau^\ast) = U_1$ for all $s \geq \tau^\ast = \tau$. It remains to check that $U_{\WS} (0 ; \tau^\ast) = U_1$, or equivalently, $U_{\SW} ( \tau^\ast ; \tau^\ast) = U_1$. Let $U_{\SW}' ( \tau^\ast ; \tau^\ast)$ denote the \emph{left} derivative of the function $U_{\SW}(\cdot; \tau^\ast)$ at $\tau^\ast$. Since \eqref{eq:cutoff} is violated, we have
\begin{equation} \label{eq:endpoint}
\begin{aligned}
    U_{\SW}' ( \tau^\ast ; \tau^\ast) 
    &= e^{-\l_0 \tau^\ast} \Brac{  U_0 \l_0 - U_1 \l_1 - e^{-\d \tau^\ast} U_1 (\d + \l_0 - \l_1) }  \\
    &\geq e^{- \l_0 \tau^\ast} \l_1 (U_0 - U_1)
    \\&> 0.
\end{aligned}
\end{equation}
Over $(0, \tau^\ast)$, the function $s \mapsto U_{\SW}'(s ; \tau^\ast)$ is a sum of at most three exponentials, so it has at most two zeros by \cref{res:zeros}. Since $U'_{\SW} (0; \tau^\ast) < 0$ and $U_{\SW}' (\tau^\ast; \tau^\ast) > 0$, it follows that $U_{\SW}'(\cdot ; \tau^\ast)$ cannot cross zero from above over $(0, \tau^\ast)$, and hence $U_{\SW} ( \cdot ; \tau^\ast)$ cannot have an interior maximizer over $[0, \tau^\ast]$. Since $\bar{s} > 0$, we must have $\bar{s} = \tau^\ast$. Thus, $U_{\SW} ( \tau^\ast ; \tau^\ast) = U_1$. 

Now we check that there is no other solution of \eqref{eq:relaxed_full}. If $\tau^\ast < \hat{\tau}$, then by \cref{claim:I}, the point mass $\d_{\tau^\ast}$ is the unique minimizer of the Lagrangian. If $\tau^\ast = \hat{\tau}$, then by \cref{claim:I} all minimizers of the Lagrangian concentrate on $[\tau^\ast, \infty)$. For any such distribution other than $\d_{\tau^\ast}$, the constraint with $s=0$ is violated. 


Next,  suppose $\tau^\ast > \hat{\tau}$. By \cref{claim:I}, the Lagrangian is minimized by any distribution concentrating on $[\hat{\tau}, \infty)$. We claim that there is a unique distribution $F$ on $[\hat{\tau}, \infty)$ that satisfies the complementary slackness condition
\begin{equation} \label{eq:CS_s}
    \int_{(s, \infty)} \Brac{ U_0 (1 - e^{-\l_0 (t- s)}) - U_1 ( 1 - e^{- (\d + \l_0) (t - s)})} \de F(t) = 0,
\end{equation}
for all $s$ in $\{ 0 \} \cup [\hat{\tau}, \infty)$.\footnote{Technically, complementary slackness only implies equality for $s= 0$ and for almost every $s \geq \hat{\tau}$, but slack in the inequality is right-continuous in $s$, so this implies equality for every $s \geq \hat{\tau}$.}
By \cref{as:sufficient}, we have $U_1 (\d + \l_0)  > U_0 \l_0$, so we can apply \cref{res:uniqueness_sum} in \Cref{sec:uniqueness_delaying_perfect} to conclude that
the distributions on $[\hat{\tau}, \infty)$ that satisfy \eqref{eq:CS_s} for every $s \geq \hat{\tau}$ are precisely the distributions $F_\pi$, for $\pi$ in $[0,1]$, given by  \begin{equation} \label{eq:exp_sol}
    F_\pi(t) = \pi + (1 - \pi) (1 - e^{-\g^\ast (t - \hat{\tau})}), \qquad t \geq \hat{\tau},
\end{equation}
where $\g^\ast$ is defined in \eqref{eq:gamma_ast_0}. Plug this distribution $F_\pi$ into condition \eqref{eq:CS_s} with $s = 0$. This equation is linear in $\pi$.  Solving gives 
\begin{equation} \label{eq:pistar}
    \pi^\ast = \frac{(U_0 - U_1) \Paren{ e^{\lambda_0 \hat{\tau}} (\delta + \lambda_0 - \lambda_1) - (\delta + 2 \lambda_0 - \l_1) } }{U_1 (\delta + \lambda_0) - U_0 \lambda_0 }.
\end{equation}
It can be shown that $\pi^\ast$ is in $(0,1)$.\footnote{Consider \eqref{eq:CS_s} at $s = 0$ with the distribution $F_\pi$. We claim that the left side is strictly negative if $\pi = 1$ and strictly positive if $\pi = 0$. For a proof, see \cref{sec:am_gm}.} We conclude that $F_{\pi^\ast}$ is the unique solution of \eqref{eq:relaxed_full}. 

\paragraph{Periodic solution} Suppose that
(a) $\l_1 \geq \l_0$ or (b) $\l_1 < \l_0$ and $\tau^\ast \leq \hat{\tau}$. We have shown that the constant $\tau^\ast$ solves some relaxation of \eqref{eq:next_inspection_A} and, furthermore, that all solutions of the relaxation concentrate on $[\bar{t}, \infty)$ for some $\bar{t} = \bar{t} ( \l_B, \l_G, \d, r, u_0, u_1) > 0$.\footnote{Recall the three different relaxations: shirk; shirk-before-work; and work-before-shirk. None of these cases is ruled out by the assumption that (a) or (b) holds.} Use the inequality 
\eqref{eq:shirk_first_binding} to conclude, as above, that $\tau^\ast$ is feasible in the original problem \eqref{eq:next_inspection_A}.\footnote{With imperfect inspections, the argument is unchanged because the agent's passage probability depends only on the total duration of shirking, not its allocation over time.}


\paragraph{Periodic--exponential solution} Suppose that $\l_0 > \l_1$ and $\tau^\ast > \hat{\tau}$. It follows that \eqref{eq:cutoff} is violated,\footnote{ If $\l_0 > \l_1$ and $\tau^\ast > \hat{\tau}$, then
\[
   e^{-\d \tau^\ast} < e^{-\d \hat{\tau}} = \frac{ \l_0 -\l_1}{\d + \l_0 - \l_1} < \frac{ U_0( \l_0 - \l_1)}{U_1 ( \d + \l_0 - \l_1)}.
\]} so we consider the work-before-shirk relaxed problem.
We have shown that the distribution $F_{\pi^\ast}$ from \eqref{eq:exp_sol} is the unique solution of the relaxed problem \eqref{eq:relaxed_full}. Then, using the assumption that $\d > 2 \l_1 - \l_0$, 
we show that $F_{\pi^\ast}$ is feasible in the original problem.

Suppose that the time $T$ until the next inspection follows the distribution $F_{\pi^\ast}$. First we check that the agent finds it weakly optimal to shirk over $[\hat{\tau}, \infty)$, no matter his action history. Over $(\hat{\tau}, \infty)$, the distribution of $T$ is memoryless, so the only state variable is the agent's belief $q_t$ that $\th_t = 0$. The agent's belief evolves according to the differential equation $\dot{q}_t = -q_t \d (1 - a_t)$. Therefore, the HJB equation reads
\[
    0 = \max_{a = 0, 1} \Set{  a u_1 + (1 - a) u_0 - q \d (1 - a) V'(q) - \l_a V(q) + \g^\ast (q U_1 - V(q))}.
\]
We verify that this HJB equation is solved by the value function 
\begin{equation} \label{eq:value_conjecture}
    V(q) = U_1 +  (q -1) (U_0 - U_1) \l_0 /\d.
\end{equation}
Plug in this value function, write $\l_a = \l_1 a +\l_0 (1 - a)$, and substitute in the expression for $\g^\ast$ from \eqref{eq:gamma_ast_0}. Simplify to get
\[
    0 = \max_{a = 0,1}~ a (q - 1) (U_0 - U_1) \l_0 ( \d + \l_0 - \l_1)/\d.
\]
Since $\d + \l_0 > \l_0 > \l_1$, this equation is satisfied. At every belief $q$, the agent weakly prefers shirking to working (strictly so if $q < 1$).

Now we consider the remaining deviations. In particular, we must consider shirk-work-shirk deviations. Recall from \eqref{eq:deviation_notation} the notation $U(a; \tau)$ for each $a \in \AA$ and $\tau \geq 0$. For $s \leq \hat{\tau}$, let
\[
    U_{\SWS}(s; F_{\pi^\ast}) = \int_{0}^{\infty} U ( 1_{[s,\hat{\tau})}; t) \de F_{\pi^\ast}(t).
\]
Define $U(a ; F_{\pi^\ast})$ analogously for each $a \in \AA$. 
Use the inequality 
\eqref{eq:shirk_first_binding}, as in previous cases, and the HJB argument above to conclude that
\begin{equation} \label{eq:SWS_bound}
   \max_{a \in \AA} U ( a ; F_{\pi^\ast}) = \max_{0 \leq s \leq \hat{\tau}} U_{\SWS}(s; F_{\pi^\ast}).
\end{equation}
In terms of the value function $V$ from \eqref{eq:value_conjecture}, we have
\begin{multline*}
   U_{\SWS}(s; F_{\pi^\ast})
    = U_0 (1 - e^{-\l_0 s}) + e^{-\l_0 s} U_1 ( 1 - e^{-\l_1 (\hat{\tau} - s)}) \\
     + e^{-\l_0 s - \l_1 (\hat{\tau} -s)} \Brac{ \pi^\ast U_1 e^{-\d s}  + (1 - \pi^\ast)V(e^{-\d s})}.
\end{multline*}
Substituting in the expression for $V(e^{-\d s})$ from \eqref{eq:value_conjecture}, we get
\[
  U_{\SWS}(s; F_{\pi^\ast}) =  U_0-(U_0-U_1)e^{-\lambda_0 s}
-
B e^{-(\lambda_0-\lambda_1)s}(1-e^{-\delta s}),
\]
where 
\[
B = e^{-\lambda_1\hat{\tau}} \Brac{ \pi^\ast U_1 + (1-\pi^\ast)\frac{\lambda_0}{\delta}(U_0-U_1)} > 0.
\]

We prove that the right side of \eqref{eq:SWS_bound} is at most $U_1$. The definition of $F_{\pi^\ast}$ ensures that 
\[
    U_{\SWS}(0;F_{\pi^\ast}) = U_{\SWS}(\hat{\tau};F_{\pi^\ast}) = U_1.
\]
Therefore, it suffices to prove that no maximizer of the function $s \mapsto U_{\SWS}(s;F_{\pi^\ast})$ over the domain $[0, \hat{\tau}]$  lies in the interior $(0, \hat{\tau})$. The argument is similar to that in \eqref{eq:suff_condition}. For each $s \in (0, \hat{\tau})$, we have
\begin{equation*} 
\begin{aligned}
U_{\SWS}'(s; F_{\pi^\ast}) &= \lambda_0(U_0-U_1)e^{-\lambda_0 s}
+
B e^{-(\lambda_0-\lambda_1)s}
\left[
(\lambda_0-\lambda_1)-(\delta+\lambda_0-\lambda_1)e^{-\delta s}
\right], \\
    U_{\SWS}''(s; F_{\pi^\ast}) 
    &= -\lambda_0^2(U_0-U_1)e^{-\lambda_0 s}
-
B e^{-(\lambda_0-\lambda_1)s}
\left[
(\lambda_0-\lambda_1)^2 - (\delta+\lambda_0-\lambda_1)^2e^{-\delta s}
\right].
\end{aligned}
\end{equation*}
For all $s \in (0,\hat{\tau})$, we have 
\begin{equation*}
\begin{aligned}
&U_{\SWS}''(s; F_{\pi^\ast})  + (\d + 2 \l_0 - 2\l_1) U_{\SWS}'(s; F_{\pi^\ast})  \\
&= \lambda_0(\delta+\lambda_0-2\lambda_1)(U_0-U_1)e^{-\lambda_0 s} 
+ B(\lambda_0-\lambda_1)(\delta+\lambda_0-\lambda_1)
e^{-(\lambda_0-\lambda_1)s}(1-e^{-\delta s}) \\
    &>0,
\end{aligned}
\end{equation*}
where for the last inequality, note that the first term is strictly positive because $\d >  2 \l_1 - \l_0$ and the second term is strictly positive because $B > 0$ and $\l_0 > \l_1$. Thus, if $U_{\SWS}'( s; F_{\pi^\ast})  = 0$, then $U_{\SWS}''(s; F_{\pi^\ast}) > 0$, so the function $U_{\SWS}(s; F_{\pi^\ast})$ cannot have an interior maximizer.

\paragraph{Comparative statics for periodic--exponential solution} In the case $\tau^\ast > \hat{\tau}$, we check that $\g^\ast$ and $\pi^\ast$ are both strictly decreasing in $\d$. For $\g^\ast$, express \eqref{eq:gamma_ast_0} as 
\[
   \frac{ \d \g^\ast}{ \l_0 (\l_0 + \d + \g^\ast)} = \frac{U_0 - U_1}{U_1}.
\]
Note that the left side is strictly increasing in $\d$ and $\g^\ast$, hence $\g^\ast$ is strictly decreasing in $\d$. 

For $\pi^\ast$, divide the numerator and denominator by $\d$ and simplify to obtain 
\begin{equation} \label{eq:decomp}
    \pi^\ast = \frac{U_0 - U_1}{U_1 - \l_0 (U_0 - U_1)/\d} N(\d),
\end{equation}
where 
\[
    N(\d) = \Paren{1 + \frac{\l_0 - \l_1}{\d}} (e^{\l_0 \hat{\tau}} -1) - \frac{\l_0}{\d}.
\]
In \eqref{eq:decomp}, the fraction is strictly positive and strictly decreasing in $\d$. Hence, $N(\d) > 0$ and it suffices to check that the function $N$ is strictly decreasing. From \eqref{eq:gamma_ast_0}, we have 
\[
    \hat{\tau} = \d^{-1} \log \Paren{ 1 + \frac{\d}{\l_0 - \l_1}},
\]
so 
\[
   \d^2 N'(\d)  = 2 \l_0 - \l_1  -A e^{\l_0 \hat{\tau}},
\]
where
\[
A = \l_0 (\l_0  - \l_1 + \d) \hat{\tau} - \l_1.
\]
For all $x > 0$, it is easily verified that 
\begin{equation} \label{eq:log}
    \log (1 + x) > \frac{2x}{2 + x} > \frac{x}{1 + x}.
\end{equation}
Taking $x = \d / (\l_0 - \l_1)$ in the outer inequality gives $(\d + \l_0 - \l_1) \hat{\tau}  > 1$, so $A > \l_0 -\l_1 > 0$. Since $e^{\l_0 \hat{\tau}} > 1 + \l_0 \hat{\tau}$, we have
\begin{equation*}
\begin{aligned}
 \d^2 N'(\d) 
 &< 2 \l_0 - \l_1 - A -\l_0 \hat{\tau} (\l_0 - \l_1) \\
 &= \l_0 [ 2 - (\d + 2 (\l_0 - \l_1)) \hat{\tau}] \\
 &< 0,
 \end{aligned}
\end{equation*}
where the last inequality follows from \eqref{eq:log} with $x = \d / (\l_0 - \l_1)$.

\paragraph{Threshold $\bar{\l}_B$} We check that there exists a threshold $\bar{\l}_B = \bar{\l}_B (\l_G, \d, r, u_0, u_1)$, with $\bar{\l}_B  > \l_G$, such that for all $(\l_B, \l_G, \d, r, u_0, u_1)$ satisfying $\d > 2 \l_G - \l_B + r$ and Assumptions~\ref{as:nontrivial}--\ref{as:sufficient}, the inequality $\l_B > \bar{\l}_B$ holds if and only if $\l_B > \l_G$ and $\tau^\ast > \hat{\tau}$. For $u_0 \leq u_1$,  \cref{as:nontrivial} implies that $\l_B < \l_G$, so we may arbitrarily choose $\bar{\l}_B (\l_G, \d, r, u_0, u_1) > \l_G$. Hereafter, we may assume $u_0 > u_1$. 

In order to state the next claim, define the following domains: 
\[
      D^\ast = \{ (\ell_0, d, y_0) \in (0, \infty)^3 : d > y_0 - \ell_0 > 0 \},  \qquad
    \hat{D} = \{ ( \ell_0, d) \in (0,\infty)^2 : \ell_0  > 1\}.
\]

\begin{claim}[Thresholds] \label{res:thresholds} \mbox{}
\begin{enumerate}
    \item There exists a continuous function $g^\ast \colon D^\ast \to (0,\infty)$ such that
    \[
        \l_1  \tau^\ast  = g^\ast (\l_0/\l_1, \d /\l_1,u_0/u_1),
    \]
   whenever $(\l_0/\l_1, \d /\l_1,u_0/u_1)$ is in $D^\ast$. Moreover, $g^\ast$ is strictly increasing in its first two arguments and strictly decreasing in its third argument. For each fixed $d, y_0 > 0$, we have $\lim_{\ell_0 \uparrow y_0} g^\ast ( \ell_0, d, y_0) = \infty$ and, if $y_0 > d$, we have $\lim_{\ell_0 \downarrow (y_0 - d)} g^\ast ( \ell_0, d, y_0) = 0$.
    \item There exists a continuous function $\hat{g} \colon \hat{D} \to (0, \infty)$ such that 
    \[
        \l_1  \hat{\tau} = \hat{g} (\l_0/\l_1, \d /\l_1),
    \]
      whenever $(\l_0/\l_1, \d /\l_1)$ is in $\hat{D}$.  Moreover, $\hat{g}$ is strictly decreasing in both arguments. For any fixed $d> 0$, we have $\lim_{\ell_0 \downarrow 1} \hat{g}(\ell_0, d) = \infty$. For any fixed $\ell_0 > 1$, we have $\lim_{d \uparrow \infty} \hat{g}(\ell_0, d) = 0$. 
\end{enumerate}
\end{claim}


For fixed $(d,y_0)$ with $d > 0$ and $y_0 > 1$, \cref{res:thresholds} (proven in \cref{sec:proofs_of_claims}) implies that over the interval $( \max \{ 1, y_0 -d \} , y_0)$, the map $\ell_0 \mapsto g^\ast ( \ell_0, d, y_0) - \hat{g} (\ell_0,d)$ is strictly increasing and crosses $0$ exactly once. Denote the unique zero by $\bar{\ell}_0 = \bar{\ell}_0 ( d, y_0)$. By \cref{res:thresholds}, the function $\bar{\ell}_0$ is strictly decreasing in $d$ and strictly increasing in $y_0$. Moreover, for each $y_0  > 1$, we have $\lim_{d \uparrow \infty} \bar{\ell}_0 ( d, y_0) = 1$. Thus, we have proven \cref{rem:threshold_B} (where $g \coloneqq \bar{\ell}_0 - 1$).

Now we return to the original variables. For fixed $(\l_G, \d,r, u_0, u_1)$ with $u_0/u_1 > 1$, define $\bar{\l}_B = \bar{\l}_B ( \l_G, \d, r, u_0, u_1)$ by 
\[
    \frac{\bar{\l}_B + r}{\l_G +r} = \bar{\ell}_0 \Paren{ \frac{\d}{\l_G + r}, \frac{u_0}{u_1} }.
\]
By construction, $\bar{\l}_B > \l_G$. 

Finally, we confirm that a modified version of \cref{res:imperfect_inspections} goes through under a more general flow payoff specification. Suppose that for some payoff parameter $\mu$, the flow payoffs can be expressed as $\hat{u}_i ( \l_G, \l_B; \mu)$ for $i = 0,1$, where $\hat{u}_i$ is continuous in $(\l_G, 
\l_B)$. In view of \cref{rem:threshold_B}, the condition 
\[
    \l_B > \bar{\l}_B \big( \l_G, \d, r, \hat{u}_0 ( \l_G, \l_B; \mu), \hat{u}_1 ( \l_G, \l_B; \mu)  \big)
\] 
in \cref{res:imperfect_inspections} can be expressed as 
\begin{equation} \label{eq:cond_g_mu}
    \frac{ \l_B + r}{\l_G + r} > 1 + g \Paren{  \frac{\d}{\l_G + r},\frac{\hat{u}_0(\l_G, \l_B; \mu)}{\hat{u}_1 (\l_G, \l_B;\mu)}},
\end{equation}
for some $(0, \infty)$-valued function $g$ that is strictly increasing in its second argument. As long as the ratio $\hat{u}_0(\l_G, \l_B; \mu)/ \hat{u}_1 (\l_G, \l_B;\mu)$ is weakly decreasing in $\l_B$ for each fixed $\l_G$ and $\mu$, we can equivalently express condition \eqref{eq:cond_g_mu} as 
\[
    \l_B > \hat{\l}_B ( \l_G, \d, r; \mu),
\]
for some function $\hat{\l}_B$ with $\hat{\l}_B > \l_G$. In particular, this holds if $W_B \leq 0$.

\newpage

\bibliographystyle{ecta}
\bibliography{References.bib}

\newpage

\section{Online appendix: Additional results} \label{sec:online_appendix_results}

\subsection{Designing transfers and inspections} \label{sec:transfers_inspections}

In the main model, we study the optimal design of inspections, given fixed monetary incentives. Here, we consider the joint design of monetary incentives and inspections.  We find that for a range of parameters, it is strictly optimal to use inspections.


For simplicity, consider the case of perfect inspections; the analysis could be extended to imperfect inspections, at the cost of additional algebra. Suppose that the inspection cost is $\k > 0$,\footnote{In the main model we could normalize $\k = 1$ since the principal faced no other costs.} and the principal can directly control the reward $W_G \geq 0$ that the agent receives upon a breakthrough.\footnote{The principal may also control a flow wage during employment. It is easy to see that for any combination of $w,W_G$ with $w>0$, the principal provides more incentives for effort at the same expected cost by setting instead $\tilde W_G = W_G + w/\l_G$ and $\tilde w = 0$.} The principal jointly minimizes the sum of the expected inspection cost and the wage bill subject to the constraint that the agent is induced to work continuously. We take the other parameters as given. 



We can directly solve this problem by analyzing how the principal's minimal expected normalized cost in the main model varies with $U_1$. So we may fix the arrival parameters $\l_G > 0$ and $\l_B \geq 0$, discount rate $r > 0$, and the payoff parameters $\bar{u}(0), \bar{u}(1), W_B$. The solution depends on $W_G$ via $U_1 = (\bar{u}(1) +  \l_G W_G) / (\l_G + r)$. Let $\ubar{U}_1 = \bar{u}(1) / (\l_G + r)$. Parallel to \cref{as:nontrivial} from the main model, here we assume that $U_0 > 0$ and $U_0 > \ubar{U}_1$. Thus, if the principal chooses $W_G = 0$, then inspections are necessary to motivate the agent to work.

The expected bonus payment is given by
\[  
    \frac{\l_G W_G}{\l_G + r} = U_1 -  \frac{ \bar{u}(1)}{\l_G +r}.
\]
Fixing the other parameters, let $C(U_1)$ denote the optimal cost from the main model; see \eqref{eq:P_Bellman}. We restrict the minimization to $U_1 > 0$ since $C(U_1) \to \infty$ as $U_1 \to 0$, and $C(U_1) \to 0$ as $U_1 \to U_0$. Extend $C$ continuously to this limit by setting $C(U_0) = 0$. No inspections are needed if the agent weakly prefers working to shirking in the absence of inspections; see \cref{res:NoInspections}. 
Therefore, the optimal joint policy is obtained by solving
\begin{equation} \label{eq:joint_problem}
    \miz_{U_1 \in (0, U_0] \cap [\ubar{U}_1, U_0]} \, \k C(U_1) + U_1 -  \frac{ \bar{u}(1)}{\l_G +r}.
\end{equation}
The function $C$ is strictly decreasing because fewer inspections are necessary if a breakthrough comes with a larger reward.
 The principal trades off a higher wage bill in the form of $U_1$ against a lower expected inspection cost in the form of $\k C(U_1)$.

In this joint problem, we say that \emph{inspections are strictly optimal} if every minimizer of \eqref{eq:joint_problem} is strictly smaller than $U_0$. From \cref{res:speeding_up_perfect} and \cref{res:delaying_perfect}, we obtain expressions for the function $C$. Straightforward algebra then yields the following characterization of when inspections are strictly optimal.

\begin{prop}[Inspections and transfers] Suppose inspections are perfect. In the joint design problem, inspections are strictly optimal if and only if $\k < \bar{\k}$, where 
    \[
    \bar{\k} = \begin{cases} 
      U_0 \frac{\l_G + r}{\l_B + r} \Paren{ \frac{\l_G + r}{\l_G - \l_B}}^{\frac{ \l_G - \l_B}{\l_B + r}} &\text{if}~\l_G > \l_B, \\
    U_0 \frac{\l_G + r}{\l_B + r} &\text{if}~\l_G \leq \l_B.
    \end{cases}
    \]
\end{prop}

\begin{proof} For the proof, we use the notation $\l_1 \coloneqq \l_G + r$ and $\l_0 \coloneqq \l_B + r$. 

First consider the case $\l_1 > \l_0$. By \cref{res:speeding_up_perfect}, the cost $C(U_1)$ is induced by the periodic inspection policy, 
with period  $\tau^\ast$ given by
\[
    e^{-\l_1 \tau^\ast} = 
    \begin{cases}
    \left(\dfrac{U_0-U_1}{U_0}\right)^{\lambda_1/\lambda_0}
    &\text{if}~ \lambda_0 U_0 \ge \lambda_1 U_1, \\
    \left(\dfrac{\lambda_1-\lambda_0}{\lambda_1}\right)^{\lambda_1/\lambda_0}
    \dfrac{(U_0-U_1)\lambda_0}{U_1(\lambda_1-\lambda_0)}
    &\text{if}~ \lambda_0 U_0 < \lambda_1 U_1.
\end{cases}
\]
The associated normalized cost is
\[
    C(U_1) = \sum_{n = 1}^{\infty} e^{-\l_1 n \tau^\ast} = \frac{e^{-\l_1 \tau^\ast}}{1 - e^{-\l_1 \tau^\ast}}.  
\]
To simplify notation, let 
\[
    \alpha = \frac{\l_0}{\l_1}
    \qquad\text{and}\qquad
    \beta = \alpha(1-\alpha)^{1/\a-1}.
\]
With this notation, some algebra gives
\[
    C(U_1) = 
    \begin{cases}
    \dfrac{
    \left(1-U_1/U_0\right)^{1/\alpha}
    }{
    1-
    \left(1-U_1/U_0\right)^{1/\alpha}
    },
    &\text{if}~ U_1\le \alpha U_0, \\[1em]
    \dfrac{
    \beta\left(U_0/U_1-1\right)
    }{
    1-
    \beta\left(U_0/U_1-1\right)
    },
    &\text{if}~ U_1>\alpha U_0.
    \end{cases}
\]
It can be shown that the function $C$ is convex on $(0, U_0]$.\footnote{

Note that $C'(U_1)=-\frac{1}{U_0}N(U_1/U_0)$, where
\[
N(x)=
\begin{cases}
\dfrac{\alpha^{-1}(1-x)^{1/\alpha-1}}
{\left[1-(1-x)^{1/\alpha}\right]^2},
& x\le \alpha,\\[1.2em]
\dfrac{\beta}{\left[(1+\beta)x-\beta\right]^2},
& x>\alpha.
\end{cases}
\]
We claim that $N$ is strictly decreasing on each region, and the two expressions agree at $x=\alpha$.  Indeed, for $x < \a$, we have
\[
\frac{N'(x)}{N(x)}
=
-\frac{1/\alpha-1}{1-x}
-
\frac{2}{\alpha}
\frac{(1-x)^{1/\alpha-1}}{1-(1-x)^{1/\alpha}}
<0.
\]
For $x > \a$, we have
\[
N'(x)
=
-\frac{2\beta(1+\beta)}
{\left((1+\beta)x-\beta\right)^3}
<0.
\]
Equality at $x = \a$ follows upon noting that
\(
(1+\beta)\alpha-\beta
=
\alpha\left[1-(1-\alpha)^{1/\alpha}\right].
\)
} By convexity, inspections are strictly optimal if and only if $\k C'(U_0) > - 1$. One can compute $C' (U_0) = -\b/ U_0$, so we obtain the desired condition $\k < U_0 / \b$. The proof is complete upon observing that $\bar{\k} = U_0 / \b$ in the case $\l_1 > \l_0$. 

Next, consider the case $\l_0 \geq \l_1$.  By \cref{res:delaying_perfect} (and a continuity argument for the case $\l_0 = \l_1$), the cost $C(U_1)$ is induced by the inspection policy with constant hazard rate $\g^\ast$ given by 
\[
\g^\ast =
\frac{\l_0(U_0-U_1)}{U_1}.
\]
Under this policy, inspections arrive at constant rate $\g^\ast$ while the project continues, so the 
expected discounted cost is given by
\[
    C(U_1) =  \int_{0}^{\infty} \g^\ast e^{-\l_1 t} \de t = \frac{\g^\ast}{\l_1} = \frac{ \l_0 (U_0-U_1)}{\l_1 U_1}.
\]
One can compute $C''(U_1)=\frac{2\l_0U_0}{\l_1U_1^3}>0$, so $C$ is convex on $(0, U_0]$.  By convexity, inspections are strictly optimal if and only if $\k C' (U_0) > -1$. One can compute $C'(U_0) = -\l_0 /(\l_1 U_0)$, so we obtain the desired condition $\k < U_0 \l_1 / \l_0$. 
\end{proof}

\subsection{Penalties for failed inspections} \label{sec:punishments}

In the main model, we normalize the agent's outside option and continuation value after termination to $0$. In this section, we illustrate how this normalization is performed if termination results in the penalty continuation value $- P$. With this penalty, the agent's incentive constraint in \eqref{eq:next_inspection_A} now reads
\begin{equation} \label{eq:P_IC}
    \E \Brac{ \int_{0}^{T} D_t(a) u(a_t) \de t + p_T(a) D_T(a)  U_1 - (1 - p_T(a)) D_T(a)  P } \leq U_1.
\end{equation}
In the main model, the final term on the left was zero, so it was omitted. To perform the normalization, we define normalized payoff parameters relative to this penalty: 
\[
 \bar{u}'(0) = \bar{u}(0)+ r P, \qquad 
 \bar{u}'(1) =  \bar{u}(1) + r P
 \]
and
\[
    W_G' =  W_G + P, \qquad W_B' = W_B + P.
\]
Thus,
\[
    u_0' = u_0 + (r + \l_B) P, 
    \qquad
    u_1' = u_1 + ( r + \l_G) P,
\]
and
\[
   U_0' = U_0 + P, \qquad U_1' = U_1 + P.
\]
Straightforward algebra shows that \eqref{eq:P_IC} is equivalent to the original constraint with the primed variables: 
\[
    \E \Brac{ \int_{0}^{T} D_t(a) u'(a_t) \de t + p_T(a) D_T(a)  U_1'} \leq U_1'.
\]

\cref{as:nontrivial} is applied to the normalized variables: $U_0' > U_1' > 0$. That is, $U_0 + P > U_1 + P > 0$, or equivalently, $U_0 > U_1 > - P$. So the ranking between $U_0$ and $U_1$ is unchanged, but now we require that the agent strictly prefers always working to the penalty payoff. Similarly, \cref{as:sufficient} is applied to the normalized variables: $\d > (\l_B + r) (U_0' - U_1')/U_1'$, that is, $\d > (\l_B + r) (U_0 - U_1)/(U_1 + P)$. With these adjustments, all of our results can be applied to the setting with penalties, upon performing this normalization.

\subsection{Robustness to small recovery rate} \label{sec:robust_small_rho}

Here we formalize the claimed robustness to perturbing $\rho$. Recall that if $u_0 > u_1$, then with the perfect inspection technology with no recovery, the agent never works after having shirked in any binding deviation. We check that the binding deviations remain unchanged as long as $\d$ is sufficiently  large and $\rho$ is sufficiently small.

\begin{thm}[Robustness to recovery]
\label{res:low recovery robustness}
Assume $u_0 > u_1$. There exist thresholds $\bar{\d} = \bar{\d} (\l_G, \l_B, r, u_0, u_1) > 2 \l_G - \l_B + r$ and $\bar{\rho} = \bar{\rho} ( \l_G, \l_B, r, u_0, u_1) > 0$ such that under the inspection technology with recovery, if $\d > \bar{\d}$ and $\rho < \bar{\rho}$, then parts \ref{res:speeding_up_imperfect} and \ref{res:delaying_imperfect} of \cref{res:imperfect_inspections} hold, with the same threshold functions $\bar{\l}_B$ and  $\bar{t}$, and the same period $\tau^\ast$.
\end{thm}

\begin{proof} We follow the proof of \cref{res:imperfect_inspections} (\cref{sec:main_proof}), indicating the appropriate modifications to accommodate a positive recovery rate $\rho$. In particular, with a positive recovery rate $\rho$, the agent's payoffs from shirk-before-work deviations take a different form. For $s \leq \tau$, the agent's probability of failing the inspection is $(1 - e^{-\d s}) e^{-\rho (\tau -s)}$, so 
\begin{equation*}
\begin{aligned}
     &U_{\SW} (s ; \tau) \\
     &= U_0 ( 1 - e^{-\l_0 s} ) + e^{-\l_0 s} U_1 ( 1 -  e^{-\l_1 (\tau - s)} )  + e^{-\l_0 s - \l_1 ( \tau - s)} U_1 \Brac{ 1 - (1 - e^{-\d s}) e^{-\rho (\tau - s)}} \\
     &= U_0  - (U_0 - U_1) e^{-\l_0 s} 
     - U_1 e^{-\l_0 s - \l_1 ( \tau- s)} (1-e^{-\d s})e^{-\rho(\tau -s)}.
\end{aligned}
\end{equation*}
Define $\tau^\ast$ as in the proof of \cref{res:imperfect_inspections}, with this new expression for $U_{\SW}$. As before (see footnote \ref{ft:well_defined}), it can be shown that $\tau^\ast$ is well-defined. Also define the period $\tau_S^\ast$ to be the largest time $t$ such that $U_S (t) = U_1$. The period $\tau_S^\ast$ does not depend on $\rho$ because the function $U_S$ does not depend on $\rho$. By construction, $\tau^\ast \leq \tau_S^\ast$. Below, we will establish equality. In the argument below, we will reference the following conditions:
\begin{align}
\label{eq:delta_rho_system}
\d + \l_0 - 2 \l_1 - 2 \rho &> 0, \\
\label{eq:delta_rho_system2}
\rho U_1 +   e^{-\d \tau^\ast} (\d + \l_0 -   \l_1 - \rho) U_1 &< u_0 - u_1.
\end{align}

\paragraph{Period $\tau_S^\ast$} We claim that if \eqref{eq:delta_rho_system} and \eqref{eq:delta_rho_system2} both hold, then $\tau^\ast = \tau_S^\ast$. It suffices to show that
$U_{\SW}' (0; \tau^\ast) < 0$ and $U_{\SW}' ( \tau^\ast; \tau^\ast) > 0$, for then the conclusion follows from \cref{res:zeros}.\footnote{Here and below, derivatives evaluated at $s = 0$ are \emph{right} derivatives and derivatives evaluated at $s = \tau^\ast$ (or $\hat{\tau}$) are \emph{left} derivatives.} We have
\begin{equation*}
\begin{aligned}
    U_{\SW}'(0 ; \tau^\ast)
   &= (U_0 - U_1) \l_0 - e^{- (\l_1 + \rho)\tau^\ast} U_1 \d, \\
   U_{\SW}''(0 ; \tau^\ast) 
    &= - (U_0 - U_1) \l_0^2 + e^{-(\l_1 + \rho) \tau^\ast} U_1 \d (\d +2 \l_0 - 2\l_1 - 2 \rho).
\end{aligned}
\end{equation*}
Recall that $U_{\SW}'(0;\tau^\ast) \leq 0$, by the definition of $\tau^\ast$. Suppose for a contradiction that $U_{\SW}'(0; \tau^\ast )  = 0$. Then 
\begin{equation} \label{eq:second_deriv_periodic}
    \begin{aligned}
    U_{\SW}'' (0;\tau^\ast) &=  U_{\SW}'' (0;\tau^\ast) + U_{\SW}'(0;\tau^\ast) ( \d + 2\l_0 - 2 \l_1 - 2 \rho) \\
    &= (U_0 - U_1) \l_0 (\d + \l_0 - 2 \l_1 - 2 \rho) \\
    &> 0,
    \end{aligned}
\end{equation}
by \eqref{eq:delta_rho_system}. Thus, $U_{\SW} (s ; \tau^\ast) > U_{\SW} (0 ; \tau^\ast) = U_1$ for $s$ sufficiently small, contrary to the definition of $\tau^\ast$. 

Second, we have
\[
   U_{\SW}' (\tau^\ast ; \tau^\ast ) 
    = e^{- \l_0 \tau^\ast} \bigl(  U_0 \l_0 - (\l_1 + \rho) U_1  - e^{-\d \tau^\ast} (\d + \l_0 -   \l_1 - \rho) U_1 \bigr),
\]
which is strictly positive by \eqref{eq:delta_rho_system2}.

\paragraph{Periodic solution} Assume \eqref{eq:delta_rho_system} and \eqref{eq:delta_rho_system2} both hold. As argued above, we have $\tau^\ast = \tau_S^\ast$. Suppose that
(a) $\l_1 \geq \l_0$ or (b) $\l_1 < \l_0$ and $\tau^\ast \leq \hat{\tau}$. Use the inequality 
\eqref{eq:shirk_first_binding} to conclude, as above, that $\tau^\ast$ is feasible in the original problem \eqref{eq:next_inspection_A}.\footnote{\label{ft:SW}In fact, with recovery, frontloading shirking has the additional benefit of increasing the agent's passage probability. If the state is $0$ with probability $q$, then after shirking for duration $\D$ and then working for duration $\D$, the state is $0$ with probability $q_{\SW} = 1 -  (1 -    q e^{- \d \D}) e^{-\rho \D}$. If instead the agent works for duration $\D$ and then shirks for duration $\D$, then the state is $0$ with probability $q_{\WS} = (1 - (1 - q)e^{-\rho \D})e^{-\d \D}$. With $\d$ and $\rho$ both strictly positive, it can be checked that $q_{\SW} > q_{\WS}$, no matter the value of $q$.}

\paragraph{Periodic--exponential solution}  Assume \eqref{eq:delta_rho_system} and \eqref{eq:delta_rho_system2} both hold.  In particular, we have $\tau^\ast = \tau_S^\ast$. Suppose that $\l_0 > \l_1$ and $\tau^\ast > \hat{\tau}$.  Assume further that 
\begin{equation} \label{eq:delta_rho_system3}
    \l_0 > \l_1 + \rho.
\end{equation}
As in the proof of \cref{res:imperfect_inspections}, the distribution $F_{\pi^\ast}$ from \eqref{eq:exp_sol} is the unique solution of the relaxed problem \eqref{eq:relaxed_full}. Note that the definition of $\pi^\ast$ does not depend on $\rho$. Here we give a sufficient condition for $F_{\pi^\ast}$ to be feasible in the original problem, now with recovery rate $\rho$.

Suppose that the time $T$ until the next inspection follows the distribution $F_{\pi^\ast}$. First we check that the agent finds it weakly optimal to shirk over $[\hat{\tau}, \infty)$, no matter his action history. As before, over $(\hat{\tau}, \infty)$, the distribution of $T$ is memoryless, so the only state variable is the agent's belief $q_t$ that $\th_t = 0$. With recovery rate $\rho$, the agent's belief evolves according to the differential equation 
\[
    \dot{q}_t = (1 - q_t) \rho a_t - q_t \d (1 - a_t). 
\]
The HJB equation reads
\begin{multline*} 
    0 = \max_{a = 0,1}~ \Bigl\{ 
    (1 -a) u_0 + a u_1  + \Brac{ (1 - q) \rho a - q \d (1 - a)} V'(q) \\
    - \l_a V(q) + \g^\ast (q U_1 - V(q)) \Bigr\}.
\end{multline*}
We verify that this HJB equation is solved by the value function
\begin{equation} \label{eq:value_function_recovery}
    V(q) = U_1 +  (q -1) (U_0 - U_1) \l_0 /\d.
\end{equation}
Note that this expression is the same as in \eqref{eq:value_conjecture}. Plug in this value function, write $\l_a = \l_1 a + \l_0 (1 - a)$, and substitute in the expression for $\g^\ast$ from \cref{res:imperfect_inspections}. Simplify to get
\[
    0 = \max_{a = 0,1}~ a (q - 1) (U_0 - U_1) \l_0 ( \d + \l_0 - \l_1 - \rho)/\d.
\]
By \eqref{eq:delta_rho_system}, this equation is satisfied. At every belief $q$, the agent weakly prefers shirking to working (strictly so if $q < 1$). 

Now we consider the remaining deviations. In particular, we must consider shirk-work-shirk deviations. Following the argument in \cref{ft:SW} and the HJB argument above, we conclude that
\begin{equation} \label{eq:SWS_bound_recovery}
    \max_{a \in \AA} U (a; F_{\pi^\ast}) = \max_{0 \leq s \leq \hat{\tau}} U_{\SWS} (s; F_{\pi^\ast}). 
\end{equation}
In terms of the value function $V$ from \eqref{eq:value_function_recovery}, we have
\begin{multline*}
    U_{\SWS} (s ; F_{\pi^\ast})
    =    U_0 (1 - e^{-\l_0 s}) + e^{-\l_0 s} U_1 ( 1 - e^{-\l_1 (\hat{\tau} - s)}) \\
    + e^{-\l_0 s - \l_1 (\hat{\tau} -s)} \Brac{ \pi^\ast q(s) U_1  + (1 - \pi^\ast) V(q(s)) },
\end{multline*}
where $q(s) = 1 -(1-e^{-\d s})e^{-\rho(\hat{\tau}-s)}$.
Substituting in the expression for $V(q(s))$ from \eqref{eq:value_function_recovery}, we get
\[
  U_{\SWS}(s; F_{\pi^\ast}) =  U_0-(U_0-U_1)e^{-\l_0 s}
-
B e^{-(\l_0-\l_1-\rho)s}(1-e^{-\d s}),
\]
where
\[
B = e^{-(\l_1+\rho)\hat{\tau}} \Brac{ \pi^\ast U_1 + (1-\pi^\ast)\frac{\l_0}{\d}(U_0-U_1)} > 0.
\]

We prove that the right side of \eqref{eq:SWS_bound_recovery} is at most \(U_1\). The definition of \(F_{\pi^\ast}\) ensures that
\[
    U_{\SWS}(0;F_{\pi^\ast}) = U_{\SWS}(\hat{\tau};F_{\pi^\ast}) = U_1.
\]
Therefore, it suffices to prove that no maximizer of the function
\(s \mapsto U_{\SWS}(s;F_{\pi^\ast})\) over the domain
\([0,\hat{\tau}]\) lies in the interior \((0,\hat{\tau})\). The argument is
similar to that in \eqref{eq:suff_condition}. For each
\(s \in (0,\hat{\tau})\), we have
\begin{equation*}
\begin{aligned}
U_{\SWS}'(s; F_{\pi^\ast})
&= \l_0(U_0-U_1)e^{-\l_0 s}
+
B e^{-(\l_0-\l_1-\rho)s}
\left[
(\l_0-\l_1-\rho)-(\d+\l_0-\l_1-\rho)e^{-\d s}
\right], \\
    U_{\SWS}''(s; F_{\pi^\ast})
    &= -\l_0^2(U_0-U_1)e^{-\l_0 s}
-
B e^{-(\l_0-\l_1-\rho)s}
\left[
(\l_0-\l_1-\rho)^2 - (\d+\l_0-\l_1-\rho)^2e^{-\d s}
\right].
\end{aligned}
\end{equation*}
For all \(s \in (0,\hat{\tau})\), we have
\begin{equation*}
\begin{aligned}
&U_{\SWS}''(s; F_{\pi^\ast})
+(\d+2\l_0-2\l_1-2\rho)U_{\SWS}'(s; F_{\pi^\ast})
\\
&= \l_0(\d+\l_0-2\l_1-2\rho)(U_0-U_1)e^{-\l_0 s}
\\
&\quad
+ B(\l_0-\l_1-\rho)(\d+\l_0-\l_1-\rho)
e^{-(\l_0-\l_1-\rho)s}(1-e^{-\d s})
\\
&>0,
\end{aligned}
\end{equation*}
where for the last inequality, note that the first term is strictly positive
by \eqref{eq:delta_rho_system} and the second term is strictly positive by \eqref{eq:delta_rho_system3} because $B>0$. Thus, if
\(U_{\SWS}'(s; F_{\pi^\ast})=0\), then
\(U_{\SWS}''(s; F_{\pi^\ast})>0\), so the function
\(U_{\SWS}(s; F_{\pi^\ast})\) cannot have an interior maximizer.

\paragraph{Solving for the thresholds} Now we find threshold functions $\bar{\d}$ and $\bar{\rho}$ such that  if $\d > \bar{\d}$ and $\rho < \bar{\rho}$, then \eqref{eq:delta_rho_system}--\eqref{eq:delta_rho_system2} hold, and whenever $\l_0 > \l_1$, \eqref{eq:delta_rho_system3} also holds. Choose $\bar{\rho} = \bar{\rho} ( \l_0, \l_1, u_0, u_1) > 0$ such that $\bar{\rho} U_1 < (u_0- u_1) /2$ and, if $\l_0 > \l_1$, also $\bar{\rho} < \l_0 - \l_1$. Write $\tau^\ast = \tau^\ast (\d, \rho) =  \tau^\ast ( \l_0, \l_1, u_0, u_1;\d, \rho)$. From the definition of $U_{\SW}$, observe that $\tau^\ast$ is weakly increasing in $\d$. Thus, for any fixed $\d_0$, we have
\[
    \lim_{\d \to \infty} e^{- \d \tau^\ast (\d, \bar{\rho})} (\d + \l_0 - \l_1) U_1 \leq  \lim_{\d \to \infty} e^{- \d \tau^\ast (\d_0, \bar{\rho})} (\d + \l_0 - \l_1) U_1 = 0.
\]
Therefore, we may choose $\bar{\d}  = \bar{\d} (\l_0, \l_1, u_0, u_1) > 2 \l_1 - \l_0  + 2 \bar{\rho}$ such that 
\[
    e^{ -\d \tau^\ast (\d, \bar{\rho})} ( \d+ \l_0 - \l_1 ) U_1 < \frac{u_0 - u_1}{2}
\]
whenever $\d \geq \bar{\d}$. 

To complete the proof, fix $\d > \bar{\d}$ and $\rho < \bar{\rho}$. For \eqref{eq:delta_rho_system}, note that 
\[
\d + \l_0 - 2 \l_1 - 2 \rho  > \bar{\d} + \l_0 - 2 \l_1 - 2 \bar{\rho} > 0.
\]
For \eqref{eq:delta_rho_system2}, we have
\[
\rho U_1 +   e^{-\d \tau^\ast} (\d + \l_0 -   \l_1 - \rho) U_1 < \bar{\rho} U_1 +  e^{-\d \tau^\ast} (\d + \l_0 -  \l_1) U_1 < u_0 - u_1, 
\]
where the last inequality uses the fact that $\tau^\ast$ is weakly decreasing in $\rho$, which is immediate from the definition of $U_{\SW}$.
\end{proof}

\subsection{Deadlines versus inspections}\label{sec:deadlines}

In innovation environments ($\l_G > \l_B$) with monetary incentives (and no inspections), a general finding is that deterministic deadlines are optimal; see the literature review (\cref{sec:lit}). In this section, we compare this finding with our results, and we discuss the differences between deadlines and inspections. 


When an inspection is conducted, the agent's continuation value depends on the probability that he passes the inspection. This probability depends on the agent's past actions. By contrast, when the deadline is reached, the project ends with certainty and the agent's continuation value is zero, independent of the agent's past actions.  
Mathematically, a deadline can be thought of as an inspection with a degenerate passage probability: $p_t(a) = 0$ for all times $t$ and action paths $a$.

To highlight the  differences between deadlines and inspections, we solve for the optimal timing of a deadline in an environment that is otherwise as close as possible to our main model. The agent's annuitized flow payoffs are $u_0$ and $u_1$, the common discount rate is $r$, and the breakthrough and breakdown rates are $\l_G$ and $\l_B$. In the main model, the principal minimizes the expected cost of inspections subject to the constraint that working continuously is a best response. At a deadline, the project is terminated (if it has not already ended in a breakthrough or breakdown). Using only a deadline, it is not feasible for the principal to induce the agent to work continuously, without terminating the project before a breakthrough or breakdown arrives. Instead, we consider the principal's payoff maximization problem, where the principal gets a lump-sum payoff of $1$ from a breakthrough. The principal designs the timing of the deadline to maximize her expected payoff. We will see below that this problem is equivalent to the following alternative problem: the principal pays a fixed cost upon hitting the deadline, and the principal minimizes the expected cost of the deadline subject to the constraint that working continuously until the deadline is a best response.

To make the problem feasible and nontrivial, assume $\l_G > \l_B$; $u_1 > u_0$; and $U_0 > U_1$.\footnote{A finite deadline increases the incentive to work only if $\l_G > \l_B$ and $u_1 > u_0$.} It can be shown that it is optimal for the principal to induce the agent to work continuously until the deadline. Thus, the principal chooses a positive random deadline $T$ to solve
    \begin{equation} \label{eq:optimal_deadline}
    \begin{aligned}
       &\text{maximize} && \E \Brac{ \int_{0}^{T} e^{-r t} \l_G e^{- \l_G t} \de t} \\
       &\text{subject to} && \E \Brac{ \int_{0}^{T} D_t (a) u(a_t) \de t } \leq \E \Brac{ \int_{0}^{T} e^{-(\l_G + r) t} u_1 \de t}, \quad a \in \mathcal{A}.
    \end{aligned}
    \end{equation}
    The objective reduces to $(1 - \E e^{-(\l_G + r ) T}) \l_G /(\l_G + r)$, so the principal equivalently minimizes $\E e^{-(\l_G + r) T}$, as in \eqref{eq:next_inspection_A} in the main text. The key difference here is that the agent's continuation value at time $T$ is zero, independent of his previous actions. 
    
    In \eqref{eq:optimal_deadline}, the binding constraints are local for any distribution of $T$. Let $\E_s = \E [ \cdot | T > s]$. For each time $s$, the local constraint takes a form similar to \eqref{eq:local_relaxed}:
    \begin{equation} \label{eq:deadline_local}
        0 \ge (u_0-u_1) - (\l_B-\l_G) \E_s \Brac{ \int_{s}^{T} e^{-(\l_G + r) (t-s)} u_1 \de t}, \qquad s \geq 0. 
    \end{equation}
    Recall that $u_1 > u_0$ and $\l_G > \l_B$. This local constraint requires that the flow benefit from working outweighs the loss from shortening the project. Compare \eqref{eq:deadline_local} with \eqref{eq:local_relaxed}. There are two important differences. First, the agent's action at time $s$ has no effect on his continuation value at the deadline $T$, so the left side of \eqref{eq:local_relaxed} vanishes in \eqref{eq:deadline_local}. In \eqref{eq:local_relaxed}, the left side reflects the marginal effect of the agent's time-$s$ action on the passage probability $p_T$. Second, the agent's expected payoff, at time $s$, from working until the end of the project depends on the time until the deadline. In \eqref{eq:local_relaxed}, this \mbox{time-$s$} continuation payoff from working is $U_1$, no matter the timing of the subsequent inspections.

Consider the relaxed problem that imposes only the time-$0$ local  constraint. The solution set of this relaxed problem consists of all random variables $T$ satisfying 
\[
    \E e^{-(\l_G + r) T} = 1 - \frac{u_1 - u_0}{ (\l_G - \l_B) U_1}. 
\]
There is a range of relaxed solutions because the agent's effective discount factor is the same under a local deviation as it is on path. It can be checked that the deterministic and exponential solutions of this relaxed problem are feasible in \eqref{eq:optimal_deadline}, and hence both are optimal. 

By contrast, in our main inspection problem, \emph{global} deviations are more attractive because the passage probability $p_t$ is strictly supermodular in the action path. In the innovation regime $(\l_G > \l_B$), the agent's discount factor becomes less convex
when he plans to shirk for a positive duration. This creates a strict force toward periodic inspections. By \cref{res:imperfect_inspections}.\ref{res:speeding_up_imperfect}, under the assumption $\d > 2 \l_G - \l_B + r$, an exponential policy is strictly suboptimal because it does not maintain a gap between consecutive inspections. 

While the deadline design problem in \eqref{eq:optimal_deadline} has many solutions, in the single-stage benchmark of their model, \cite{Green2016} find that a deterministic deadline is \emph{uniquely} optimal for incentivizing breakthroughs. There are no breakdowns or discounting. The principal also designs time-varying bonuses. Mathematically, the principal can pay a cost to increase the agent's flow payoff $u_1(t)$.
To build intuition for this result, consider the deterministic and exponential solutions of \eqref{eq:optimal_deadline}. Under the exponential solution, all the local constraints hold with equality. Under the deterministic solution, only the local constraint at $s=0$ holds with equality. When the principal designs time-varying bonus payments, she can take advantage of the slack in the subsequent local constraints by decreasing bonus payments for later breakthroughs. This force is specific to the innovation setting ($\l_G > \l_B$).

\newpage

\section{Online appendix: Additional proofs} \label{sec:online_appendix_proofs}

\subsection{Proof of Theorem~\ref{res:high_recovery_exponential}} \label{sec:proof_recovery}

We solve the relaxed problem that requires that all local deviations are unprofitable. Then we verify that the solution of this relaxed problem is feasible in the original problem.

\paragraph{Local deviations}
For $s,h,t \geq 0$, let $U_s(h;t)$ denote the agent's expected payoff if the principal inspects at time $t$ and the agent plans to shirk over $[s, s + h)$ and work otherwise. Using the notation from \eqref{eq:deviation_notation}, we have $U_s(h;t)  = U( 1_{[0,s) \cup [s+ h,\infty)}; t)$.  For $s + h \leq t$, the agent's probability of failing the inspection is $(1-e^{-\d h})e^{-\rho(t-s - h)}$, so 
\begin{equation*}
\begin{aligned}
    U_s ( h ; t)
    &= U_1 (1 - e^{-\l_1 s}) 
    + e^{-\l_1 s} U_0 (1 - e^{-\l_0 h})
    + e^{-\l_1 s - \l_0 h} U_1 ( 1 - e^{-\l_1 (t - s - h)} ) \\
    &\quad + e^{-\l_1 (t  - h) - \l_0 h} U_1 \Brac{ 1-(1-e^{-\d h})e^{-\rho(t - s- h)}}.
\end{aligned}
\end{equation*}
For $s < t$, differentiate with respect to $h$ and evaluate at $h = 0$. After simplifying, we have
\begin{equation} \label{eq:s0}
    U_s'(0 ; t) = e^{-\l_1 s} \Brac{ (U_0 - U_1) \l_0 - e^{-(\l_1 + \rho) (t -s)} U_1 \d }.
\end{equation}
For $s \geq t$, we have $U_{s} (h ; t) = U_1$ for all $h \geq 0$, so $U_{s}'(0 ; t) = 0$. 

\paragraph{Relaxed problem: local deviations}
Consider the relaxed problem of choosing a positive random variable $T$ to solve
\begin{equation*}
\begin{aligned}
    &\text{minimize} &&\E e^{-\l_1 T} \\
    &\text{subject to} && \E U_{s}'(0 ; T) \leq 0, \quad s \geq 0. 
\end{aligned}
\end{equation*}
To see that this constraint is necessary, recall that for all times $s,h \geq 0$, we must have $\E U_{s} (h ; T)  \leq U_1 = \E U_s (0 ; T)$. Now differentiate under the integral sign, using the dominated convergence theorem.

After substituting in the expression for $U_s'$ above, we see that this problem is equivalent to choosing a distribution $F$ on $(0, \infty)$ to solve
\begin{equation} \label{eq:relaxed_recovery}
\begin{aligned}
&\text{minimize} && \int_{(0, \infty)} e^{-\l_1 t} \de F(t) \\
&\text{subject to} &&\int_{ (s, \infty)} \Brac{U_1 \d e^{-(\l_1 +\rho)(t-s)} - (U_0-U_1) \l_0} \de F(t) \geq 0, \quad s \geq 0.
\end{aligned}
\end{equation}
Note that this is an alternative formulation of problem \eqref{eq:local_relaxed} in the main text. This problem \eqref{eq:relaxed_recovery} takes the form of \eqref{eq:program_lemma} with 
\[
   A = \frac{ (U_0 - U_1) \l_0}{U_1 \d}, \qquad \a = \l_1 + \rho, \qquad \b = \l_1.
\]
By Assumptions~\ref{as:nontrivial}--\ref{as:sufficient}, we have $0 < A < 1$. We have assumed $\rho > 0$, so $\a > \b > 0$. Therefore, we can apply \cref{res:exponential_sol} to conclude that the unique solution is the exponential distribution with hazard rate $\g^\ast = \a A / (1 - A)$, which reduces to the expression in the theorem statement. 

\paragraph{Remaining deviations} It remains to check that if the principal uses the exponential policy with hazard rate $\g^\ast$, then it is optimal for the agent to work until the inspection. Since the distribution of time until the  next inspection is memoryless, the only state variable is the agent's belief $q_t$ that $\th_t = 0$. The agent's belief evolves according to the differential equation 
\[
    \dot{q}_t = (1 - q_t) \rho a_t - q_t \d (1 - a_t). 
\]
The HJB equation reads
\begin{multline} \label{eq:HJB_rho}
    0 = \max_{a = 0,1}~ \Bigl\{ 
     a u_1 + (1 -a) u_0 + \Brac{ (1 - q) \rho a - q \d (1 - a)} V'(q) \\
    - \l_a V(q) + \g^\ast (q U_1 - V(q)) \Bigr\}.
\end{multline}
We verify that this HJB equation is solved by the function 
\[
    V(q) = U_1 +  (q - 1) (U_0 - U_1) \l_0 /\d. 
\]
Plug in this value function, write $\l_a = \l_1 a +\l_0 (1 - a)$, and substitute in the expression for $\g^\ast$ from \cref{res:high_recovery_exponential}. Simplify to get
\[
    0 = \max_{a = 0,1}~(a - 1) (q - 1) (U_0 - U_1) \l_0 ( \d + \l_0 - \l_1 - \rho)/\d.
\]
If $\l_1  + \rho \geq \d + \l_0$, then this equation is satisfied. In this case, at every belief $q$, the agent weakly prefers working to shirking (strictly so if $q < 1$ and $\l_1  + \rho > \d + \l_0$).

\subsection{Verifying claims in the proof of Theorem~\ref{res:imperfect_inspections}} \label{sec:proofs_of_claims}

\paragraph{Proof of \cref{claim:LS}}

Differentiating $\bar{L}_S$ gives
\begin{equation} \label{eq:LS_deriv}
    \begin{aligned}
        \bar{L}_S'(x) &= \l_1^{-1} x^{(\l_0 - \l_1)/\l_1} \Brac{  U_0 \l_0 -U_1 (\d + \l_0) x^{\d / \l_1}}, \\ 
\bar{L}_S''(x) &= \l_1^{-2}  x^{(\l_0-2 \l_1)/\l_1} \Brac{  U_0 \l_0 (\l_0-\l_1) - U_1 ( \d + \l_0) (\d + \l_0-\l_1) x^{\d /\l_1}}.
    \end{aligned}
\end{equation}
Over $[0,1]$, the derivative $\bar{L}_S'$ is strictly single-crossing from above.\footnote{In the main text, we reversed the direction of the horizontal axis $[0,1]$ when plotting $\bar{L}_S$ and $\bar{L}$. In the proofs, we assume that the interval $[0,1]$ has its standard orientation.} Therefore, $\bar{L}_S$ is strictly quasiconcave. Its unique maximizer, $x_{0,S}$, is given by
\[
    x_{0,S}^{\d / \l_1} = \frac{U_0 \l_0}{U_1 (\d + \l_0)}.
\]
By \cref{as:sufficient}, we have $0 < x_{0,S} < 1$. We separate into two cases.

First suppose $\l_1 \geq \l_0$. By assumption, $\d + \l_0 > 2 \l_1 > \l_1$, so  by \eqref{eq:LS_deriv}, the function $\bar{L}_S$ is strictly concave over $[0,1]$. 

Next suppose $\l_1 < \l_0$. From \eqref{eq:LS_deriv},  the second derivative $\bar{L}_S''$ is single-crossing from above. For $x \geq x_c$, we have 
\[
 U_1 (\d + \l_0 - \l_1 ) x^{\d /\l_1} \geq U_0 (\l_0 - \l_1) > 0,
\]
so \eqref{eq:LS_deriv} implies that $\bar{L}_S''(x) < 0$. From \eqref{eq:LS_deriv}, it is straightforward to show that
\[
   x_c  \bar{L}_S' ( x_c) = 
 \bar{L}_S (x_c) =  \bar{L}_S(x_c) - \bar{L}_S (0).
\]
Therefore, $\cav \bar{L}_S$ is affine over $[0, x_c]$ and agrees with $\bar{L}_S$ over $[x_c,1]$.

\paragraph{Proof of \cref{claim:U}}
  
We have
\[
    \bar{L} (x)
    =
    \begin{cases}
        \bar{L}_S (x) &\text{if}~ x \geq \bar{x}, \\
        \bar{L}_S ( \bar{x}) -U_1 \bar{x}^{\l_0/\l_1} (1 - x/\bar{x})(1 - \bar{x}^{\d / \l_1}) &\text{if}~ x < \bar{x}.
    \end{cases}
\]
By \cref{claim:LS}, the function $\bar{L}_S$ is strictly quasiconcave and has interior maximizer $x_{0,S}$. There are two cases. 

First suppose $\bar{x} \geq x_{0,S}$. In this case, \cref{claim:U} holds with $x_0 = \bar{x}$ since  $\bar{L}$ is affine and strictly increasing over $[0, \bar{x}]$, and $\bar{L}_S$ is strictly decreasing over $[\bar{x}, 1]$. 

Next suppose $\bar{x} < x_{0,S}$. In this case, \cref{claim:U} holds with $x_0 = x_{0,S}$. Clearly, $\bar{L}$ is affine over $[0,\bar{x}]$. Over the interval $[x^\ast, x_0]$, the function $\bar{L}_S$ is strictly concave by \cref{claim:LS}.\footnote{In particular, if $\l_0 > \l_1$, then $x^\ast \geq x_c$ by \eqref{eq:cutoff}.} Over $[\bar{x}, x_0]$, the function $\bar{L}$ coincides with $\bar{L}_S$, so if $\bar{x} < x_0$, then $\bar{L}_S$ is strictly concave over $[\bar{x}, x_0]$. Finally, to show that $\bar{L}$ is concave over $[0,x_0]$, we check that concavity is preserved at the kink. Suppose not. Then the left and right derivatives of $\bar{L}$ at $\bar{x}$ satisfy $\bar{L}' (\bar{x}-) < \bar{L}' (\bar{x} +) =  \bar{L}_S' (\bar{x})$. Over $[x^\ast, \bar{x}]$, the function $\bar{L}$ is affine and
$\bar{L}_S$ is strictly concave, so $\bar{L}(x^\ast) > \bar{L}_S ( x^\ast)$, hence $U_{\SW} ( \bar{s} ; \tau^\ast) < U_{S} ( \tau^\ast)$, contrary to the definitions of $\bar{s}$ and $\tau^\ast$. 

\paragraph{Proof of \cref{claim:I}} 

The definition of $\bar{\h}$ eliminates the first line in \eqref{eq:I_long}. If $\tau^\ast \geq \hat{\tau}$, then $\tau = \hat{\tau}$, so the second and third lines of \eqref{eq:I_long} vanish as well.  If $\tau^\ast < \hat{\tau}$, then $\tau= \tau^\ast$. In this case, it can be checked that the derivative of \eqref{eq:I_long} is zero at $t = \tau$ and is strictly positive over $(\tau, \infty)$.\footnote{In \eqref{eq:I_long}, since $\tau < \hat{\tau}$, the coefficient on $e^{-\l_0 t}$ is negative and the coefficient on $e^{-(\d + \l_0) t}$ is positive. After differentiating, these signs flip, so the derivative becomes positive for $t > \tau$.} In both cases, it can be shown that $I$ is strictly decreasing over $[0, \tau]$. Since $h(t;t) = 0$ for all $t$, the integrand $I$ is differentiable at $t = \tau$, and we have $I'(\tau) = 0$. To prove that $I'(t) < 0$ for $t < \tau$, we equivalently show that  $e^{\l_1 t} I'(t) < 0$ for $t < \tau$. Since $I'(\tau) = 0$, it suffices to prove that $e^{\l_1 t} I'(t)$ is strictly increasing over $[0, \tau]$. For $t < \tau$ we have
\[
   \bigl (e^{\l_1 t} I'(t) \bigr)' =  e^{-( \l_0 - \l_1) t} \h_0 \Brac{U_1 (\d + \l_0) e^{-\d t} (\d + \l_0 - \l_1) - U_0 \l_0 (\l_0 - \l_1)} > 0,
\]
where the inequality holds because $U_1 (\d + \l_0) > U_0 \l_0$ (by \cref{as:sufficient}) and $e^{-\d t} (\d + \l_0 - \l_1) > \l_0 - \l_1$ (since $t < \tau \leq \hat{\tau}$).

\paragraph{Proof of Claim~\ref{res:thresholds}} Define new variables: $\ell_0 = \l_0/\l_1$; $d = \d /\l_1$; and $y_0 = u_0/u_1$. In terms of these variables, Assumptions \ref{as:nontrivial}--\ref{as:sufficient} are jointly equivalent to the inequality $d > y_0 - \ell_0 > 0$. This inequality defines the domain $D^\ast$.

First consider $\tau^\ast$. In \eqref{eq:SW_Bellman_imperfect}, set $s' = \l_1 s$ and $t' = \l_1 t$. Then $\l_1 \tau^\ast$ is the largest time $t'$ such that
\begin{equation} \label{eq:time_prime}
    \max_{s' \in [0, t']} U_{\SW} ( s'/\l_1; t'/\l_1) \leq U_1.
\end{equation}
After some algebra, the inequality \eqref{eq:time_prime} can equivalently be expressed in terms of  $(\ell_0,d,y_0)$ as
\begin{equation} \label{eq:max_new_var}
    \max_{s' \in [0, t']}  \Set{ \frac{y_0}{\ell_0} (1 - e^{-\ell_0 s'}) + e^{-\ell_0 s'} \Paren{1 - e^{-(t' - s')}(1 - e^{-d s'}) } } \leq  1.
\end{equation}
Let $g^\ast (\ell_0, d, y_0)$ be the largest $t'$ for which \eqref{eq:max_new_var} holds. For fixed $t' > 0$ and $s'$ in $(0,t']$ the term in brackets is strictly decreasing in $\ell_0$ and $d$ and strictly increasing in $y_0$.\footnote{For the dependence on $\ell_0$,  observe that $(1 -e^{-\ell_0 s'})/(\ell_0 s')$ is the slope of the secant line to the convex function $\exp$ over the interval $[- \ell_0 s', 0]$.} It follows that $g^\ast$ is continuous.  Moreover, the derivative of the term in brackets, with respect to $s'$, evaluated at $s' = 0$, is 
 $y_0 - \ell_0 - d e^{-t'}$, which is also strictly decreasing in $\ell_0$ and $d$ and strictly increasing in $y_0$. We conclude that $g^\ast (\ell_0, d, y_0)$ is strictly increasing in $\ell_0$ and $d$, and strictly decreasing in $y_0$.\footnote{We must analyze the derivative in order to establish that these comparative statics are strict; see the argument in \cref{ft:deriv}.} Fix $d, y_0 > 0$. For each $t' > 0$, there exists $\ell_0$ in $(0,y_0)$ such that \eqref{eq:max_new_var} holds. Thus, $\lim_{\ell_0 \uparrow y_0} g^\ast ( \ell_0, d, y_0) = \infty$. If $y_0 > d$, then \eqref{eq:max_new_var} fails whenever $\ell_0 < y_0 - d e^{- t'}$, so we have $\lim_{\ell_0 \downarrow (y_0 - d)} g^\ast ( \ell_0, d, y_0) = 0$.

Now consider $\hat{\tau}$. For $\ell_0 > 1$, the definition of $\hat{\tau}$ in \eqref{eq:gamma_ast_0} can be expressed as
\begin{equation} \label{eq:hatg}
  \l_1 \hat{\tau} =  \frac{1}{d} \log \Paren{ 1 + \frac{d}{\ell_0 - 1}}.
\end{equation}
Define $\hat{g} (\ell_0, d)$ to be the right side of \eqref{eq:hatg}. Thus, $\hat{g} (\ell_0, d)$ is strictly decreasing in $\ell_0$ and $d$, and satisfies the claimed limits.\footnote{For the dependence on $d$,  observe that $(\ell_0 - 1) \l_1 \hat{\tau}$ is the slope of the secant line to the $\log$ function over the interval $[1, 1 + d/(\ell_0 - 1)]$.}

\subsection{Verifying the point mass in  Theorem~\ref{res:imperfect_inspections}} \label{sec:am_gm}

Consider \eqref{eq:CS_s} at $s = 0$. Plug in the distribution $F_\pi$ from \eqref{eq:exp_sol} and rearrange to get
\begin{equation} \label{eq:CS_app}
 \pi h(\hat{\tau})  + (1 - \pi) 
   \int_{(\hat{\tau}, \infty)} h(t) \g^\ast e^{-\g^\ast ( t- \hat{\tau})} \de t = 0,
\end{equation}
where the function $h$ is defined by $h(t) =  U_0 (1 - e^{-\l_0 t }) - U_1 ( 1 - e^{- (\d + \l_0) t})$. Note that 
\[
    h'(t) =  e^{- \l_0 t} \Brac{  U_0 \l_0 - U_1 (\d + \l_0)  e^{-\d t}}.
\]

First, we claim that the left side of \eqref{eq:CS_app} is strictly negative at $\pi = 1$. At $\pi = 1$, the left side equals $h(\hat{\tau})$. 
By \eqref{eq:integrand_ineq}, we have $h'(t) > 0$ for $t \geq \hat{\tau}$. By assumption, $\hat{\tau} < \tau^\ast$, so $h(\hat{\tau}) < h(\tau^\ast) \leq 0$, where the last inequality holds because $\tau^\ast$ is feasible in  \eqref{eq:relaxed_full}.

Next, we claim that the left side of \eqref{eq:CS_app} is strictly positive at $\pi = 0$. We have $h(0) = 0$, and we showed above that $h(\hat{\tau}) < 0$. Since $h'$ is strictly single-crossing from below, we conclude that $h(t) < 0$ for all $0 < t < \hat{\tau}$. Therefore, 
\begin{equation*}
\begin{aligned}
    \int_{(\hat{\tau}, \infty)} h(t) \g^\ast e^{-\g^\ast ( t- \hat{\tau})} \de t &= e^{\g^\ast \hat{\tau}} \Brac{ \int_{0}^{\infty} h(t) \g^\ast e^{-\g^\ast t} \de t - \int_{0}^{\hat{\tau}} h(t) \g^\ast e^{-\g^\ast t} \de t } \\
    &> e^{\g^\ast \hat{\tau}}  \int_{0}^{\infty} h(t) \g^\ast e^{-\g^\ast t} \de t  \\
    &= 0, 
\end{aligned}
\end{equation*}
where the last equality follows from the definition of $\g^\ast$.

\subsection{Proof of Lemma~\ref{res:exponential_sol}} \label{sec:proof_exponential_lemma}

Attach a nonnegative multiplier $\h_0$ to the time-$0$ constraint and a nonnegative, integrable density multiplier $\h(s)$ to the time-$s$ constraint, for $s > 0$. The Lagrangian becomes
\begin{multline*}
    L(F;\h_0, \h)
    = 
    \int_{ (0,\infty)}  e^{-\b t} \de F(t)
    - \h_0 \int_{(0,\infty)} (e^{-\a t} - A)  \de F(t) \\
    - \int_{0}^{\infty} \Brac{\int_{(s,\infty)} ( e^{-\a (t - s)} -A ) \de F(t) } \h(s) \de s.
\end{multline*}
Change the order of integration in the double integral to get
\[
   L(F; \h_0, \h) = \int_{(0, \infty)} I(t) \de F(t),
\]
where
\[
    I(t) = e^{-\b t} - \h_0 (e^{-\a t} - A)  - \int_{0}^{t} \h(s) (e^{- \a (t - s)} - A)\de s.
\]
Let $\h(s) = \bar{\h} e^{-\b s}$ for some nonnegative coefficient $\bar{\h}$ to be determined below. Substitute in this expression, integrate, and group like terms to get
\[
    I(t) = e^{- \b t} \Brac{1 -  \bar{\h} \Paren{ \frac{A}{\b} + \frac{1}{\a - \b}}} + e^{- \a t} \Brac{  - \h_0 + \frac{\bar{\h}}{\a - \b} } + \Paren{ \h_0 + \frac{ \bar{\h}}{\b}} A .
\]
To make the bracketed terms vanish, take
\[
    \bar{\h} = \Paren{ \frac{A}{\b} + \frac{1}{\a - \b}}^{-1}, \qquad \h_0 = \frac{1}{\a - \b} \Paren{ \frac{A}{\b} + \frac{1}{\a - \b}}^{-1}.
\]
These multipliers are nonnegative since $\a > \b > 0$ and $A > 0$. With these multipliers, the Lagrangian reduces to a constant. Therefore, a distribution $F$ over $(0, \infty)$ solves \eqref{eq:program_lemma} if and only if $F$ satisfies every inequality constraint in \eqref{eq:program_lemma} with equality.\footnote{Technically, complementary slackness implies equality only for almost every $s  \geq 0$, but it can be shown that this implies equality at every $s \geq 0$.} Since $0 < A < 1$, we can apply \cref{res:uniqueness} from \cref{sec:uniqueness_delaying_perfect} (with coefficient $1/A$ in the integrand) to conclude that the unique solution is the exponential distribution with hazard rate $\g^\ast = \a A / (1 - A)$.

\subsection{Uniqueness lemmas} \label{sec:uniqueness_delaying_perfect}

The proofs of uniqueness rely on the following lemmas. 


\begin{lem}[Unique fixed point---single exponential] \label{res:uniqueness} Fix $A > 1$ and $\a >0$. For each $\pi$ in $[0,1)$, there exists exactly one distribution $F$ on $[0, \infty)$ with $F(0)= \pi$ satisfying
\begin{equation} \label{eq:binding}
    \int_{(s, \infty)} A e^{-\a ( t - s)} \de F(t) = 1 - F(s),
\end{equation}
for all $s \geq 0$. Namely, $F(t) = \pi + (1-\pi)(1 - e^{-\g t})$, for all $t \geq 0$, where $\g = \a / (A - 1)$. 
\end{lem}


\begin{proof}
Let $F$ be a cumulative distribution function on $[0, \infty)$ that satisfies this system. To simplify notation, let $\bar{F} = 1 - F$. 
For each $s \geq 0$, the layer-cake representation and Tonelli's theorem give
\begin{align*}
    \int_{(s,\infty)} A e^{-\a(t-s)} \de F(t)
    &= \int_{(s,\infty)}
       \int_t^\infty A\a e^{-\a(x-s)} \de x\,\de F(t) \\
    &= \int_s^\infty A\a e^{-\a(x-s)}
       [F(x)-F(s)] \de x \\
    &= A\bar{F}(s)
       - \int_s^\infty A\a e^{-\a(t-s)}
       \bar{F}(t) \de t.
\end{align*}
Substitute this equality into \eqref{eq:binding} to get
\begin{equation} \label{eq:AC_equality}
    (A-1)\bar{F}(s)
    =
    \int_s^\infty A\a e^{-\a(t-s)}
    \bar{F}(t) \de t.
\end{equation}
The right side is continuous in $s$, so $\bar{F}$ is continuous. For each $s > 0$, differentiating \eqref{eq:AC_equality} gives
\[
    (A - 1) \bar{F}'(s) = - A \a \bar{F} (s) + \a (A - 1) \bar{F}(s),
\]
which simplifies to $\bar{F}'(s) = - \g \bar{F}(s)$. Since
$\bar{F}(0) = 1- \pi$, we conclude that $\bar{F}(t) = (1 - \pi) e^{-\g t}$ for all $t \geq 0$. It is easily verified that this distribution satisfies the system (since $A \g / (\a + \g) = 1$). 
\end{proof}

\begin{lem}[Unique fixed point---sum of exponentials] \label{res:uniqueness_sum} Fix positive numbers $A,B, \a, \b$ with $A - B = 1$ and $\b B > \a A$.\footnote{Provided that $A > B$, the condition $A - B = 1$ is a normalization. This normalization simplifies the expression for $\g$.} For each $\pi$ in $[0,1]$, there exists exactly one distribution $F$ on $[0, \infty)$ with $F(0)= \pi$ that satisfies
\begin{equation} \label{eq:binding_sum}
    \int_{(s, \infty)} \Brac{ A (1 - e^{-\a ( t- s)})
    -  B (1 - e^{-\b (t -s)})}\de F(t) = 0,
\end{equation}
for all $s \geq 0$. Namely, $F(t) = \pi + (1 - \pi)(1 - e^{-\g t})$, for all $t \geq 0$, where $\g = \a \b / (\b B - \a A)$. 
\end{lem}


\begin{proof} We may assume $\pi < 1$; otherwise the result is trivial. Let $F$ be a cumulative distribution function on $[0, \infty)$ that satisfies this system.  The integrand is continuous in $(s,t)$ and vanishes when $s = t$. Therefore, we can calculate the right-derivative of the left side with respect to $s$ by differentiating under the integral (by dominated convergence) and ignoring the change in the left endpoint. Thus,
\begin{equation} \label{eq:binding_zero}
    \int_{(s, \infty)} \Brac{ - \a A e^{-\a ( t - s)}
    +  \b B e^{-\b (t -s)} }\de F(t) = 0,
\end{equation}
for all $s \geq 0$. Multiply \eqref{eq:binding_sum} by $\b$ and subtract \eqref{eq:binding_zero}. Simplify using the equality $A - B = 1$ to conclude that
\[
   \int_{(s, \infty)}  \frac{(\b - \a)A}{\b} e^{-\a ( t - s)}  \de F(t)  = 1 - F(s),
\]
for all $s \geq 0$. Since $A = B + 1$ and $\b B > \a A$, it follows that $\b A > \a A + \b$. Therefore, $(\b - \a) A/\b > 1$,
so we can apply \cref{res:uniqueness} to complete the proof, noting that
\[
    \g
    =
    \frac{\a}{ (\b - \a) A/\b - 1} = \frac{ \a \b}{\b A - \a A - \b}
    = \frac{\a \b}{\b B - \a A}. \qedhere
\]
\end{proof}

\end{document}